\documentclass{article}

\usepackage[T1]{fontenc}
\usepackage{palatino}
\usepackage{fullpage}
\usepackage{graphicx}
\usepackage{amsmath}
\usepackage{amsfonts}
\usepackage{mathrsfs}
\usepackage{amssymb}
\usepackage[dvipsnames,svgnames,x11names,hyperref]{xcolor}
\definecolor{lkcol}{RGB}{130,70,200}
\usepackage[hidelinks,colorlinks=true,linkcolor=lkcol,citecolor=blue,linktocpage]{hyperref}
\usepackage{cite}
\usepackage{csquotes}
\usepackage{caption}
\usepackage{subcaption}
\usepackage{nicematrix}
\usepackage{mathtools}
\usepackage{tabularray}
\usepackage{float}
\usepackage{amsthm}
\usepackage{dsfont}
\usepackage{calc}
\usepackage{fancyhdr}
\usepackage[boxsize=5pt]{ytableau}
\usepackage{titling}
\usepackage{youngtab}
\usepackage{enumitem}
\usepackage{array,booktabs}
\usepackage{rotating}
\usepackage{longtable}
\usepackage{stmaryrd}

\numberwithin{equation}{section}

\newcommand{\IC}{\mathbb{C}}

\newcommand{\IR}{\mathbb{R}}
\newcommand{\IZ}{\mathbb{Z}}

\newcommand{\CS}{\mathcal{S}}
\newcommand{\xh}{\hat{x}}
\newcommand{\yh}{\hat{y}}
\newcommand{\mq}{\mathfrak{q}}
\newcommand{\FG}{\mathrm{FG}}
\newcommand{\FN}{\mathrm{FN}}

\newcommand{\Sk}{\mathrm{Sk}}

\renewcommand{\flat}{\mathrm{flat}}

\renewcommand{\Re}{{\rm{Re}}}

\def\be{\begin{equation}}
\def\ee{\end{equation}}

\def\IR{{\mathbb{R}}}

\def\IZ{{\mathbb{Z}}}

\def\IC{{\mathbb{C}}}

\def\CA{\mathcal{A}}

\def\CL{\mathcal{L}}

\def\CB{{\mathcal{B}}}
\def\CD{{\mathcal{D}}}

\def\CM{{\mathcal{M}}}
\def\CN{{\mathcal{N}}}

\def\CP{{\mathcal{P}}}
\def\CU{\mathcal{U}}

\def\CW{{\mathcal{W}}}

\def\CW{\mathcal{W}}

\def\tr{{\rm{tr}}}

\def\wr{{\mathfrak{wr}}}

\def\Tr{{\rm{Tr}}}

\def\scL{\mathscr{L}}
\def\scX{\mathscr{X}}

\def\fg{\mathfrak{g}}

\def\fY{\Gamma}

\def\sfD{\mathsf{D}}
\def\sfE{\mathsf{E}}

\def\fOmega{\overline{\underline{\Omega}}}
\newcommand{\SH}{{\mathbf{S}\ddot{\mathbf{H}}}}
\renewcommand{\H}{{{\mathbf{H}}}}

\def\rk{\operatorname{rk}}

\newcommand{\pic}[2]{\raisebox{-.5\height}{\includegraphics[scale=#2]{#1}}}

	\makeatletter 
	\newcommand{\Label}{
		\@Labeli
	}
	\newcommand\@Labeli{\@ifnextchar\endlabels{\@Labelend}{\@Labelii}}
	\newcommand\@Labelii[3]{
		\pinlabel {#1} [tr] at #2 #3
		\@Labeli 
	}
	\newcommand\@Labelend[1]{
	}
	\makeatother

\newcommand\OC{\pic{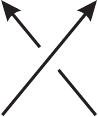}{.50}}
\newcommand\UC{\pic{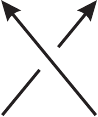} {.50}}
\newcommand\SP{\pic{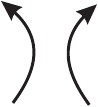} {.50}}
\newcommand\UK{\pic{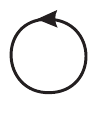} {.50}}
\newcommand\PT{\pic{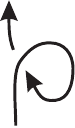} {.50}}
\newcommand\NT{\pic{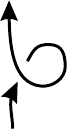} {.50}}
\newcommand\ST{\pic{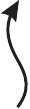} {.50}}

\newcommand\SD{\pic{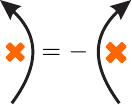} {.55}}
\newcommand\PS{\pic{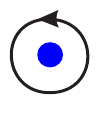} {.50}}

\newtheorem{remark}{Remark}

\usepackage{authblk}

\title{
Spherical DAHA as an algebra of framed BPS states
\\[30pt]
}
\author[1]{Kunal Gupta}
\author[1,2,3]{Pietro Longhi}
\affil[1]{Department of Physics and Astronomy, Uppsala University, Box 516, 751 20 Uppsala, Sweden}
\affil[2]{Department of Mathematics, Uppsala University, Box 480, 751 06 Uppsala, Sweden}
\affil[3]{Centre for Geometry and Physics, Uppsala University, Box 516, 751 20 Uppsala, Sweden}
\date{}                     
\fancypagestyle{firstpage}
{
    \fancyhead[L]{}    
    \fancyhead[R]{UUITP-22/26}
}

\begin{document}

\maketitle
\thispagestyle{firstpage}

\begin{abstract}
Line operators of 4d $\CN=2$ theories of class $\CS$ of type $A_{N-1}$ in a half Omega background provide a physical model for $SL_N$ skein algebras.
In this paper we extend this correspondence to $GL_N$ skein algebras and prove that the $GL_2$ skein algebra of the punctured torus is isomorphic to the $GL_2$ spherical double affine Hecke algebra ${{\mathbf{S}\ddot{\mathbf{H}}}}^{q,t}_2$.
The construction is based on $\mathfrak{q}$-nonabelianization for the $SU(2)$ $\mathcal{N}=2^*$ gauge theory, which realizes the skein algebra as a quantum torus algebra associated with the Seiberg-Witten curve.
Different regions of the Coulomb branch yield distinct presentations of ${{\mathbf{S}\ddot{\mathbf{H}}}}^{q,t}_2$. These range from the Macdonald $q$-difference module in Fenchel-Nielsen (weak coupling) charts to cluster-type realizations in Fock-Goncharov (strong coupling) charts.
Transitions between these descriptions are governed by the vanilla BPS spectrum of the $\mathcal{N}=2^*$ theory via framed wall-crossing, which provides a unified physical framework for several representations of spherical DAHA.

\end{abstract}

\newpage

\tableofcontents

\section{Introduction}

This paper explores a physical realization of spherical Double Affine Hecke Algebras (DAHAs) as skein algebras of surfaces within the context of 4d $\CN=2$ quantum field theory.
Various relations between DAHAs and skein algebras of the punctured torus have been proposed from a mathematical perspective \cite{2017arXiv170806024J, morton2021dahas, gunningham2024skeins}.
The construction developed in this work takes a different approach, based on identifying faithful $q$-difference modules of DAHAs to skein algebras of ramified coverings of the punctured torus defined in physics by Seiberg-Witten curves.
In fact, the approach explored in this work extends directly to the much broader setting of skein algebras of arbitrary Riemann surfaces with punctures, such as genus-two algebras introduced in \cite{Arthamonov:2017oxw, Hikami:2019jaw}. This class of algebras is expected to arise in physics in the setting of 4d $\CN=2$ QFTs defined by a twisted compactification of the 6d $(2,0)$ theory of type $A_{N-1}$ on a Riemann surface $C$, also known as theories of class $\CS$ \cite{Gaiotto:2009we, Gaiotto:2009hg}.

Taking $C$ to be the punctured torus, we construct an explicit isomorphism between the $GL_2$ skein algebra of $C$
and the spherical DAHA of $GL_2$
\be\label{eq:isomorphism-intro}
	\Sk(C\times I, GL_2) \simeq \SH_2^{q,t}\,,
\ee 
with the identification $q^{\frac12} = -\mq^{-1}$, $t = ({\mq}{\mu})^{-1}$.
Here $\mq$ is the quantum parameter for the $GL_2$ skein algebra, and $\mu+\mu^{-1}$ corresponds to the value of a counterclockwise loop around the puncture, 
related to a global symmetry of the class $\CS$ theory. 
Our approach is based on the low energy description of the class $\CS$ theory, which provides a characterization of $\Sk(C\times I, GL_2)$ based on the  $GL_1$ skein algebra of a surface $\Sigma$ of genus 2 with two punctures (the Seiberg-Witten curve).
We show that this matches exactly the $q$-difference module of $\SH_2^{q,t}$ generated by Macdonald difference operators
\be\label{eq:isomorphism-intro-IR}
	\Sk(\Sigma\times I, GL_1) \simeq \text{$q$-difference module of }\SH_2^{q,t}\,.
\ee
The $GL_1$ skein algebra of a genus 2 surface is a quantum torus algebra generated by operators obeying $\hat y_j\hat x_i = q^{\delta_{ij}} \hat x_i\hat y_j$. We establish that the generators of the $GL_2$ skein algebra of the punctured torus along the meridian $m$, longitude $l$, and their twice-around counterparts $m^2, l^2$ take the following form, up to a similarity transformation (see Section \ref{sec:line-ops-DAHA} for details)
\be\label{eq:Mac-From-U-intro}
\begin{split}
	\hat{\CL}^{\FN}_{m}\,  & \sim \xh_1 + \xh_2 ,
    	\\[5pt]
	\hat{\CL}^{\FN}_{l}\, & \sim
	\frac{t \xh_1 - \xh_2}{ \xh_1 - \xh_2} \yh_1
	+\frac{t \xh_2 - \xh_1}{ \xh_2 - \xh_1} \yh_2 ,
\end{split}
\qquad\qquad
\begin{split}
	\frac{\hat{\CL}^{\FN}_{m^2} + q^{\frac{1}{2}} (\hat\CL^{\FN}_m)^2}{q^{-\frac{1}{2}}+q^{\frac{1}{2}}}
	\,
	& \sim \hat{x}_1\hat{x}_2,
	\\
	\frac{\hat{\CL}^{\FN}_{l^2} + q^{-\frac{1}{2}} (\hat\CL^{\FN}_l)^2}{q^{-\frac{1}{2}}+q^{\frac{1}{2}}}
	\,
	& \sim t\,\hat{y}_1\hat{y}_2 .
\end{split}
\ee
After choosing a polarization in which $\hat x_i$ act as multiplication operators, and $\hat y_j$ as $q$-shift operators, the expressions on the right hand side recover the faithful representation of spherical DAHA based on Macdonald $q$-difference operators \cite{cherednik2005double}.
These relations, together with their counterparts for negative modes, establish \eqref{eq:isomorphism-intro}.

In addition to \eqref{eq:Mac-From-U-intro} we also obtain other representations of $\SH_2^{q,t}$ in terms of the quantum torus algebra associated to $\Sigma$, such as
\be\label{eq:Mac-From-U-intro-FG}
\begin{split}
	\hat{\CL}^{\FG}_{m} 
	= \xh_1
    + \xh_2\left(1  - \frac{q}{t} \,\frac{\yh_2}{\yh_1} \right),
    	\\[5pt] 
	\hat{\CL}^{\FG}_{l}
	= 
	\yh_2 + \left(1
    - t \,\frac{\xh_1}{\xh_2} \right) \yh_1,
\end{split}
\qquad\qquad
\begin{split}
	\frac{\hat{\CL}^{\FG}_{m^2} + q^{\frac{1}{2}} (\hat\CL^{\FG}_m)^2}{q^{-\frac{1}{2}}+q^{\frac{1}{2}}}
	= \hat{x}_1 \hat{x}_2,
	\\
	\frac{\hat{\CL}^{\FG}_{l^2} + q^{-\frac{1}{2}} (\hat\CL^{\FG}_l)^2}{q^{-\frac{1}{2}}+q^{\frac{1}{2}}}
	= \hat{y}_1 \hat{y}_2.
\end{split}
\ee
These are closely related to recent characterizations of $GL_2$ spherical DAHA based on Halln\"as-Ruijsenaars eigenfunctions for the quantum open Toda chains \cite{di2024ruijsenaars}.
Remarkably, the low energy physics of 4d $\CN=2$ QFT interpolates between this and the Macdonald presentation in \eqref{eq:Mac-From-U-intro}, providing both an explicit construction of these algebras and a unifying picture for different presentations. 
To explain how physics provides this unifying framework we recall next key facts about the relevant class $\CS$ theories.

When a 4d $\CN=2$ theory is considered on $S^1\times \IR^3$, its Coulomb branch $\CM$ is modeled by the $SL_N$-character variety of $C$ and admits a set of local coordinates defined by expectation values of loop operators wrapping the circle \cite{Drukker:2009tz, Alday:2009fs, Drukker:2009id}.
Our work is concerned with the algebra $\CA$ obtained from quantization of the ring of functions on $\CM$ defined by placing the theory in a half-Omega background \cite{Gaiotto:2010be}. More specifically, we consider a certain \textit{extension} of the physical moduli space $\CM$ corresponding to the $GL_N$ character variety. Quantization then gives rise to the surface skein algebra $\Sk(C\times I, GL_N)$, an extension of the physical algebra of loop operators $\CA$.

Various relationships between skein algebras, DAHA, and character varieties have appeared in the literature. In particular, the isomorphism between the $SL_2$ skein algebra of the punctured torus and spherical DAHA of type $A_1$ \cite{frohman2000skein, bullock2000multiplicative, koornwinder2008zhedanov, terwilliger2011universal} has generated renewed interest in recent years, in connection to positivity properties \cite{Bousseau:2020qgo}, brane quantization \cite{Gukov:2022gei} and alternative mathematical characterizations of $\CM$ \cite{Allegretti:2024svn, Allegretti:2024idu} by Braverman-Finkelberg-Nakajima \cite{Nakajima:2015txa, Braverman:2016wma, Braverman:2016pwk}.
Our result provides an extension of this isomorphism to the $GL_2$ skein algebra on one side and to the $GL_2$ spherical DAHA on the other.

To establish the relation \eqref{eq:isomorphism-intro} we leverage properties of the underlying \textit{physical} theory -- the $\CN=2^*$ theory with gauge group $SU(2)$ \cite{Gaiotto:2009gz}.
A useful characterization of the physical subalgebra $\CA$ is provided by the quantum UV-IR map, which provides a model based on a quantum torus algebra of IR line operators defined by the spectrum of \emph{framed BPS states} \cite{Gaiotto:2010be}.
A priori, the domain of application of the quantum UV-IR map is restricted to the physical algebra of line operators, modeled by the $SL_2$ skein algebra of $C$. 
The extension to $GL_2$ is enabled by the geometric realization of the quantum UV-IR via $\mq$-nonabelianization \cite{Neitzke:2020jik}.
This framework establishes an isomorphism between $\Sk(C\times I, GL_2) $ and the $GL_1$ skein algebra of a ramified 2-fold covering of $C$.
We apply the $\mq$-nonabelianization framework to the Seiberg-Witten curve of $\CN=2^*$ theory, and obtain several models for the left-hand side of \eqref{eq:isomorphism-intro} based on a quantum torus.

More specifically, we consider different physical regions in Coulomb moduli space, corresponding to different shapes of the Seiberg-Witten curve.
Just like ordinary BPS states, the spectrum of framed BPS states (and their extension defined by $\mq$-nonabelianization) exhibits wall-crossing jumps. 
Overall, the regions of moduli space can be classified into two qualitatively distinct types:
\begin{itemize}
\item
In weak coupling regions, denoted \textit{Fenchel-Nielsen} charts, $\mq$-nonabelianization reproduces the well-known presentation of spherical DAHA based on Macdonald difference operators given in \eqref{eq:Mac-From-U-intro}.
\item
In strong coupling regions, denoted \textit{Fock-Goncharov} charts, we obtain quantum torus presentations closely related to the embedding of $\SH_2^{q,t}$ into the quantized coordinate ring of the $GL_2$ character variety of $C$ obtained in \cite{di2024ruijsenaars}. The $\mq$-nonabelianization map constructed in these regions produces \eqref{eq:Mac-From-U-intro-FG}.
\end{itemize}
A key advantage of using the \textit{physical} Seiberg-Witten curve of $SU(2)$ $\CN=2^*$ theory is that it encodes the vanilla BPS spectrum of the 4d theory. In particular, vanilla BPS states control framed wall-crossing phenomena which interpolate between the Macdonald presentation of spherical DAHA in Fenchel-Nielsen charts and the dual presentations in Fock-Goncharov charts.

\paragraph{Acknowledgements}
We thank Misha Bershtein, Tobias Ekholm, David Jordan, Maxim Zabzine, and Yegor Zenkevich for helpful discussions.
The work of KG is supported by the Knut and Alice Wallenberg Foundation grant
KAW2021.0170, and by the Olle Engkvists Stiftelse Grant 2180108.
The work of P.L. is supported by the Knut and Alice Wallenberg Foundation,
KAW2020.0307 Wallenberg Scholar and by the Swedish Research Council, VR 2022-06593, Centre of Excellence in Geometry and Physics at Uppsala University.

\section{Line operators in class $\CS$ theories}\label{sec:}

\subsection{Coulomb branch and moduli spaces of flat connections
}\label{sec:}

The vacuum manifold of a 4d $\CN=2$ QFT on $S^1\times \IR^3$ features a Coulomb branch $\CM$, whose points parameterize expectation values of half-BPS loop operators wrapping $S^1$ \cite{Seiberg:1996nz, Gaiotto:2010be}.
More precisely, loop operators provide local holomorphic coordinates on $\CM$, which is hyperk\"ahler, with respect to a complex structure parameterized by $\zeta\in \IC^*$ whose phase $\theta = \arg \zeta$ encodes the unbroken supercharges.
In complex structures $\zeta=0$ and $\infty$ instead $\CM$ acquires the structures of a Lagrangian torus fibration over the Coulomb branch $\CB$ of the theory on $\mathbb{R}^4$. 
This structure can be deduced from the low-energy description of the 4d QFT on $S^1\times \IR^3$ as a three dimensional $\CN=4$ sigma model with target space $\CM$. From this perspective the base $\CB$ parameterizes vevs of 4d vectormultiplet scalars, while torus fibers parameterize vevs of 3d periodic scalars corresponding to electric and magnetic holonomies \cite{Gaiotto:2010okc}.

In the context of theories of class $\CS$, namely QFTs defined by a twisted reduction of the 6d $(2,0)$ theory of type $\fg$ on a Riemann surface $C$ \cite{Gaiotto:2009we, Gaiotto:2009hg}, $\CM$ coincides with moduli spaces of flat connections on Riemann surfaces
\be\label{eq:UV-IR-isomorphism}
	\CM\simeq \CM^{\text{UV}}_{\flat} \simeq \CM^{\text{IR}}_{\flat}\,. 
\ee
We recall how each of these is defined. Let $G$ be a reductive group and consider a $G$-local system on a surface $S$ of genus $g$ with $n$ punctures, with unconstrained monodromy at punctures. 
The character variety 
\be\label{eq:char-var-def}
	\CM_{G}(S_{g,n}) = \mathrm{Hom}(\pi_1(S_{g,n}),G) \sslash G\,.
\ee  
has generic dimension
\be\label{eq:M-char-dim}
	\dim_{\IC}\CM_{G}(S_{g,n})= (2g+n-2) \dim G + \dim Z(G)
\ee
where $Z(G)$ denotes the center of $G$. 
The UV moduli space is then a subspace of the $GL_{N}$ character variety of the UV curve $C$
\be\label{eq:M-UV-inclusion}
	\CM^{\text{UV}}_{\flat} \subseteq\CM_{GL_N}(C)\,.
\ee
The IR moduli space is instead a certain subspace of a $GL_1$ character variety
\be\label{eq:GL1-char-var}
	\CM^{\text{IR}}_{\flat} \subseteq \tilde \CM_{GL_1}(\Sigma) 
	= \mathrm{Hom}(H_1(\Sigma',\IZ),\IC^*) \sslash \IC^*\,.
\ee
Here $\Sigma' := \Sigma\setminus (\text{ramification locus})$ denotes the Seiberg-Witten curve viewed as a ramified covering of $C$ (the covering map is part of the definition for theories of class $\CS$), and with the ramification locus removed. Moreover, the $\mathrm{Hom}$ map is restricted to have holonomy $-1$ around each point in the ramification locus, see \cite{Hollands:2013qza}.
If $\Sigma$ has genus $g$ and $n$ punctures, then \eqref{eq:M-char-dim} implies $\dim \tilde \CM_{GL_1}(\Sigma) =2g+n-1$.
The subspace $\CM^{\text{IR}}_{\flat}$ is defined by restricting \eqref{eq:GL1-char-var} to homomorphisms from a certain \textit{physical sublattice} $\fY \subseteq H_1(\Sigma',\IZ)$ of rank equal to the dimension of the UV moduli space
\be\label{eq:IR-moduli-dimension}
	\rk \fY \equiv \dim \CM^{\text{IR}}_{\flat}  =\dim \CM^{\text{UV}}_{\flat} \,.
\ee
Physically $\rk \fY = 2r+f$ where $r$ is the rank of the physical Coulomb branch and $f$ is the rank of the flavour symmetry group of the QFT.

\subsection{The UV-IR map defined by framed BPS states}\label{sec:UV-IR-framed}
The isomorphism \eqref{eq:UV-IR-isomorphism} between UV and IR moduli spaces can be made explicit by choosing appropriate coordinates on both $\CM_{\flat}^{\text{UV/IR}}$, namely expectation values of UV and IR loop operators.
We recall that
\begin{itemize}
\item
A 4d $\CN=2$ QFT of class $\CS$ is equipped with a set of UV line operators whose classification is discussed in \cite{Kapustin:2005py, Kapustin:2006pk, Kapustin:2006hi, Gaiotto:2010be, Aharony:2013hda} for theories with a Lagrangian description, and in  \cite{Drukker:2009tz, Xie:2013vfa, Tachikawa:2013hya, Coman:2015lna, Bhardwaj:2021pfz} for non-Lagrangian QFTs. 
For the purposes of this work, a UV line operator $\scL_a$ will be labeled by elements of the \textit{skein algebra} of the UV curve $C$, see below for its definition. 
\item
In a generic Coulomb vacuum the 4d QFT flows to an abelian $U(1)^r$ gauge theory with global flavor symmetry $U(1)^f$.
IR line operators $\scX_\eta$ are therefore graded by electromagnetic and flavor charges valued in the physical lattice $\eta\in \fY\simeq \IZ^{2r+f}$.
\end{itemize}

Expectation values of these operators, taken to wrap the compactification circle, provide local coordinates on the UV and IR moduli spaces
\be\label{eq:coordinate-systems}
\begin{split}
	&\langle \scL_a\rangle_{S^1}  \ : \quad \ \text{local coordinates on }\CM_{\flat}^{\text{UV}} \\
	&\langle \scX_\eta\rangle_{S^1} \ : \quad \ \text{local coordinates on }\CM_{\flat}^{\text{IR}} \\
\end{split}
\ee
In terms of these, the isomorphism $\CM_{\flat}^{\text{UV}}\simeq \CM_{\flat}^{\text{IR}}$ 
takes the following remarkable form
\be\label{eq:UV-IR-relation-classical}
	\langle \scL_a\rangle_{S^1} = \sum_{\eta} \fOmega(a,\eta,\theta) \, \langle \scX_\eta\rangle_{S^1}\,.
\ee
Here $\fOmega(a,\eta,\theta)$ are integer-valued coefficients which correspond, in QFT, to protected indices of \textit{framed} BPS states, see \cite{Gaiotto:2010be}. 
We also recall the meaning of $\theta$: this is a phase that encodes the half-BPS supersymmetry algebra preserved by $\scL_a$ in the UV and by $\scX_\eta$ in the IR.

Perhaps the most important property of the UV-IR relation \eqref{eq:UV-IR-relation-classical} is how it depends on $\theta= \arg \zeta$. Let us fix a point in $\CM$.
While $\langle \scL_a\rangle_{S^1}$ depend smoothly (holomorphically) on $\zeta$, both $\fOmega(a,\eta,\theta)$ and $\langle \scX_\eta\rangle_{S^1}$ exhibit mutually compensating jumps at values of $\theta$ that correspond to central charges of `vanilla' BPS states of the theory, see \cite[Section 6]{Gaiotto:2010be}.  This phenomenon is known as `framed' wall-crossing, and one of its main applications is to compute vanilla BPS states by using UV loop operators as probes.\footnote{
More generally, moving around in $\CM$ (and specifically over its base $\CB$, the 4d Coulomb branch), both $\fOmega(a,\eta,\theta)$ and $X_{\eta}$ can jump by the same framed wall-crossing phenomenon: $\fOmega$ are piecewise constant while $X_\eta$ are piecewise holomorphic.
}

\subsection{Spectral networks realization of the UV-IR map}\label{sec:spectral-networks-and-nonabelianization}

Computing the UV-IR map \eqref{eq:UV-IR-relation-classical} directly in QFT is, in general, challenging because it involves determining protected indices $\fOmega(a,\eta,\theta)$ of framed BPS states, see \cite{Moore:2015szp, Moore:2015qyu, Brennan:2018ura} for a discussion.
In the case of class $\CS$ theories this problem can be bypassed by means of the geometric characterizations \eqref{eq:coordinate-systems} of $\langle \scL_a\rangle_{S^1}$ and $\langle \scX_\eta\rangle_{S^1}$ as local coordinates on moduli spaces of flat connections on the UV and IR curves $C$ and $\Sigma$, respectively.

Recall that the Seiberg-Witten curve $\Sigma$ of a class $\CS$ theory coincides with the Hitchin spectral curve, which is a ramified covering of the UV curve $C$ embedded in the cotangent bundle.
In this work we will be concerned with ramified coverings of degree $2$
\be
	\Sigma = \{\lambda^2 - \phi_2 =0\} \subset T^*C
\ee
where the Seiberg-Witten differential $\lambda = y\, dz$ coincides with the Liouville 1-form and $\phi_2\in \CB$ a quadratic differential corresponding to a choice of Coulomb vacuum.

A choice of phase $\theta$ defines a foliation of $C$ determined by the geometry of $\Sigma$ via 
\begin{equation}
    \operatorname{Im}(e^{-i\theta}\sqrt{\phi_2}) = 0,
\end{equation}
Leaves can be given both an orientation {and} a labeling $(+-)$ or $(-+)$, modulo the identification under simultaneous reversal of orientation and switch of labeling. 
Denoting by $t\in \mathbb{R}$ a local parameter along a leaf, if $\Re(e^{-i\theta}(y_+-y_-)\frac{dz}{dt}) > 0$, then the leaf is directed along $\arg \frac{dz}{dt}$ and has label $(+-)$ (or is directed along $\arg -\frac{dz}{dt}$ and has label $(-+)$).

\begin{figure}[h!]
    \centering
        \includegraphics[width=0.3\textwidth]{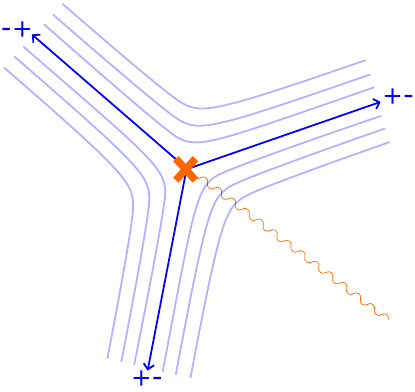}
        \caption{Thin blue lines denote generic leaves of a foliation defined by a quadratic differential in the neighbourhood of a branch point. Thickened lines denote directed critical leaves, i.e. trajectories of the spectral network.}
        \label{fig:bpleaves}
\end{figure}

The spectral network $\CW(\theta)$ is the set of leaves that start or end on a branch point, and we will refer to these simply as \emph{trajectories}.
Branch points correspond to zeroes of $\phi_2$, the behaviour of the foliation and of trajectories near a branch point is shown in Figure \ref{fig:bpleaves}. For generic values of $\theta$ all trajectories end at punctures on $C$.

A spectral network determines an explicit relation between $GL_N$ local systems on $C$ and almost-flat $GL_1$ local systems on an $N$-fold covering $\Sigma$, known as the \textit{nonabelianization map}.\footnote{\label{foot:gl1-conventions}
We remark that there are two conventions, depending on what type of abelian connections one works with on the IR curve $\Sigma$. 
For the purposes of this paper we will work with \textit{almost} flat connections \cite{Hollands:2013qza}, see the definition in \eqref{eq:GL1-char-var}. There is also a formulation based on \textit{twisted} flat connections, see the original work \cite{Gaiotto:2012rg}.
Holonomies of the former obey the untwisted product rule \eqref{eq:classical-IR-algebra}, while holonomies of the latter obey a twisted product rule, see \cite[Section 6.3]{Gaiotto:2012rg}.
The physical interpretation of both types of variables is discussed in \cite[Section 6.4]{Gaiotto:2010be}, where these are identified with vevs of IR loop operators on $S^1\times \mathbb{R}^3$ with different $\mathfrak{su}(2)){
text{spin}}\oplus \mathfrak{su}(2)_R$ twists along the circle.
}
We sketch the construction of this map for the case of interest to us ($N=2$), and refer to the original references \cite{Gaiotto:2012rg, Hollands:2013qza} for further details.

\begin{figure}[h!]
    \centering
        \includegraphics[width=0.8\textwidth]{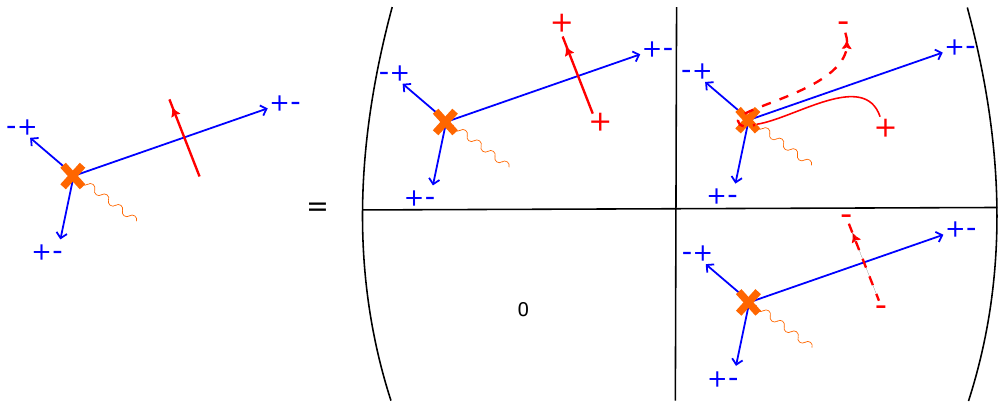}
        \caption{The solid red line on the left hand side shows the path $\varrho$, the solid/dashed line on the right hand side denotes the detour paths on the $+/-$ branch that we get after breaking $\varrho$ and concatenating with the lifts of the spectral network. $F_{-+}(\varrho)$ is zero because there is no detour path that starts from $-$ branch and ends at $+$ branch.}
        \label{fig:detourexample}
\end{figure}

To an \textit{open} directed path $\varrho$ on $C$, nonabelianization associates a $2\times 2$ matrix $F(\varrho)$ that can be regarded as the parallel transport of a $GL_2$ flat connection on $C$ in some choice of gauge.
Each entry $F_{ij}(\varrho)$ of this matrix is computed by taking a sum of all $GL_1$ parallel transports $X_\xi$ along paths $\xi$ on the cover $\Sigma$ that start on sheet $i$ and end on sheet $j$.
Each path $\xi$ is constructed from $\varrho$ by breaking the latter at points where it intersects $\CW(\theta)$ and concatenating lifts of $\varrho$ with oriented lifts of $\CW(\theta)$ to sheets of  $\Sigma$.
Each such concatenation is also known as a \textit{detour path} of $\varrho$, and comes with a certain integer-valued weight, see Figure \ref{fig:detourexample} for an example.

This definition of $F(\varrho)$ applies also to closed directed paths on $C$ with a basepoint, viewed as open paths.

To a \textit{closed} path $\varrho_a$ (without basepoint) one may associate gauge invariant functions of $F$, such as the trace coordinates 
\be
	\CL_a := \Tr F(\varrho_a)\,.
\ee
By the definition of $\CL_a$ this trace is a generating series of \textit{closed} paths on $\Sigma$, and we will denote by $X_\gamma$ the $GL_1$ holonomy along a closed path in homology class $\gamma\in H_1(\Sigma', \IZ)$. 
Nonabelianization therefore relates \textit{trace coordinates} of $GL_2$ local systems on $C$ to formal generating series of $GL_1$ holonomies on $\Sigma'$, establishing an isomorphism 
\be\label{eq:nonabelianization-isomrphism-GL2}
	\CM_{GL_2}(C) \simeq \tilde\CM_{GL_1}(\Sigma)\,.
\ee
See \eqref{eq:GL1-char-var} for the definition of the right hand side.

The application of this map to physics comes through the geometric dictionary for line operators of class $\CS$ theories.
A closed path on $\Sigma$ in homology class $\eta\in \fY$ classifies the charge of an IR line operator $\scX_\eta$, and the its $GL_1$ holonomy coincides with the normalized expectation value.
Similarly a closed path $\varrho_a\subset C$ defines a UV line operator $\scL_a$, and its $GL_2$ trace coordinate coincide with the normalized expectation value, see footnote \ref{foot:gl1-conventions} and \cite{Drukker:2009tz, Alday:2009fs, Drukker:2009id}
\be\label{eq:UV-IR-nonabelianization}
	X_\eta = \langle \scX_\eta\rangle_{S^1} ,\qquad
	\CL_a = \langle \scL_{a}\rangle_{S^1}\,.
\ee
Combining with \eqref{eq:nonabelianization-classical}, this gives an explicit realization of the relation \eqref{eq:UV-IR-relation-classical} where framed BPS indices are identified with the expansion coefficients of the sum over closed cycles
\be\label{eq:nonabelianization-classical}
	\CL_{a} = \sum_\gamma \fOmega(a,\gamma,\theta) \, X_{\gamma}\,.
\ee

\begin{remark}[Nonabelianization vs UV-IR map]\label{rmk:physical-holonomies}
Nonabelianization relates $GL_2$ local systems on $C$ to almost-flat $GL_1$ local systems on $\Sigma$. 
In particular the latter are parameterized by holonomies along generators for the whole homology lattice $H_1(\Sigma',\IZ)$.

The application of nonabelianization to the computation of the physical UV-IR map \eqref{eq:UV-IR-relation-classical} involves a restriction to the $SL_2$ character variety, as this is the relevant moduli space for a class $\CS$ theory of type $A_1$. 
On the IR side, this is corresponds to a restriction from $ \tilde \CM_{GL_1}(\Sigma)$ to the subspace $\CM_{\flat}^{\mathrm{IR}}$ locally parameterized by holonomies $X_\eta$ associated to generators of the physical charge lattice $\eta\in \fY$.
\end{remark}

One way to think about nonabelianization is that \eqref{eq:nonabelianization-classical} provides a cluster chart on $\CM_{GL_2}(C)$ with coordinate functions $X_\gamma\in \IC^*$ the $GL_1$ holonomies. 
Specifically, a point $\nu \in \tilde \CM_{GL_1}(\Sigma)$ with coordinates $\{X_\gamma(\nu)\}_\gamma$
is mapped to a point  $\mu \in \CM_{GL_2}(C)$ with coordinates $\{\CL_a(\mu)\}_a$ determined by $X_\gamma(\nu)$ via \eqref{eq:nonabelianization-classical}.
This map is invertible, and therefore one may regard it as assigning $\IC^*$-valued coordinates $X_{\eta}$ to a point $\mu\in \CM_{GL_2}(C)$. 
However, an important subtlety arises at this point.

\begin{remark}[Physical vs formal variables and active vs passive changes of coordinates]
The physical vevs $\langle \scX_\eta\rangle$ (respectively $\langle \scL_\eta\rangle$) are holomorphic functions of $\zeta$ (piecewise holomorphic in the case of the former). 
For each $\zeta\in \IC^*$ they parameterize a certain family of $GL_1$ local systems on $\Sigma$ (respectively, $GL_2$ local systems on $C$). 
In contrast, $X_\eta$ and $\CL_a$ are independent of $\zeta$. 
Thus, the identification \eqref{eq:UV-IR-nonabelianization} should really be understood to hold at some reference value of $\zeta = \zeta_{*}$, that we are free to fix. 
However, coefficients $\fOmega$ in \eqref{eq:UV-IR-relation-classical} and \eqref{eq:nonabelianization-classical} are truly the same and feature the same jumps at certain values of the phase $\theta$.
This means that a point $\in \tilde \CM_{GL_1}(\Sigma)$ with coordinates $X_{\gamma}$ is mapped by \eqref{eq:nonabelianization-classical} to a point $\mu\in \CM_{GL_2}(C)$ with coordinates $\CL_a$ determined by $\fOmega(a,\eta,\theta)$ for some value of $\theta$, but the same $\nu$ is mapped by \eqref{eq:nonabelianization-classical} to a different point $\mu'\in \CM_{GL_2}(C)$ with coordinates $\CL_a'$ determined by $\fOmega(a,\eta,\theta')$ at another value of the phase.
Compared with the behavior of the physical UV-IR relation \eqref{eq:UV-IR-relation-classical}, we can regard this as switching from a passive coordinate transformation to an active one. This matches the conventions of spectral networks from \cite{Gaiotto:2012rg}, as we show next.
\end{remark}

Since the integer-valued coefficients only depend on the topological type of the spectral network $\CW(\theta)$, different topological types correspond to different coordinate charts and produce different expressions for coordinates on $\CM_{GL_2}(C)$
\be
	\CL_{a} = \sum_{\gamma} \fOmega(a,\gamma,\theta) \, X_{\gamma} \,,
	\qquad
	\CL_{a}' = \sum_{\gamma} \fOmega(a,\gamma,\theta') \, X_{\gamma}\,.
\ee
It is shown in \cite{Gaiotto:2012rg} that any two such expansions are related by a Kontsevich-Soibelman transformation on the spectral coordinates
\be
	\CL_{a}'  = \prod_{\eta}^{\curvearrowleft} \mathcal{K}_{\eta}^{\Omega(\eta)} \, \CL_{a}\,.
\ee
Here $\curvearrowleft$ denotes that the product over homology cycles $\eta\in \fY$  is ordered by increasing values of $\arg Z_\eta$ to the left, $\Omega(\eta)\in \IZ$ is a given data corresponding to the BPS index\footnote{For this reason Kontsevich-Soibelman transformations are associated with physical charges $\eta\in \fY$.
The quadratic refinement $\sigma(\eta)$ is equal to $-1$ for hypermultiplets and $+1$ for vector multiplets and it enters because we are working with the untwisted product, see Eq \eqref{eq:classical-IR-algebra} and \cite{Gaiotto:2009hg}.}
 (defined by a choice of $\Sigma$), and $\mathcal{K}_\eta$ acts on $X_\gamma$ as the change of coordinates \cite{Kontsevich:2008fj}
\begin{equation}\label{eq:KS-transformation}
    \mathcal{K}_\eta X_{\gamma} = (1-
    \sigma(\eta)X_\eta)^{\langle\eta,\gamma\rangle}X_{\gamma}\,.
\end{equation}
In the case of degree-2 coverings there are two kinds of charts, defined respectively by Fock-Goncharov networks which are dual to WKB triangulations \cite{Gaiotto:2009hg}, and by Fenchel-Nielsen networks which are dual to pants decomposition of $C$ \cite{Hollands:2013qza}. Coordinate transformations between different types will be discussed below.

\section{Algebras of line operators}\label{sec:algebra}

In this section we review algebras of UV and IR loop operators in QFT, their geometric realization through the quantization of character varieties as skein algebras of surfaces, and the $q$-nonabelianization map relating UV and IR skein algebras.

\subsection{Quantized algebra of line operators}\label{sec:deformation-quantization}

Correlation functions of half-BPS loop operators are independent of the position of insertions, because (broken) translation generators of the 3d Poincar\'e algebra are $Q$-exact with respect to unbroken supercharges, see \cite{Kapustin:2006hi, Kapustin:2007wm}.
In \cite{Gaiotto:2010be} this observation was leveraged to introduce an algebraic structure on the set of UV line operators
of a 4d $\CN=2$ QFT.
Its definition is based on the relation between spaces of framed BPS states (whose Euler characteristics are the framed BPS indices $\fOmega$ in \eqref{eq:UV-IR-relation-classical}) of single line operators $\scL_a, \scL_b$ and that of their simultaneous insertion $\scL_a\circ \scL_b$.\footnote{
Note that this product should not be confused with the OPE product, as remarked in \cite[Section 3.6]{Gaiotto:2010be}. We also remark that another definition of algebraic structures on loop operators arises from the circle compactification. 
In the three dimensional $\mathcal{N}=4$ sigma model with $\CM$ as the target space, these loop operators are interpreted as local operators with an associated ring structure \cite{Cremonesi:2013lqa,Nakajima:2015txa, Braverman:2016wma}.
}

This algebraic structure can be conveniently expressed at the level of generating series of framed BPS indices, 
which is precisely the output of the UV-IR map \eqref{eq:UV-IR-relation-classical}.
First observe that Gauss' law of charge conservation enforces a simple algebraic structure on the set of IR loop operators, namely $\scX_\eta\cdot \scX_{\eta'}\sim \scX_{\eta+\eta'}$. 
At the level of expectation values this is encoded by the product of $GL_1$ holonomies
\be\label{eq:classical-IR-algebra}
	X_{\eta} X_{\eta'} =  X_{\eta+\eta'}
\ee
Using this relation, the product of two UV line operators is defined by taking the product of their expectation values
\be\label{eq:classical-UV-algebra}
	\CL_{a}\cdot \CL_{b} = \sum_{c} f_{ab}^c \, \CL_{c}\,.
\ee
Here the fusion coefficients $f_{ab}^c = f_{ba}^c$ are defined by the decomposition of the product into a choice of basis for the space of line operators.
In practice these are determined by the framed BPS indices $\fOmega(a,\eta,\theta)$ that appear in \eqref{eq:UV-IR-relation-classical}.

We will be concerned with a certain noncommutative deformation of the algebraic structures \eqref{eq:classical-IR-algebra} and \eqref{eq:classical-UV-algebra} discussed in \cite{Gaiotto:2010be}.
In the context of 4d $\CN=2$ QFT these arise by turning on a half-Omega background, as this effectively forces operators to lie along a common line $\IR\subset\IR^3$, and this gives a notion of ordering. 
\be\label{eq:deformation-quantization}
\begin{split}
	X_{\eta}\ \mapsto\ \hat X_{\eta} & \ \in\  \CA^{\text{IR}}
	\\
	\scL_a \ \mapsto \ \hat\scL_a & \ \in\  \CA^{\text{UV}}
\end{split}
\ee
Geometrically these algebras arise from deformation quantization of coordinate rings on $\CM$, respectively viewed as $\CM_\flat^{\text{IR/UV}}$. 
Recall that $\CM \simeq \CM_{\flat}^{\text{IR}}$ is hyperk\"ahler and that it admits a $\IC^*$ family of Poisson structures 
which is inherited (via embedding) by the Poisson structure on $ \tilde \CM_{GL_1}(\Sigma_{g,n})$
\be\label{eq:Poisson-structure}
	\{\log X_\gamma,\log X_{\gamma'}\} = \langle\gamma,\gamma'\rangle\,,
\ee
where $ \langle\gamma,\gamma'\rangle$ denotes intersection pairing on $H_1(\Sigma',\IZ)$.
Deformation quantization then gives rise to the quantum torus algebra\footnote{The variable $\mq$ used in our work coincides with $-y$ from \cite{Gaiotto:2010be}. For this reason we write the Protected Spin Character in \eqref{eq:UV-IR-relation-quantum} and later as a function of $-\mq$.}
\be\label{eq:quantum-IR-algebra}
	\hat X_{\eta} \cdot \hat X_{\eta'} = (-\mq)^{\langle\eta,\eta'\rangle} \hat X_{\eta+\eta'} ,
\ee
Note that the classical limit \eqref{eq:classical-IR-algebra} corresponds to the specialization
\be\label{eq:q-classical}
	\mq\to -1\,.
\ee

The deformed algebra of UV line operators involves the spaces of framed BPS states.
It is argued in \cite{Gaiotto:2010be} that the UV-IR relation \eqref{eq:UV-IR-relation-classical} generalizes in presence of the half-Omega background to a relation between generators of $\CA^{\text{UV}}$ and $\CA^{\text{IR}}$
\be\label{eq:UV-IR-relation-quantum}
	\hat\CL_a = \sum_{\eta} \fOmega(a,\eta,\theta;-\mq)\, \hat X_{\eta}\,.
\ee
Here  $\fOmega(a,\eta,\theta;-\mq)\in \IZ[\mq^{\pm 1}]$ are the \textit{framed protected spin characters} and decompose into linear combinations of $\mathfrak{su}(2)$ characters, under an assumption known as the `no-exotics' condition \cite{Gaiotto:2010be}.
This expansion of UV line operators $\hat\CL_a$ together with the quantum torus algebra \eqref{eq:quantum-IR-algebra} induces a deformation of \eqref{eq:classical-UV-algebra}
\be\label{eq:quantum-UV-algebra}
	\hat\CL_{a}\cdot \hat\CL_{b} = \sum_{c} f_{ab}^c(\mq) \, \hat\CL_{c},
\ee
whose structure coefficients $f_{ab}^c(\mq)$ are determined by the Laurent polynomials $\fOmega(a,\eta,\theta;-\mq)$.

\subsection{Twisted skein algebras}\label{sec:skein-algebras}

As mentioned earlier, the geometric counterparts of the algebraic structures $\CA^{\text{UV/IR}}$ correspond to quantization of character varieties of Riemann surfaces. 
A useful framework to discuss the emerging algebraic structures is provided by the setting of skein algebras of surfaces, associated respectively to the UV and IR curves $C$ and $\Sigma$.  We recall key definitions.

The HOMFLYPT skein module $\Sk(M)$ of a three-manifold $M$ is the set of equivalence classes of links under framed isotopy, modulo the local relations\footnote{Skein modules of 3-manifolds arise in physics in the context of Chern-Simons theory as the space of inequivalent Wilson line configurations \cite{Witten:1988hf}.
Relations \eqref{eq:skein-rel-HOMFLY} arise from quantization of $U(N)_k$ Chern-Simons theory, with specializations $\mq=e^{2\pi i / (k+N)}$ and $a=\mq^{N}$.
More precisely, this gives rise to the $GL_N$-skein module $\Sk(M,GL_N)$.}
\be\label{eq:skein-rel-HOMFLY}
\begin{split}
	\OC - \UC &=(\mq-\mq^{-1})\ \SP ,\\[2mm]
	 \UK = \frac{a-a^{-1}}{\mq-\mq^{-1}}\,, \qquad
	\PT&=a \ \ST\ ,\qquad \NT =a^{-1}\ \ST\, .
\end{split}
\ee
The $GL_N$ skein module corresponds to a specialization to $a=\mq^N$ together with additional relations, see~\cite{gunningham2023finiteness}.
In particular the $GL_1$ skein module $\Sk(M,GL_1)$ can be defined via the following relations
\be\label{eq:skein-rel-gl1}
\begin{split}
	\mq^{-1} \,\OC  &=  \SP = \mq \, \UC \,,
	\qquad
	 \UK = 1\,.
\end{split}
\ee
Indeed, these relations imply both the first relation in \eqref{eq:skein-rel-HOMFLY} as well as
\be
	\PT=\mq \ \ST\ , \qquad \NT =\mq^{-1}\ \ST\,.
\ee
Similarly $SL_N$ and $GL_N$ skein modules can be described as quotients of $\Sk(M)$ after specializing $a=\mq^{N}$.

When the 3-manifold is a product of a surface times an interval $S\times I$ the skein module acquires a natural algebraic structure, known as the skein algebra of the surface $S$. 
The product of any two elements $\psi, \psi' \in \Sk(S\times I)$ is defined by stacking $\psi'$ above $\psi$ along the interval.
The algebras of UV and IR loop operators \eqref{eq:deformation-quantization} have a geometric presentation (in the case of class $\CS$ theories of type $A_1$, which is the setting we mainly study) as $SL_2$ and $GL_1$ skein algebras of the UV and IR curves
\be\label{eq:skein-alg-table}
\begin{split}
	\CA^{\text{UV}} \ &\subset\  \Sk(C \times I,GL_2)  \\
	\CA^{\text{IR}} \ &\subset\  \Sk(\Sigma \times I,GL_1)  \\
\end{split}
\ee
There are two additional technical points concerning the definition of the right hand sides of \eqref{eq:skein-alg-table} \cite{Neitzke:2020jik}:
\begin{itemize}
\item
To quantize the moduli space $\tilde \CM_{GL_1}(\Sigma)$ in \eqref{eq:GL1-char-var}, 
the $GL_1$ skein module of $\Sigma\times I$ needs to be enriched with \textit{sign defects} inserted at the ramification locus of $\Sigma$ and extending along $I$. Across a sign defect we impose the additional skein relation 
\be\nonumber
	\SD
\ee
\item
The product by stacking along the $I$ direction is twisted by a sign. Given two links $L, L' \subset C\times I$, let $\epsilon(L, L')=\pm1$ denote the number of intersections of their projections to $C$, mod 2. The product on $\Sk(C\times I, GL_2)$ is twisted by $\epsilon(L, L')$.
Similarly, given links $\tilde L, \tilde L'$ in $\Sigma\times I$ we denote by $\pi(\tilde L), \pi(\tilde L')$ their projections to $C\times I$. The product on $\Sk(\Sigma\times I, GL_1)$ is twisted by $\epsilon(\pi(\tilde L), \pi(\tilde L'))$.
\end{itemize}
With the definitions of twisted skein algebras in place, we can understand the inclusions in \eqref{eq:skein-alg-table} as the restriction to $SL_2$ skein algebra for $\CA^{\text{UV}}$, and to skein elements in homology classes that belong to $\fY$ for $\CA^{\text{IR}}$.

More precisely there is a canonical isomorphism between the restriction of the $GL_1$ skein algebra of $\Sigma$ 
and the quantum torus algebra \eqref{eq:quantum-IR-algebra}, see \cite[Section 3]{Neitzke:2020jik}
\be\label{eq:skein-q-char-isomorph}
\begin{split}
	\iota: \ \Sk(\Sigma\times I, GL_1) \quad & \quad \mathop{\longrightarrow}^{\sim}\qquad \CA^{\text{IR}} \\
	[\tilde L] \qquad & \ \quad\  {\mapsto}\quad \ (-1)^{n'(\tilde L)} \mq^{\wr(\tilde L)} \hat X_{\gamma(\tilde L)} \\
\end{split}
\ee
where $\tilde L$ is a path in $\Sigma'\times I$, $[\tilde L]$ denote its equivalence class in the $GL_1$ skein module, $\gamma(\tilde L)$ denotes its homology class, $\wr$ is the \textit{writhe} (the signed number of self-crossings of the projection to $\Sigma'$) and $n'$ is the number of non-local crossings (crossings of the projection to $C$ that are not crossings on $\Sigma$).

The detailed structure of skein algebras varies greatly with the topology of the associated surface. In this work we will be mainly concerned with the $GL_2$ skein algebra of the punctured torus,\footnote{
The HOMFLYPT skein algebra of the punctured torus is discussed in \cite{morton2021dahas}. See also \cite{morton2017homflypt, Ekholm:2024ceb} for useful background and relations to physics.}
and for this purpose we will need to introduce additional skein relations. 
In particular we will identify a loop around the puncture with a certain extension of the ground ring 
\be\label{eq:GL2-puncture-skein}
	\PS = \mu+\mu^{-1}
\ee
Similarly, we will also consider the $GL_1$ of a genus-2 surface with two punctures, viewed as a double covering of the puncture torus. In this case we will impose the relations
\be\label{eq:GL1-puncture-skein}
	\PS = \mu^{\pm 1}
\ee
with positive sign on sheet $1$ and negative sign on sheet 2.


\subsection{$\mq$-nonabelianization construction of the quantum UV-IR map}\label{sec:q-nonabelianization}

The $\mq$-nonabelianization of promotes the isomorphism \eqref{eq:nonabelianization-isomrphism-GL2} between character varieties to an isomorphism of twisted skein algebras
\be\label{eq:skein-algebra-isomorphism}
	\Sk(C \times I,GL_2) \simeq \Sk(\Sigma \times I,GL_1)   \,.
\ee
In particular, through the restriction \eqref{eq:skein-alg-table} this descends to a relation between physical UV and IR line operators, providing a geometric construction of the quantum UV-IR map \eqref{eq:UV-IR-relation-quantum}.

In this section we review the definition of $\mq$-nonabelianization map following \cite{Neitzke:2020jik} .
See also \cite{Galakhov:2014xba, Gabella:2016zxu, Neitzke:2021gxr, Ekholm:2025anq, Hollands:2026rhz} for related works and applications.

\paragraph{Loops with a basepoint as 3d loops.}
The first and most fundamental difference between classical and quantum nonabelianization is that the latter takes place in one dimension higher. Instead of paths on the UV curve $C$, we need to consider elements of $\Sk(C\times I,GL_2)$. 
For bookkeeping purposes, we will represent a closed oriented path $\varrho\subset C\times I$ by an oriented path in $C$, together with a convention for encoding its uplift to $C\times I$.\footnote{The resulting uplift will not be unique, but its equivalence class is such that the rules of $q$-nonabelianization are uniquely specified.} 
For this purpose we endow $\varrho$ with a choice of \textit{basepoint} $\bullet\in \varrho$, and consider a lift of $\varrho$ that decreases monotonically along the direction specified by its orientation. To close up the path we include a vertically upwards strand at $\bullet$. 
This choice of conventions does not capture the most general type of links in $C\times I$, but it captures the kind or oriented loops that will concern us in this work. It has the advantage that the rules of $q$-nonabelianization simplify somewhat for this class of loops.

\paragraph{Lifts of $\varrho$ to $\Sigma\times I$.}
Having fixed an oriented loop with basepoint, the next step is to define its lifts to $\Sigma\times I$. For the class of loops which we consider, it turns out that all lifts to $\Sigma\times I$ can again be encoded as loops with a basepoint directly on $\Sigma$, with the same convention that they go down monotonically along the interval, with a sudden jump at the lift $\bullet_i$ of the basepoint to some appropriate sheet $i$ of $\Sigma$.
The $q$-nonabelianization map of \cite{Neitzke:2020jik} relates $[\varrho] \in \Sk(C\times I,GL_2)$ 
to a specific linear combination of `lifts' $[\tilde \varrho_i] \in \Sk(\Sigma \times I,GL_1)$ built by combining three types of local pieces
\begin{itemize}
\item
Direct lifts of segments $\varrho$ to $\Sigma$, corresponding to preimages under the covering map $\Sigma\to C$
\be\nonumber
	\pic{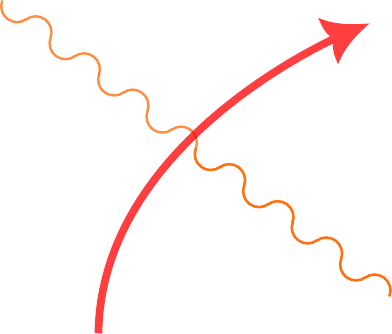}{.3} \xrightarrow[]{\text{lift from $C$ to $\Sigma$}} \pic{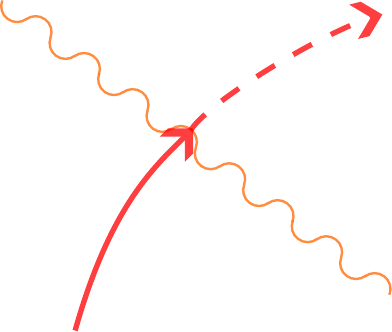}{.3}
\ee
\item  
Detours may arise when the path $\varrho$ intersects a trajectory of $\CW(\theta)$ on $C$. In particular if $\varrho$ intersects an $ij$ trajectory, a detour will be a path obtained by lifting the trajectory from the point of intersection to the branch point of origin. 
This was shown in Fig \ref{fig:detourexample}.
\item 
Exchange paths contribute whenever two segments of $\varrho$ intersect the same leaf of the foliation in $M$.
Here by a foliation on $M$ we mean a horizontal, constant along $I$, lift of the foliation on $C$ to $M=C\times I$.
Therefore exchange lifts only occur when the two points of $\varrho$ that are connected by a leaf are also at the same height along $I$.
Since we work with monotonically declining paths, exchange paths in our setting can only occur between the (mildly resolved) vertical upward segment located at the basepoint $\bullet$ and some other point along $\varrho$. 
Therefore to determine the presence (or absence) of exchange paths we simply draw the leaf of the foliation that passes through $\bullet$ and check whether it intersects $\varrho$ elsewhere. If it does, then an exchange path contributes a new type of oriented lift as shown in \eqref{eq:exchange-lift}.
\be\label{eq:exchange-lift}
	\pic{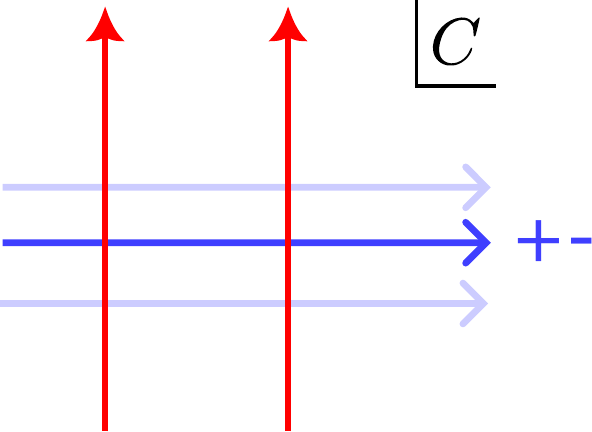}{.3} 
    \xrightarrow[]{\text{lift from $C\times I$ to $\Sigma \times I$}}
    \pic{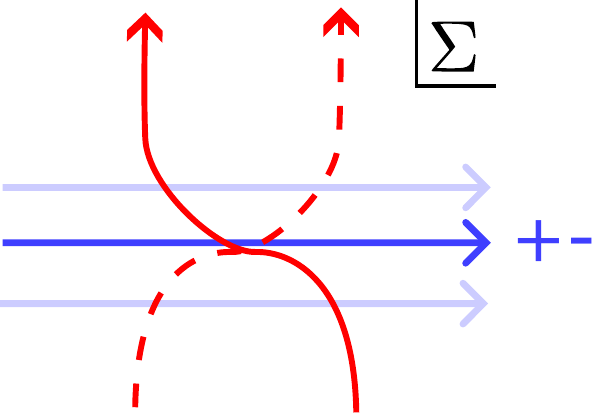}{.3}
\ee
\end{itemize}
The $\mq$-nonabelianization map
relates $\varrho_a$ to a specific linear combination of all lifts obtained in this way\footnote{More precisely, the map relates equivalence classes of paths in the respective skein modules, and should be written as $\hat\CL_{[\varrho]} =  \sum_{i} \alpha([\tilde \varrho_i])\, [\tilde \varrho_i]$. Here and below we lighten notation and omit the $[ \ \cdot\, ]$.}
\be\label{eq:q-nonabelianization}
	\hat\CL_a  = \sum_{\tilde \varrho} \alpha(\tilde \varrho)\, (-1)^{n'(\tilde \varrho)} \mq^{\wr(\tilde\varrho)} \hat X_{\tilde \varrho}\ \  \in \ \ \Sk(\Sigma\times I,GL_1)
\ee
where the definition of the weights $\alpha(\tilde \varrho)$ is summarized in Appendix \ref{app:q-nonabelianization-weights}
and we replaced the skein element $[\tilde\varrho]$ by $(-1)^{n'(\tilde \varrho)} \mq^{\wr(\tilde\varrho)} \hat X_{\tilde \varrho}$. For this purpose we recall the canonical isomorphism \eqref{eq:skein-q-char-isomorph} between $\Sk(\Sigma\times I,GL_1)$ and the quantum torus algebra \eqref{eq:quantum-IR-algebra} arising from deformation quantization of the coordinate ring on $\tilde \CM_{GL_1}(\Sigma)$.
Here $\wr(\tilde\varrho)$ is the \textit{writhe} of $\tilde\varrho$, which is the signed sum of positive and negative crossings. In our setting $\tilde\varrho$ is naturally endowed with a basepoint (the lift of $\bullet$, inducing an ordering on the loop ), and the writhe can be computed by the diagrammatic rules as shown in Eq \eqref{eq:writherule}. The factor $n'(\tilde\varrho)$ is the number of non-local crossings: those crossings in the projection of $\tilde\varrho$ to $C$ that are not crossings of its projection to $\Sigma$. This is also shown in Eq \eqref{eq:writherule}.

\be\label{eq:writherule}
    \pic{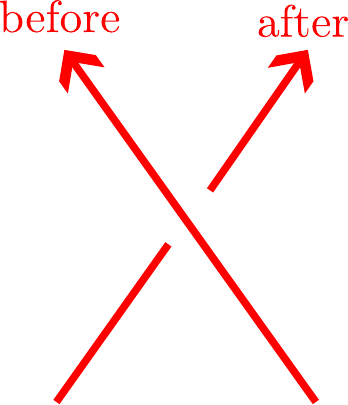}{.3} = \pic{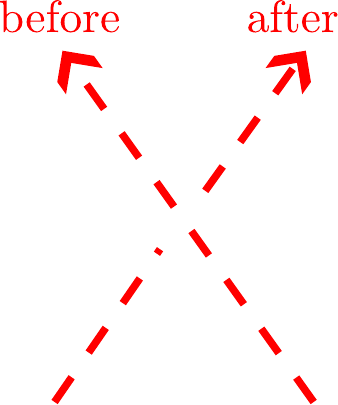}{.3} = \mq^{-1}, \quad \pic{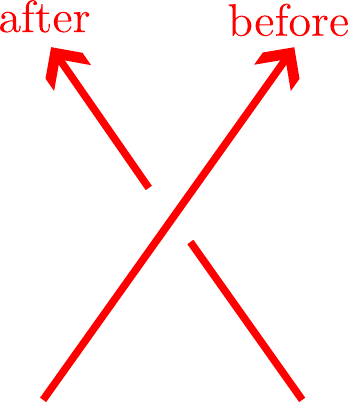}{.3} = \pic{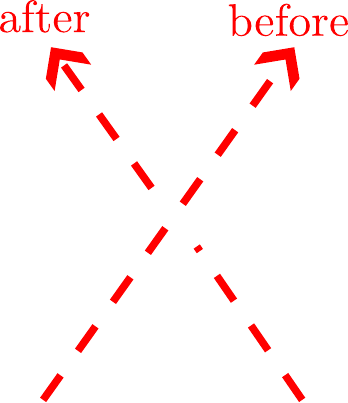}{.3} = \mq^{-1} ,\quad \pic{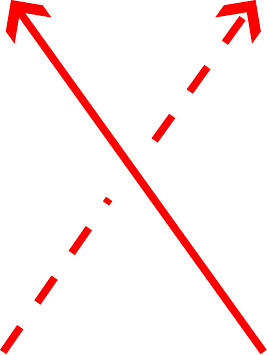}{.3} = \pic{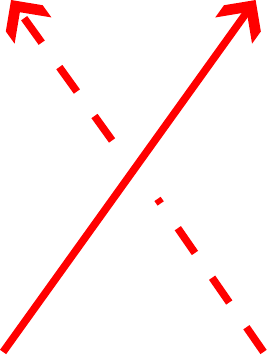}{.3} = -1
\ee

This completes the definition of the $\mq$-nonabelianization map \eqref{eq:q-nonabelianization}, and provides an explicit
realization of the quantum UV-IR relation \eqref{eq:UV-IR-relation-quantum} where the sum over IR line operators arises from summing over lifts with appropriate weights. 
In the classical limit defined by \eqref{eq:q-classical} this map recovers \eqref{eq:nonabelianization-classical}.

As in the classical case, $\mq$-nonabelianization depends on the topological type of the underlying spectral network.
Two spectral networks with different topological types map an element $[\varrho_a] \in \Sk(C\times I, GL_2)$ to different $\hat\CL_a, \hat\CL'_a\in \Sk(\Sigma\times I, GL_1)$.
It is argued in \cite{Neitzke:2020jik} that the two expressions are related by a sequence of \textit{motivic} Kontsevich-Soibelman transformations
\be\label{eq:motivic-change-of-CLa}
	\hat \CL_{a}'  = \prod_{\eta}^{\curvearrowleft} \hat{\mathcal{K}}_{\eta}^{\Omega(\eta;\mq)} \, \hat \CL_{a}\,.
\ee
These consist of a generalization of \eqref{eq:KS-transformation} with values in 
the quantum torus algebra isomorphic to $ \Sk(\Sigma\times I, GL_1)$, see \cite{Kontsevich:2008fj, Dimofte:2009bv, Dimofte:2009tm}
\begin{equation}
    \hat{\mathcal{K}}^{\Omega(\eta;\mq)}_\eta \hat{X}_{\gamma} = \mathbb{U}_\eta^{-1} \hat{X}_{\gamma}\mathbb{U}, \qquad 
    \mathbb{U}_\eta = \prod_{m}\prod_{k\geq 1}(1-\mq^{2k+m-1}\hat{X}_{\eta})^{a_m(\eta)},
    \label{eq:qdefKS}
\end{equation}
where $\Omega(\eta;\mq) = \sum_m \mq^m a_m(\eta)$ is the protected spin character~\cite{Gaiotto:2010be}, a refinement of the BPS index.

\section{Classical UV-IR map for $SL_2$ local systems}
\label{sec:UV-IR-class}

In this section we study the classical UV-IR map for line operators of the $\CN=2^*$ gauge theory.
Physically this corresponds to the computation of expectation values of Wilson and 't Hooft loops.
From a mathematical perspective, we establish an explicit isomorphism between the $SL_2$ character variety on the punctured torus and a certain subspace of the $GL_1$ character variety of the Hitchin spectral curve.
We construct the UV-IR map both in Fock-Goncharov charts, matching results of previous works, and in Fenchel-Nielsen charts, which have not been considered previously.

\subsection{Class $\CS$ construction of $SU(2)$ $\CN=2^*$ theory}\label{sec:class-S-construction}

The $\CN=2^*$ theory is a deformation of $\CN=4$ Yang-Mills theory generated by introducing a mass for the adjoint $\CN=2$ hypermultiplet that is part of the $\CN=4$ vector multiplet.
The Seiberg-Witten solution of its Coulomb branch dynamics is related to the elliptic Calogero-Moser integrable system \cite{Donagi:1995cf, Gorsky:1995zq, Martinec:1995by, DHoker:1997hut}.

The $\CN=2^*$ theory with gauge group $SU(2)$ can be realized as a twisted compactification of the 6d $(2,0)$ theory of type $A_{1}$ on a punctured torus $C$ \cite{Gaiotto:2009gz, Gaiotto:2009we}. 
The moduli space $\CM$ is identified with the space of solutions to Hitchin's equations.
A complete formulation of these equations involves specifying the behaviour of fields at the puncture, and the Higgs field is required to have a simple pole, whose residue $m\sigma_3$ encodes the mass deformation.

\begin{figure}[h!]
        \centering
        \includegraphics[width=0.4\textwidth]{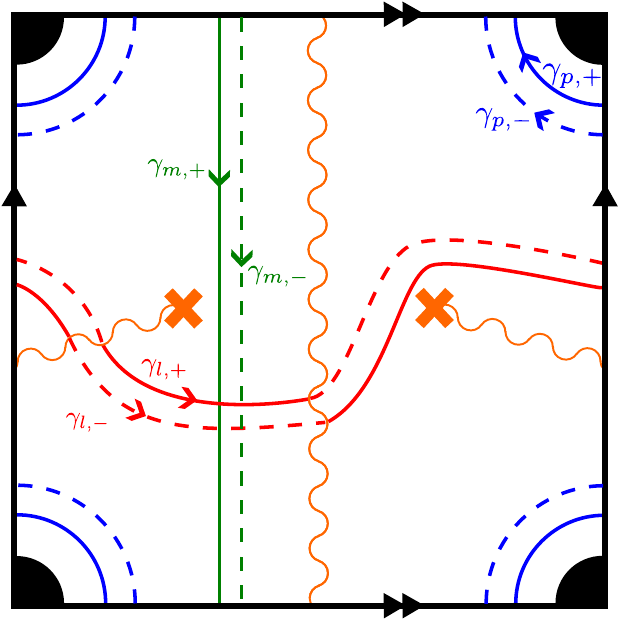}
        \caption{A choice of trivialization for the curve described by Eq \eqref{eq:SW-curve-N2star} with $\tau = i$, $u = 2$ and $m=1$, with a homology basis $\gamma_{l,\pm}, \gamma_{m,\pm}$ and $\gamma_{p,\pm}$. The orange crosses denote the branch points, the orange zig-zag line denotes the branch cuts, and the black quarter circles on the corners denote the puncture. The solid/dashed lines which of the $+/-$ branches of the curve the cycles are on. This will be our convention throughout.} 
    \label{fig:triv}
\end{figure}

The Coulomb branch $\CB$ is identified with the base of Hitchin's fibration $\CM\to \CB$ parameterizing solutions for the Higgs field. Each point in $\CB$ defines a spectral curve $\Sigma\subset T^*C$.
Denoting by $\lambda_{\text{SW}} = y \, dz$ the Liouville 1-form on $T^*C$, the spectral curve can be presented as
\be\label{eq:SW-curve-N2star}
	y^2 = u + m^2 \wp(z|\tau) 
\ee
where $u$ is a local coordinate on $\CB$ and $\tau$ is the complex structure of $C$.
This is a 2-sheeted covering of $C$ with ramification at two branch points. The puncture at $z=0$ (modulo $\IZ\oplus \tau \IZ$) lifts to two punctures on $\Sigma$.
A choice of trivialization for the projection map $\Sigma\to C$ is shown in Figure \ref{fig:triv}. 
Here we also show a set of cycles in $H_1(\Sigma',\IZ)$ that gives a complete set of local coordinates for the character variety \eqref{eq:GL1-char-var}, owing to the constraints on holonomies of cycles that encircle ramification points. 
Since $\Sigma$ has genus $g=2$ and has $n=2$ punctures, its first homology lattice has rank $2g + n-1 = 5$.
In fact, the set of cycles shown in Figure \ref{fig:triv} is overcomplete due to the relation $\gamma_{p,-} +\gamma_{p,+} = 0$, arising from the fact that the sum of circles around punctures is a boundary.
The only nontrivial intersection pairings among these homology cycles are
\be\label{eq:homology-generators-pairings}
	\langle\gamma_{l,\pm},\gamma_{m,\pm}\rangle=-1
\ee
while all others vanish. 

\subsection{UV and IR coordinate systems}

The physical moduli space of the theory $\CM_{\flat}^{\text{UV}}$ corresponds to the $SL_2$ character variety of the UV curve, whose dimension is given by formula \eqref{eq:M-char-dim} specialized to $g=n=1$
\be\label{eq:sl2-char-var-UV}
\begin{split}
	\dim_{\IC}\CM_{\flat}^{\text{UV}} = \dim_{\IC}\CM_{SL_2}(C_{1,1}) & = \dim(SL_2)+\dim(Z(SL_2))=3+0\,.
\end{split}
\ee
Combining this with \eqref{eq:IR-moduli-dimension} implies that the IR moduli space must also be of dimension $3$.
In fact $\CM_\flat^{\text{IR}}\subseteq \tilde \CM_{GL_1}(\Sigma)$ is defined by restricting from $H_1(\Sigma',\IZ)$ to the physical charge lattice $\fY$, see Remark \ref{rmk:physical-holonomies}.
For class $\CS$ theories of type $A_1$ this is universally characterized as
the anti-invariant sublattice under the $\IZ_2$ involution induced by deck transformations of $\Sigma\to C$\cite{Gaiotto:2009hg} 
\be\label{eq:physical-lattice-def}
	\Gamma = H_1(\Sigma',\IZ)^{-}\subset H_1(\Sigma', \IZ)
\ee

\begin{figure}[h!]
    \centering
    \includegraphics[width=0.4\linewidth]{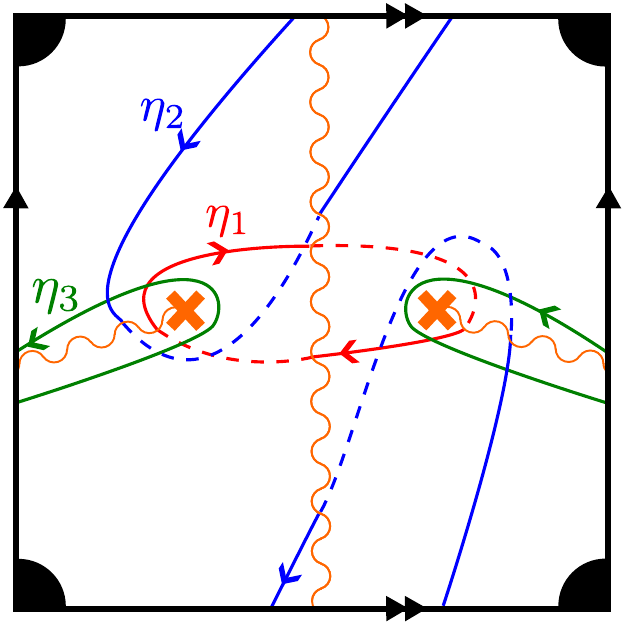}
    \caption{Basis $\eta_1, \eta_2,\eta_3$ of the physical charge lattice $\fY$. 
    } 
    \label{fig:gammabasis}
\end{figure}

To describe this explicitly we can use the basis for $H_1(\Sigma',\IZ)$ in Figure \ref{fig:triv}. The deck transformation $\iota$ induces the following map on homology
\be
	\iota_* (\gamma_{\alpha,\pm}) = \gamma_{l,\mp}\,,
	\qquad
	\alpha = l,m,p\,.
\ee
Therefore a basis of anti-invariant cycles can be chosen as
\be\label{eq:goodsl2reps}
	2\eta_m = \gamma_{m,+} - \gamma_{m,-}  \,,
	\qquad
	2\eta_l = \gamma_{l,+} - \gamma_{l,-} \,,
	\qquad
	2\eta_p = \gamma_{p,+} - \gamma_{p,-}\,.
\ee
Clearly, $\eta_{m},\eta_{l},\eta_{p}$ is not an integral basis. An alternative basis commonly used in the literature is
\be\label{eq:basis-change-sl2}
	\eta_1 = 2\eta_l + 2\eta_m -\eta_p  ,\qquad
	\eta_2 = \eta_p - 2\eta_l ,\qquad
	\eta_3 = \eta_p - 2\eta_m \,.
\ee
These generators of $\fY$ have pairing $\langle \eta_i, \eta_{i+1}\rangle = -2$ with cyclic periodicity, as can be seen from their presentation in Figure \ref{fig:gammabasis}. Note that this gives a physical lattice of rank $3$, as expected.

Local coordinates on the physical moduli space $\CM_\flat^{\text{IR}}$ are provided by $GL_1$ holonomies, which for the two bases of $\fY$ are related as follows%
\begin{equation}
    X_{\eta_1} = X_{2\eta_l+2\eta_m-\eta_p }, \qquad X_{\eta_2} = X_{\eta_p-2\eta_l}, \qquad X_{\eta_3} = X_{\eta_p-2\eta_m }. 
    \label{eq:cyclebasischange}
\end{equation}
Note that their product gives the flavor fugacity associated with the puncture
$
	X_{\eta_1+\eta_2+\eta_3}=X_{\eta_p}\,.
$.

Trace coordinates on the UV moduli space $\CM_{\flat}^{\text{UV}}\simeq \CL_{SL_2}(C)$ are associated to simple curves $\ell$ on $C$, which classify UV line operators \cite{Drukker:2009tz}. 
Let $l,m,p$ denote generators of $\pi_1(C)$ corresponding to longitude, meridian and the puncture, and note that the fundamental group is defined by the following relation
\be\label{eq:fund-group}
	\pi_1(C) = \langle l,m,p| lml^{-1}m^{-1} p=1\rangle\,,
\ee
which fixes $p$ as the group commutator of $l,m$. 
The character variety is therefore parameterized by $M_l, M_m\in SL_2(\IC)$ modulo simultaneous conjugation, giving a space of dimension $3+3-3=3$ as expected.
Introducing gauge-invariant trace coordinates
\be\label{eq:sl2-trace-coordinates}
	\CL_{l} = \tr M_l,\qquad
	\CL_{m} = \tr M_m, \qquad
	\CL_{p} = \tr M_p = \tr (M_m M_l M_m^{-1} M_l^{-1}), \qquad
	\CL_{d} = \tr (M_l M_m)\,,
\ee
the $SL_2$ trace relations give a description of $\CM_{\flat}^{\text{UV}}$ as the Fricke cubic in $\IC^4$, see e.g. \cite{goldman2009trace}
\be\label{eq:Fricke}
	\CL_{l}^2 +\CL_{m}^2+\CL_{d}^2 - \CL_{l}\CL_{m}\CL_{d}-2 = \CL_p\,.
\ee
Our task next will be to find explicit relations between $\CL_{\alpha}$'s and $X_\eta$'s.
Since the trace coordinate at the puncture is directly related to the mass fugacity
\be
	\CL_p = X_{\eta_p}+X_{-\eta_p}\,,
\ee
and since $\CL_d$ is related to $\CL_{a}, \CL_{b}, \CL_{p}$ by \eqref{eq:Fricke}, we will only need to focus on longitudinal and meridian coordinates $\CL_{l}, \CL_{m}$.

\begin{figure}[h!]
    \centering
     \begin{subfigure}[b]{0.48\textwidth}
         \centering
         \includegraphics[width=0.7\linewidth]{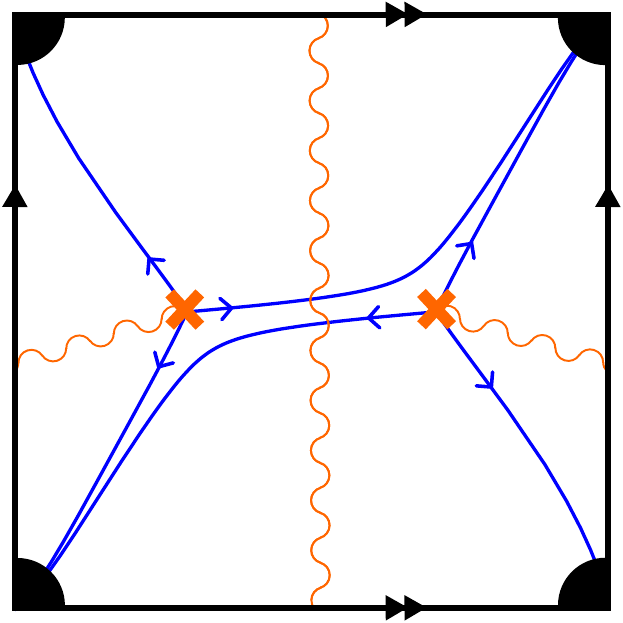}
    	\caption{Fock-Goncharov network $\CW(0.13)$.}
    	\label{fig:FG-network}
     \end{subfigure}
     \hfill 
     \begin{subfigure}[b]{0.48\textwidth}
         \centering
         \includegraphics[width=0.7\linewidth]{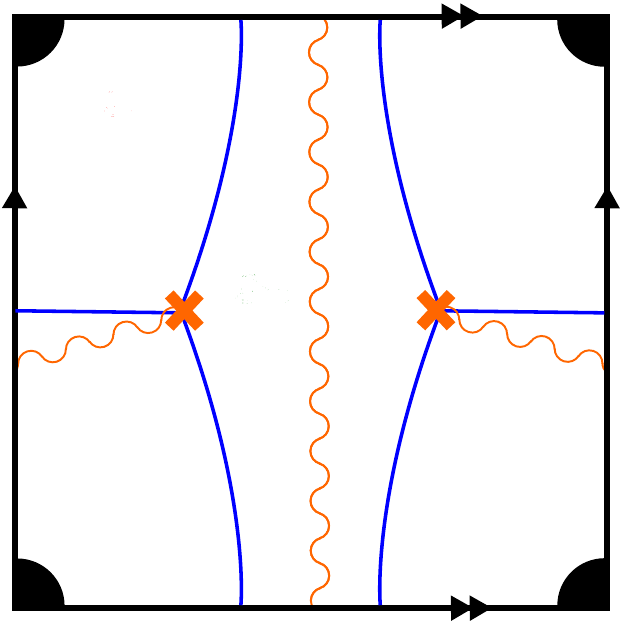}
    	\caption{Fenchel-Nielsen network $\CW(\frac{\pi}{2})$.}
    	\label{fig:FN-network}
     \end{subfigure}
\end{figure}

Recall from Section \ref{sec:UV-IR-framed} that the UV-IR relations depend on a choice of phase $\theta$ encoding the half-BPS subalgebra unbroken by line operators.
In the geometric realization of this map this dependence on $\theta$ translates in the dependence of the nonabelianization map \eqref{eq:nonabelianization-classical} on the topological type of spectral network, see Section \ref{sec:spectral-networks-and-nonabelianization}.
To study spectral networks we need to fix a choice of spectral curve $\Sigma$, and from now onwards we will consider the following point in moduli space
\be\label{eq:moduli-choice}
	\tau=i, \qquad
	u=2, \qquad
	m=1.
\ee
Plotting the spectral network for $\theta_0= 0.13$ radians we find the Fock-Goncharov network shown in Figure \ref{fig:FG-network}.
Plotting the spectral network for $\theta= \frac{\pi}{2}$ radians we find the Fenchel-Nielsen network shown in Figure \ref{fig:FN-network}.
For other, generic, values of $\theta$ the network is always of Fock-Goncharov type, at this point in moduli space.


\subsection{Fock-Goncharov chart} 
\begin{figure}[h!]
    \centering
    \includegraphics[width=0.35\linewidth]{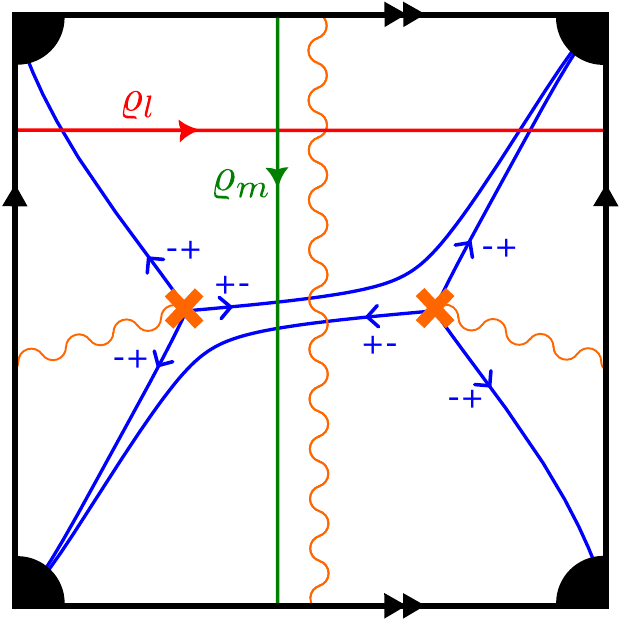}
    \caption{The Fock-Goncharov network plotted at phase $\theta_0 = 0.13$ radians is shown in blue, while the $\varrho_l$ and $\varrho_m$ denote the closed trajectories on $C$.}
    \label{fig:n2starFG}
\end{figure}

We first consider a spectral network of Fock-Goncharov type shown in Figure \ref{fig:n2starFG}.
A direct application of the nonabelianization map yields the following expressions for UV line operators 
\begin{equation}\label{eq:FGclassUVIRmap-lmp}
\begin{split}
    \CL^{\FG}_{m} &= X_{\eta_m} + \frac{1}{X_{\eta_m}}\left(1+X_{\eta_p-2\eta_l}\right),
    \\
    \CL^{\FG}_{l} &= X_{\eta_l}\left(1+X_{2\eta_m-\eta_p}\right) + \frac{1}{X_{\eta_l}},
\end{split}
\end{equation}
Here the superscript ${}^{\FG}$ labels the chart defined by the spectral network used to construct the nonabelianization map, see Figure \ref{fig:n2starFG}. 
We leave details of these computations to the interested reader, noting that these results can also be obtained as a specialization of the more general computation carried out later for ${GL}_2$ local systems, see Remark \ref{rmk:gl2-to-sl2-specialization} for details.  As a sanity check, note that using \eqref{eq:cyclebasischange} gives the equivalent expressions
\begin{equation}
\begin{split}
    \CL^{\FG}_{m} &= \pm \left(X_{\frac12(\eta_2+\eta_1)}+X_{\frac12(-\eta_2-\eta_1)}+X_{\frac12(\eta_2-\eta_1)}\right),
	\\
    \CL^{\FG}_{l} &= \pm \left(X_{\frac12(\eta_1+\eta_3)}+X _{\frac12(-\eta_1-\eta_3)}+X _{\frac12(\eta_1-\eta_3)}\right),
\end{split}
\end{equation}
These expressions match earlier results by \cite{Gang:2017ojg, Gaiotto:2010be} up to a sign ambiguity arising from non-invertibility of \eqref{eq:cyclebasischange}, see also remarks in \cite[Appendix A]{Gaiotto:2009hg}.

\subsection{Fenchel-Nielsen charts} 
We now compute the same UV loop operators in Fenchel-Nielsen charts.
A Fenchel-Nielsen network such as shown in Figure \ref{fig:FN-network} cannot be used for nonabelianization because it is degenerate. Instead one must choose a resolution by perturbing slightly away from the critical phase $\theta=\frac{\pi}{2}$. 
There are two possible choices
$\theta_\pm = \frac{\pi}{2} \pm \epsilon$, with $(0<\epsilon\ll 1)$,
which give different resolutions known as \textit{American} and \textit{British} for $\theta_-$ and $\theta_+$ respectively. Each gives a different result, and the two are related to each other by a specific coordinate change of coordinates as we will see below.

For the American resolution shown in Figure \ref{fig:n2starFNA}, a direct application of the nonabelianization map yields the following expressions for UV line operators
\begin{equation}\label{eq:N2-star-FN-UV-IR-classical-american-lmp}
    \begin{split}
 	\CL_{m}^{\FN_{-}}   & = X_{\eta_m} + \frac{1}{X_{\eta_m}},
    \\
    \CL_{l}^{\FN_{-}} &=  X_{\eta_l}\left(1+X_{2\eta_m-\eta_p}\right) 
    + {X_{-\eta_l}} \frac{1+X_{\eta_p+2\eta_m}}{(1-X_{2\eta_m})^2} 
    .
    \end{split}
\end{equation}
The corresponding expressions for the British resolution shown Figure \ref{fig:n2starFNB} are instead
\begin{equation}\label{eq:N2-star-FN-UV-IR-classical-british-lmp}
\begin{split}
    \CL_{m}^{\FN_{+}}  &=  X_{\eta_m} + \frac{1}{X_{\eta_m}},
    \\
    \CL_{l}^{\FN_{+}}  &=
X_{\eta_l}
\frac{
1+X_{-\eta_p-2\eta_m}}{
(1-X_{-2\eta_m})^2}
+{X_{-\eta_l}}\left(1+{X_{\eta_p-2\eta_m}}\right) 
    .
    \end{split}
\end{equation}
\begin{figure}[h!]
    \centering
    \includegraphics[width=0.8\linewidth]{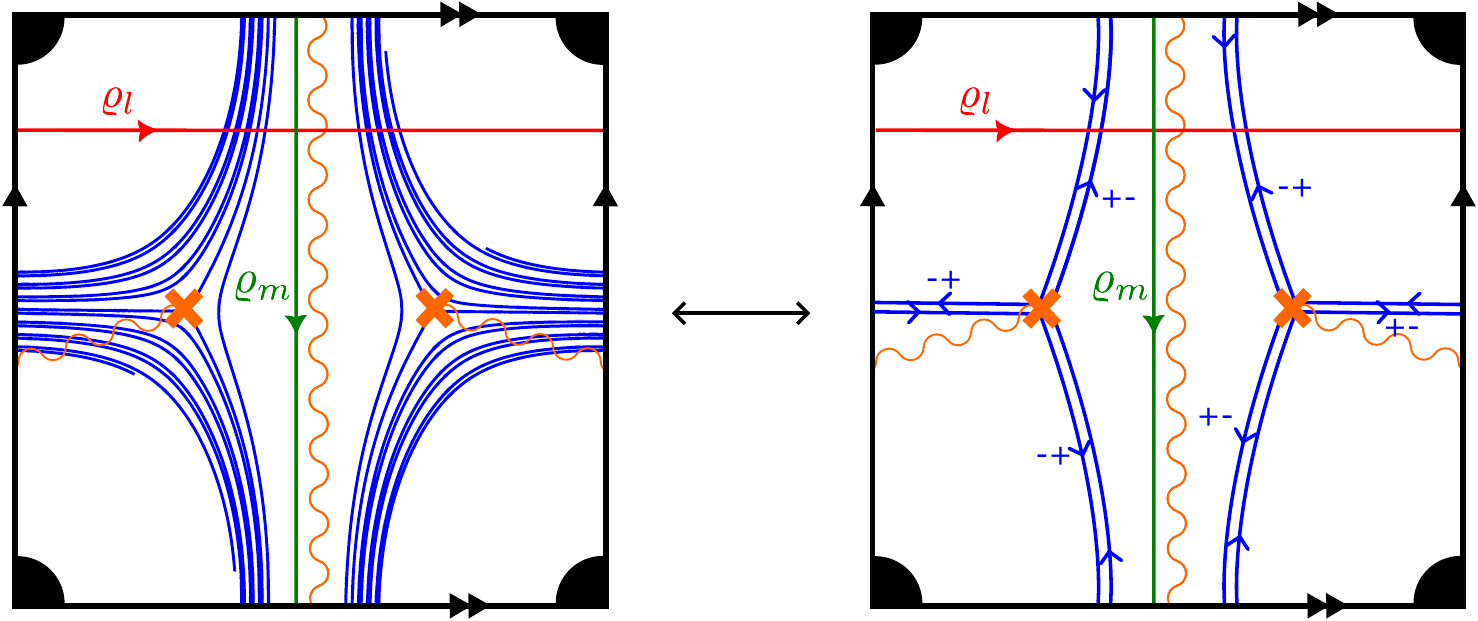}    
    \caption{The picture on the left is the actual network plotted at phase $\theta_- = \frac{\pi}{2}-0.015$ radians, while the picture on right is a compact representation of the infinitely degenerate network in the American resolution.}
    \label{fig:n2starFNA}
\end{figure}

\begin{figure}[h!]
    \centering
    \includegraphics[width=0.8\linewidth]{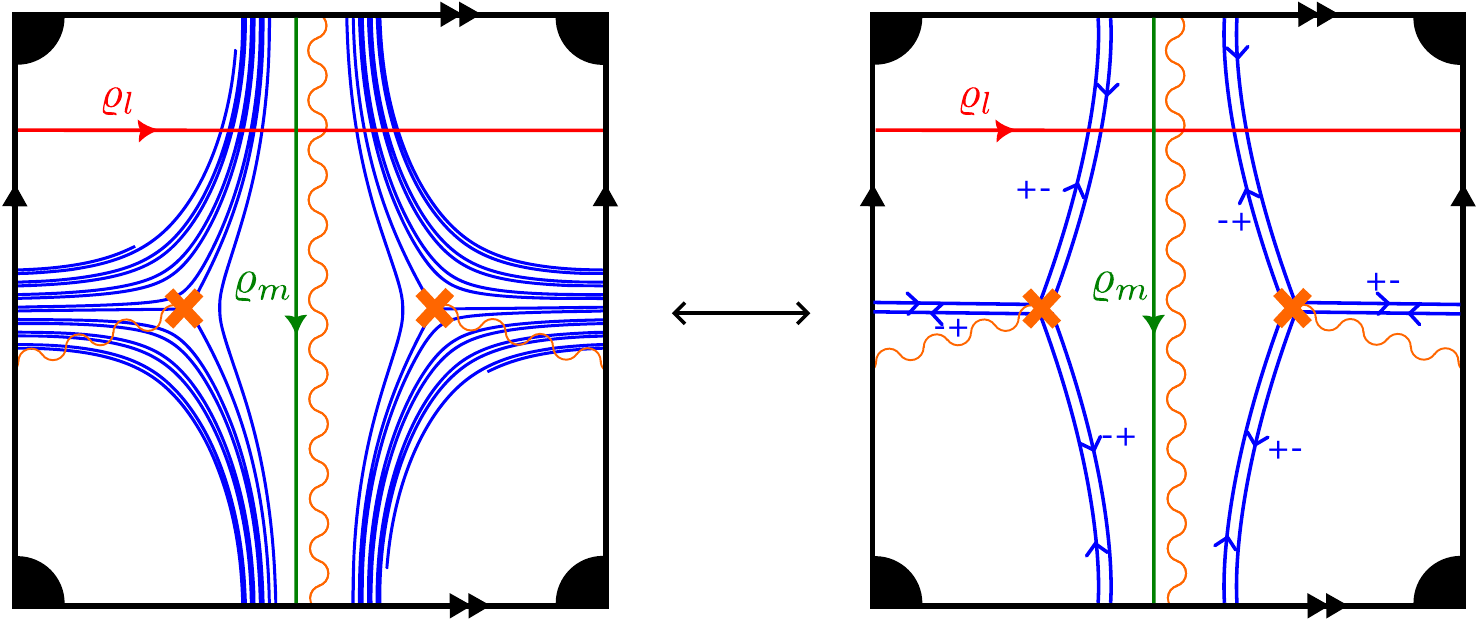}    
    \caption{The picture on the left is the actual network plotted at phase $\theta_+ = \frac{\pi}{2}+0.015$ radians, while the picture on right is a compact representation of the infinitely degenerate network in the British resolution.}
    \label{fig:n2starFNB}
\end{figure}

Again, we leave details of these computations to the interested reader noting that they correspond to a specialization of more general results obtained below, see Remark \ref{rmk:gl2-to-sl2-specialization}.

\subsection{Relations among charts}
The three different expressions for $\CL_{l},\CL_{m}$ obtained so far are all equivalent.
Their mutual relationship is encoded by the phenomenon of \textit{framed wall-crossing} \cite{Gaiotto:2010be}, which in the context of spectral networks is associated to topological jumps of the trajectories, denoted \textit{K-walls} in \cite{Gaiotto:2012rg}.
Each jump of the network corresponds to a transition between distinct coordinate charts, with transition functions expressed by a Kontsevich-Soibelman change of coordinates \eqref{eq:KS-transformation}.

To relate coordinates $\CL_a^{\FG},\CL_a^{\FN_\pm}$ appearing in the three expressions \eqref{eq:FGclassUVIRmap-lmp}, \eqref{eq:N2-star-FN-UV-IR-classical-american-lmp} and \eqref{eq:N2-star-FN-UV-IR-classical-british-lmp},  it is necessary to know which degenerations of the spectral network arise in the intervals among the three phases $\theta_0,\theta_-,\theta_+$. 
These jumps always occur at phases of central charges of stable BPS states of the $\CN=2^*$ theory, which depends on a choice of point in the Coulomb branch of the theory. Happily, this information is known for the case of interest to us.

The point in moduli space \eqref{eq:moduli-choice} that we are considering corresponds to the one studied in \cite[Section 1.5 \& Section 5]{Longhi:2015ivt}, where the full BPS spectrum is also computed (see also \cite{Longhi:2016wtv, ruter2021novel}).
First we note that at this point in moduli space the central charges of the relevant physical charges are ordered as follows 
\be
	\arg Z_{\eta_2} < \arg Z_{\eta_3}=\arg Z_{\eta_1+\eta_2} < \arg Z_{\eta_1}\,.
\ee
The full BPS spectrum is summarized by the following table of BPS indices
\be\label{eq:BPS-spectrum}
\begin{split}
	&\Omega(\eta_1 + k(\eta_1+\eta_2)) = \Omega(\eta_2 + k(\eta_1+\eta_2)) = 1
	\qquad(k\geq 0)\,,
	\\
	& \Omega(\eta_p \pm (\eta_1+\eta_2) ) =  1 \,,
	\qquad
	\Omega(\eta_1+\eta_2) = -2\,,
\end{split}
\ee
together with CPT-conjugate states, where $\eta_p = \eta_1+\eta_2+\eta_3$ from \eqref{eq:basis-change-sl2}.\footnote{The hypermultiplets with charges $\eta_p\pm(\gamma_1+\gamma_2)$ can be regarded as the two components of the adjoint hypermultiplet that are counterparts of the $W$-bosons, after breaking $\CN=4$ supersymmetry to $\CN=2$ by turning on the mass deformation. Indeed, in the massless limit $Z_{\eta_p}=0$ these two hypermultiplets pair up with the BPS $W$-boson contribution and give $\Omega(\gamma_1+\gamma_2)=1+1-2=0$. This is expected, by recombination of short multiplets into long ones due to $\CN=4$ supersymmetry. Correspondingly, the spectrum should also include a component of the adjoint hypermultiplet corresponding to the Cartan generator, naively $\Omega(\eta_p)=1$. This is omitted because it is pure flavour and decouples from the rest of the spectrum, but can be detected by framed wall-crossing with defects, see expressions in \cite[Section 4.5]{Longhi:2016wtv}.}

Next, we note that the phases $\theta_0$ of Fock-Goncharov and $\theta_\pm$ of Fenchel-Nielsen networks are related to those of BPS central charges as follows
\be
	0=\arg Z_{-\eta_1} < \theta_0 <\arg Z_{\eta_2} \,,
	\qquad
	\theta_\pm \approx \arg Z_{\eta_3} = \frac{\pi}{2}\,.
\ee
This entails the following.
\begin{enumerate}[label=(\arabic*)]
    \item The Fock-Goncharov expressions are related to the Fenchel-Nielsen ones in American resolution by the infinite phase-ordered sequence of jumps due to BPS states with charges $\eta_2+n(\eta_1+\eta_2)$
    \begin{equation}      \label{eq:classFGFN}
    \begin{split}
   	    \CL_a^{\FN_-}
	    & = \prod_{n\geq 0}^{\curvearrowleft} \mathcal{K}_{\eta_2+n(\eta_1+\eta_2)} \, 	\CL_a^{\FG}\,,
    \end{split}
    \end{equation}
    where $\curvearrowleft$ means that $n$ grows from right to left.
    This amounts to performing the following variable substitutions in $\CL_a^{FG}$
    \be\label{eq:classFGFN-explicit}
    	X_{\eta_m} \to X_{\eta_m}  \frac{X_{2\eta_l} (1-X_{2\eta_m} )^2+X_{\eta_p} }{X_{2\eta_l} (1-X_{2\eta_m} )^2+X_{2\eta_m+\eta_p} }
        ,
	\qquad
	X_{\eta_l} \to X_{\eta_l}\left(
	1+\frac{X_{2\eta_m+\eta_p}}{X_{2\eta_l} (1-X_{2\eta_m} )^2}\right),
	\qquad
	X_{\eta_p} \to X_{\eta_p}.
    \ee
    It can be checked that \eqref{eq:classFGFN-explicit} relates expressions \eqref{eq:FGclassUVIRmap-lmp} and \eqref{eq:N2-star-FN-UV-IR-classical-american-lmp} for UV line operators.
    
    \item 
    The Fenchel-Nielsen expressoins $\CL_a^{\FN_\pm}$ are related to each other by the jump of the network at $\theta=\frac{\pi}{2}$ which is due to the vectormultiplet $\Omega(\eta_1+\eta_2)=-2$ as well as the pair of hypermultiplets $\Omega(\eta_p \pm (\eta_1+\eta_2)) = 1$
\begin{equation}\label{eq:FN-jump-classical}
    \begin{split}
        \CL_{a}^{\FN_+} & = \mathcal{K}_{\eta_3} \mathcal{K}_{\eta_1+\eta_2}^{-2}\mathcal{K}_{2\eta_1+2\eta_2+\eta_3} \CL_{a}^{\FN_-} 
        \\
        & \mathop{=}^{\eqref{eq:basis-change-sl2}} \mathcal{K}_{\eta_p-2\eta_m} \mathcal{K}_{2\eta_m }^{-2}\mathcal{K}_{2\eta_m +\eta_p} \CL_{a}^{\FN_-}\,.
    \end{split}
\end{equation}
This amounts to the following substitutions in expressions for $\CL_a^{\FN_+}$ 
\be\label{eq:classical-FN-jump-explicit}
\begin{split}
	&
	X_{\eta_m } \to  X_{\eta_m } 
	\,,
	\qquad
	X_{\eta_l} \to 
	\frac{
	(1+X_{2\eta_m+\eta_p })}{(1+X_{\eta_p-2\eta_m } )^{} 
	(1-X_{2\eta_m } )^{2}
	}  X_{\eta_l} 
	\,,
	\qquad
	X_{\eta_p} \to X_{\eta_p} 
	\,.
\end{split}
\ee
It can be checked that this relation gives an exact match of \eqref{eq:N2-star-FN-UV-IR-classical-american-lmp} and \eqref{eq:N2-star-FN-UV-IR-classical-british-lmp}.

\end{enumerate}

\section{$\mq$-nonabelianization for ${GL}_2$ local systems }\label{sec:UV-IR-quant}

In this section we generalize our previous discussion in two ways: we switch from $SL_2$ to $GL_2$ local systems on the punctured torus, and we consider the deformation quantization of its coordinate ring. 
We therefore set out to study the $GL_2$ skein algebra of the punctured torus. In particular, we will compute the isomorphism \eqref{eq:skein-algebra-isomorphism} that gives a presentation in terms of the $GL_1$ skein algebra on a ramified covering surface.

While the moduli spaces and related algebraic structures studied in this section are broader than the physically relevant moduli spaces $\CM_{\flat}^{\text{UV/IR}}$ and line operator algebras $\CA^{\text{UV/IR}}$ of the $\CN=2^*$ theory, 
all results obtained below can be easily specialized to statements about the latter, see \eqref{eq:M-UV-inclusion}-\eqref{eq:GL1-char-var} and \eqref{eq:skein-alg-table}.

\subsection{UV and IR skein algebras}\label{sec:gl2-coordinate-systems}
The moduli space of ${GL}_2$ local systems on the punctured torus has dimension given by \eqref{eq:M-char-dim} specialized to $g=n=1$
\be
\begin{split}
	\dim_{\IC}\CM_{GL_2}(C) & = \dim(GL_2)+\dim(Z(GL_2))=4+1\,.
\end{split}
\ee
We will study its relation to $GL_1$ local systems on a ramified covering $\Sigma\to C$.
From \eqref{eq:nonabelianization-isomrphism-GL2} it follows that we should consider a covering $\Sigma$ with $\dim \tilde \CM_{GL_1}(\Sigma) = 5$.
In fact, a convenient choice of covering with this property is precisely provided by the \textit{same} spectral curve \eqref{eq:SW-curve-N2star} that we studied for the $SL_2$ system. 
Generators for $H_1(\Sigma,\IZ)$ are given in Figure \ref{fig:triv}, modulo the relation $\gamma_{p,+}+\gamma_{p,-}=0$.
More precisely, we recall that these are really defined as defined on $\Sigma'$, and that holonomies around ramification points are fixed to $-1$, see \eqref{eq:GL1-char-var}.
Local coordinates on $\tilde \CM_{GL_1}(\Sigma)$ are then provided by holonomies $X_{\gamma_{l,\pm}}, X_{\gamma_{m,\pm}}, X_{\gamma_{p,\pm}}$, modulo the relation $X_{\gamma_{p,+}}X_{\gamma_{p,-}}=1$. 
Moreover, choosing moduli for $\Sigma$ again as in \eqref{eq:moduli-choice} we deduce that
spectral networks data will be exactly the same as before, for all values of $\theta$. 
Once again, we will consider both Fock-Goncharov and Fenchel-Nielsen charts.%
\footnote{The fact that we can use the same spectral networks as in the $SL_2$ case is not surprising. Spectral networks naturally relate $GL_2$ local systems on $C$ to $GL_1$ local systems on $\Sigma$. The physical setting of $\CN=2^*$ theory just corresponds to a restriction of this relation, whereas here we consider its full scope.}


To set up nonabelianization, we first identify a suitable set of coordinates on $\CM_{GL_2}(C)$.
From the defining relation of the fundamental group \eqref{eq:fund-group} it is clear that character variety is parameterized by holonomy matrices $M_l, M_m\in GL_2$ modulo simultaneous conjugation $(M_l, M_m)\sim (gM_lg^{-1}, gM_mg^{-1})$. 
Since $\dim GL_2=4$ and since the generic stabilizer corresponds to $g\in\IC^*$ which has dimension 1,  the generic conjugacy orbit has dimension $3$ and the dimension of the character variety is indeed $4+4-3=5$ as expected.
Comparing with the 3-dimensional $SL_2$ character variety, the additional dimensions correspond to the determinants of meridian and longitudinal holonomy matrices.
Introducing the gauge-invariant coordinates
\be\label{eq:gl2-trace-coordinates}
\begin{split}
	\CL_{l} &= \tr M_l,\qquad
	\ \ \ \CL_{m} = \tr M_m, \qquad
	\ \ \ \CL_{d} = \tr (M_l M_m),
	\\
	\CD_{l} &= \det M_l,\qquad
	\CD_{m} = \det M_m,\qquad
	\CL_{p} = \tr M_p = \tr (M_m M_l M_m^{-1} M_l^{-1})\,.
\end{split}
\ee
the $GL_2$ trace-determinant relations give a description of $\CM_{GL_2}(C)$ as an algebraic variety\footnote{
We remark that the ${GL}_{2}$ character variety can be viewed as a local fibration over $(\IC^*)^2$ parameterized by $(\CD_l,\CD_m)$. The fibers of this map are deformations of the Fricke cubic \eqref{eq:Fricke}, which corresponds to the distinguished fibre at $(1,1)$.
A quantum counterpart of this relation can be written down in terms of Wilson lines in fully antisymmetric representations replacing the determinants, see Remark \ref{rmk:q-det}.
} in $\IC^6$
\be\label{eq:Fricke-gl2}
	\CL_{l}^2\CD_m +\CL_{m}^2\CD_l+\CL_{d}^2 - \CL_{l}\CL_{m}\CL_{d}-2\CD_{l}\CD_m = \CD_{l}\CD_m \,\CL_p\,.
\ee

Our task involves expressing these coordinates in terms of abelian holonomies $X_\gamma$.
The trace coordinate at the puncture is directly related to the flavor fugacity by $\CL_p = X_{\eta_p}+X_{-\eta_p}$, and 
since $\CL_d$ is fixed by \eqref{eq:Fricke-gl2} we only need to compute four of the remaining functions. 
Note that determinants may be expressed in terms of trace coordinates as 
$\CD_j = \frac{1}{2}(\CL_{j}^2 - \CL_{j^2})$ where $j=m,l$ and $\CL_{l^2} = \tr M_l^2, \CL_{m^2} = \tr M_{m}^2$
correspond to paths winding twice around longitude and meridian.
We will therefore focus on longitudinal and meridian coordinates and their second-order counterparts $\CL_{l}, \CL_{m}, \CL_{l^2}, \CL_{m^2}$ in the following.


More precisely, we will be interested in obtaining explicit relations between generators of the $GL_2$ skein algebra of $C$ and generators $\hat X_{\gamma}$  of the $GL_1$ skein algebra of $\Sigma$ by expressing 
\be
\begin{split}
	\Sk(C\times I, GL_2) \ \ & \to  \ \ \Sk(\Sigma\times I, GL_1) \\
	[\varrho_a] \ \ & \mapsto  \ \ \hat \CL_a \qquad \qquad (a=l,m,l^2, m^2)
\end{split}
\ee
via the $\mq$-nonabelianization map \eqref{eq:q-nonabelianization}.
We will compute this both in Fock-Goncharov and Fenchel-Nielsen networks encountered earlier. 
The derivation is somewhat laborious, and readers interested only in the main results can jump to Subsection \ref{sec:FG-FN-Results} where we provide a summary. 
For later convenience, we record here the quantum torus relations among generators of $\Sk(\Sigma\times I, GL_1)$
\be
	\hat{X}_{\gamma_{l,i}} \hat{X}_{\gamma_{m,j}} = \mq^{-2\delta_{ij}} \hat{X}_{\gamma_{m,j}} \hat{X}_{\gamma_{l,i}}\,,
	\qquad
	[\hat{X}_{\gamma_{l,i}},\hat{X}_{\gamma_{p,j}}]=[\hat{X}_{\gamma_{m,i}},\hat{X}_{\gamma_{p,j}}]=0 \,.
    \label{eq:quantumtorusalgebra}
\ee
Puncture holonomies $\hat{X}_{\gamma_{p,\pm}}$ are in the center of the algebra, and we simply denote  $X_{\gamma_p} \equiv  \hat{X}_{\gamma_p,+} = (\hat{X}_{\gamma_{p,-}})^{-1}$.  
Both on the base and on the covering we impose relations \eqref{eq:GL2-puncture-skein} and \eqref{eq:GL1-puncture-skein}, with $\mu :=  X_{\gamma_p}$, namely 
\be\label{eq:puncture-GL2-skein}
	\hat\CL_p = X_{\gamma_p}+ X_{\gamma_p}^{-1}\,.
\ee

\subsection{Fock-Goncharov chart}\label{sec:quantum-FG-chart}
We will now compute the $\mq$-nonabelianization map in the Fock-Goncharov chart of Figure \ref{fig:n2starFG} for ${GL}_2$ skein algebra of the surface $C$. 
More specifically, we will be interested in computing the images $\hat\CL_m^{\FG},\hat\CL_l^{\FG},$ $\hat\CL_{m^2}^{\FG}, \hat\CL_{l^2}^{\FG}\in \Sk(\Sigma\times I, GL_1)$ of the four generators of $Sk(C\times I,GL_2)$ corresponding to $m,l,m^2,l^2$, for reasons explained in Section \ref{sec:gl2-coordinate-systems}. 
We choose a specific representative path $\varrho_j$ for each of these as follows
\allowdisplaybreaks
\be
\begin{split}\label{eq:rho-def-gl2}
    \varrho_{m}   = \pic{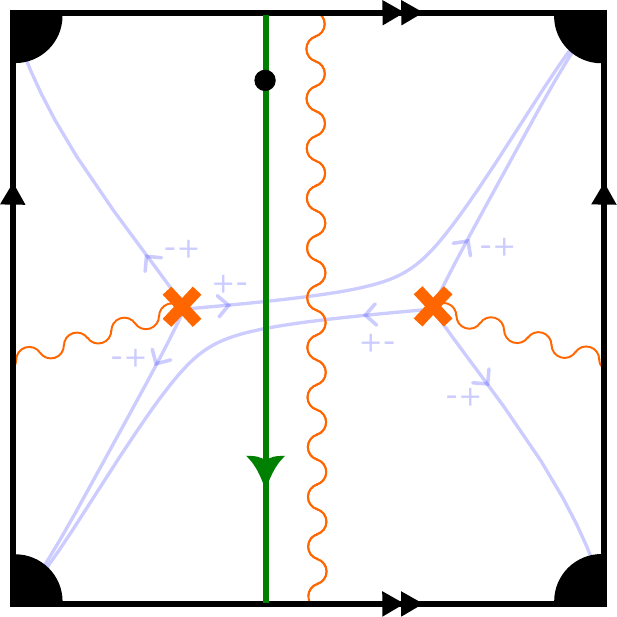}{.4},  &\qquad\qquad \ \,
    \varrho_{l}   = \pic{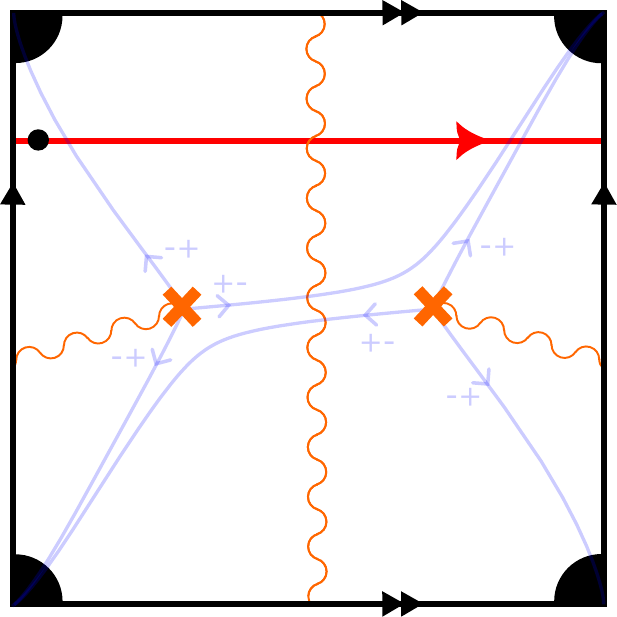}{.4},  \\
    \varrho_{m^2} = \pic{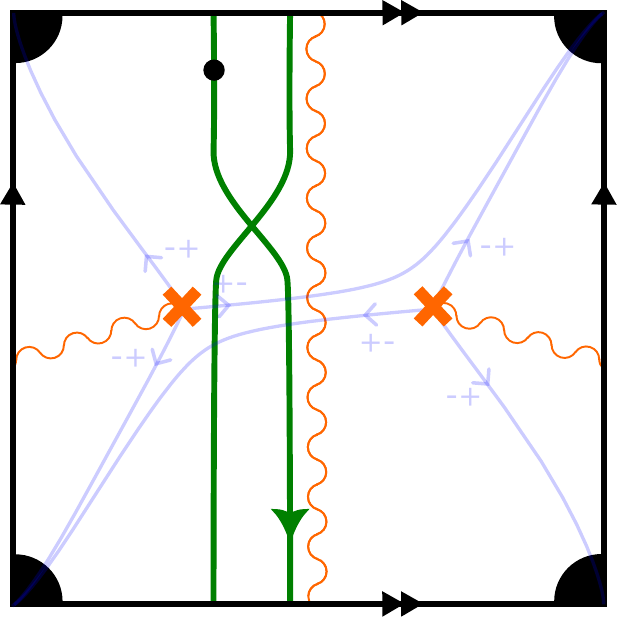}{.4}, &\qquad\qquad
    \varrho_{l^2} = \pic{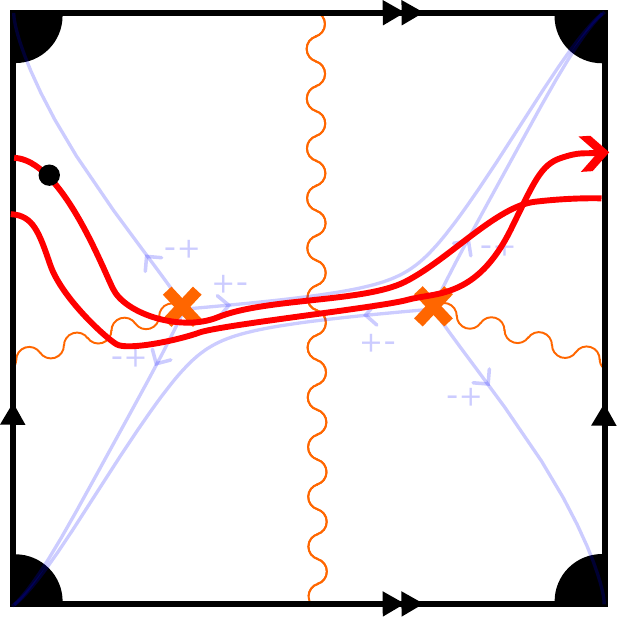}{.4}.
\end{split}
\ee
As explained in Section \ref{sec:q-nonabelianization}, we encode each path in $C\times I$ by drawing its projection on $C$ together with a choice of basepoint. We next compute the expansion of each of these in terms of $\Sk(\Sigma\times I,GL_1)$ generators.

\paragraph{The operator $\hat\CL_m^{\FG}$.} We start by listing the lifts of $\varrho_m$ from $C$ to $\Sigma$.
\begin{equation*}
    \tilde{\varrho}_{m}^{(1)} = \pic{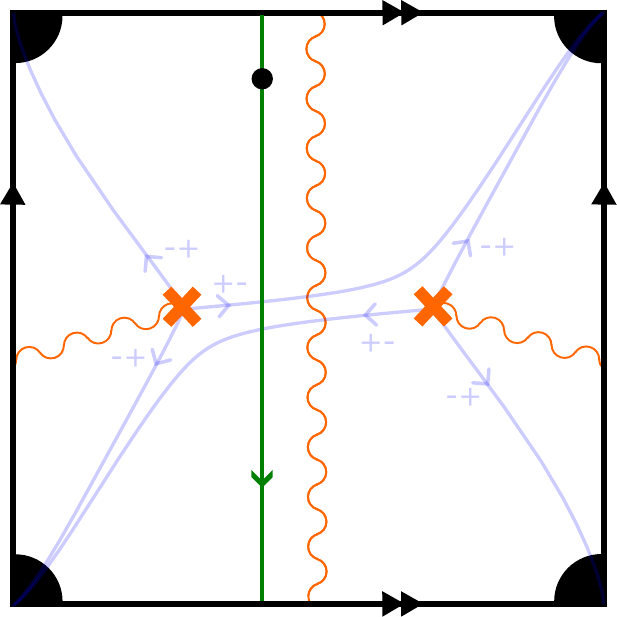}{.3}, \qquad
    \tilde{\varrho}_{m}^{(2)} = \pic{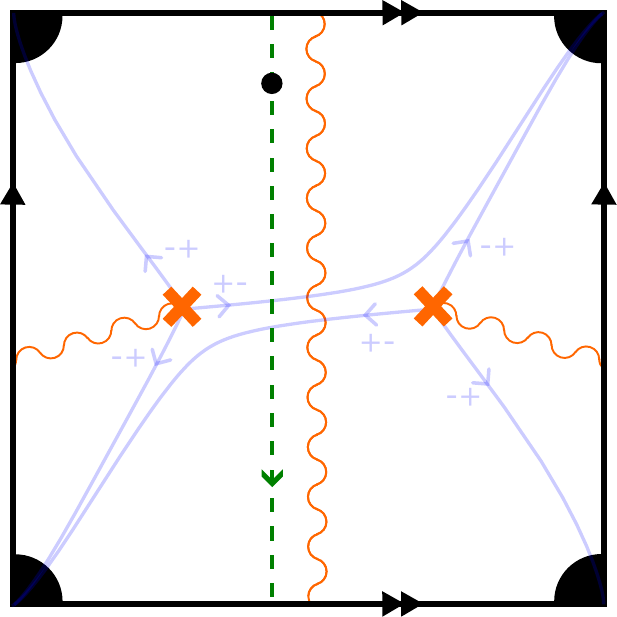}{.3}, \qquad
    \tilde{\varrho}_{m}^{(3)} = \pic{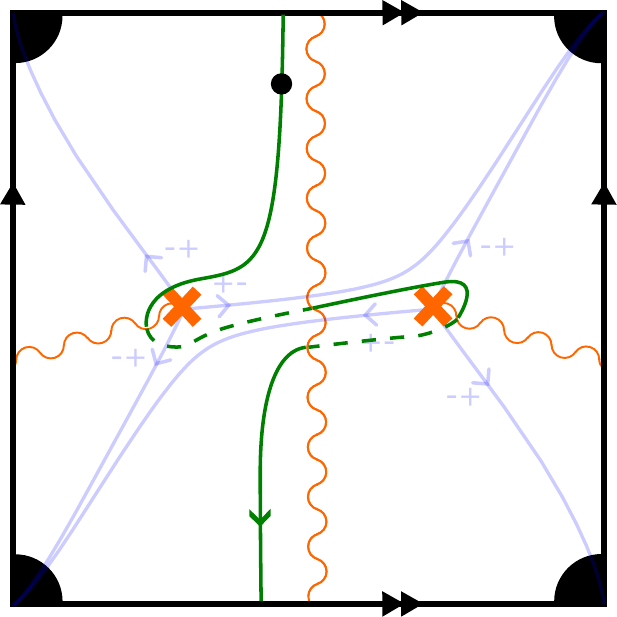}{.3}.
\end{equation*}
Next, we collect their weights and writhe information in the following table.
\begin{center}
\begin{tabular}{|c|c|c|c|c|c|c|} 
 \hline
 $\tilde{\varrho}$ & $\alpha_f(\tilde{\varrho})$ & $\alpha_e(\tilde{\varrho})$ & $\alpha_d(\tilde{\varrho})$ & $\alpha_w(\tilde{\varrho})$ & $(-1)^{n'(\tilde{\varrho})} \mq^{\mathfrak{wr}(\tilde{\varrho})} \hat{X}_{\tilde{\varrho}}$ & Total normal ordered product\\
 \hline\hline
 $\tilde{\varrho}_{m}^{(1)}$ & 1 & 1  &  1 & 1  & $\hat{X}_{\gamma_{m,+}}$  & $\hat{X}_{\gamma_{m,+}}$   \\ 
 $\tilde{\varrho}_{m}^{(2)}$ & 1  & 1   & 1 & 1  & $\hat{X}_{\gamma_{m,-}}$  & $\hat{X}_{\gamma_{m,-}}$    \\
 $\tilde{\varrho}_{m}^{(3)}$ & 1 &  1 & $1$  & $ 1$  &  $\hat{X}_{\gamma_{m,-}+\gamma_{l,-}-\gamma_{l,+}+\gamma_p}$  & $-\mq^{-1} \hat{X}_{\gamma_{m,-}} \hat{X}_{\gamma_{l,-}-\gamma_{l,+}+\gamma_p}$\\
 \hline
\end{tabular}
\end{center}
For convenience we have listed the normal-ordered product of $\hat X_\gamma$, defined by moving all $\hat X_{\gamma_{m,\pm}}$ to the left and $\hat X_{\gamma_{l,\pm}}$ to the right, in the last column.
Summing up all contributions gives the following element of $\Sk(\Sigma\times I, GL_1)$
\begin{equation}
    \hat\CL^{\FG}_m = 
    \hat{X}_{\gamma_{m,+}}
    + \hat{X}_{\gamma_{m,-}}\left(1 - \mq^{-1}\,\hat{X}_{\gamma_{l,-}-\gamma_{l,+}+\gamma_p}\right).
    \label{eq:FG-Lm}
\end{equation}

\paragraph{The operator $\hat\CL_l^{\FG}$.} 
The lifts of $\varrho_l$ to $\Sigma$ are given by
\begin{equation*}
    \tilde{\varrho}_{l}^{(1)} = \pic{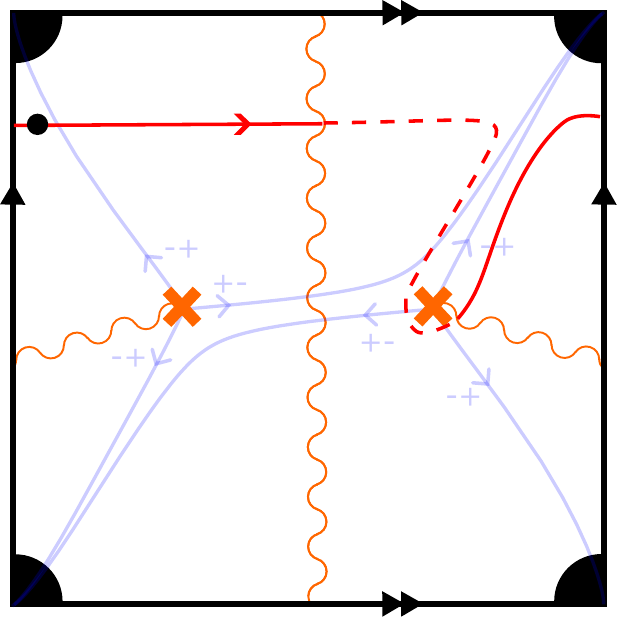}{.3}, \qquad
    \tilde{\varrho}_{l}^{(2)} = \pic{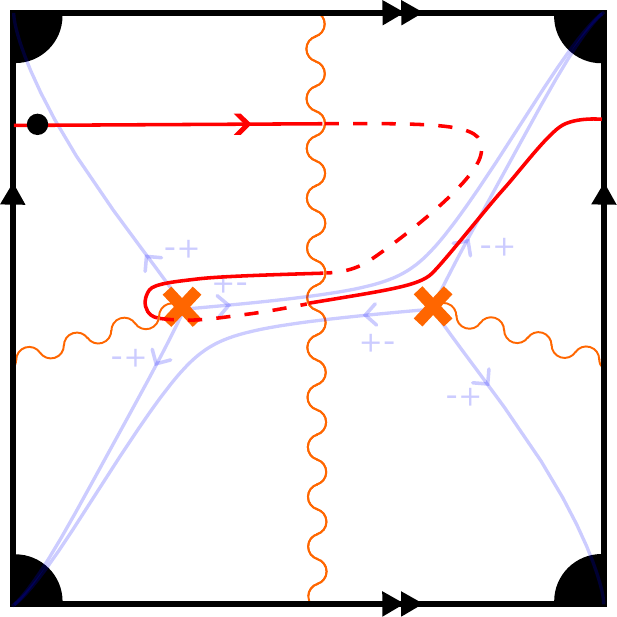}{.3}, \qquad
    \tilde{\varrho}_{l}^{(3)} = \pic{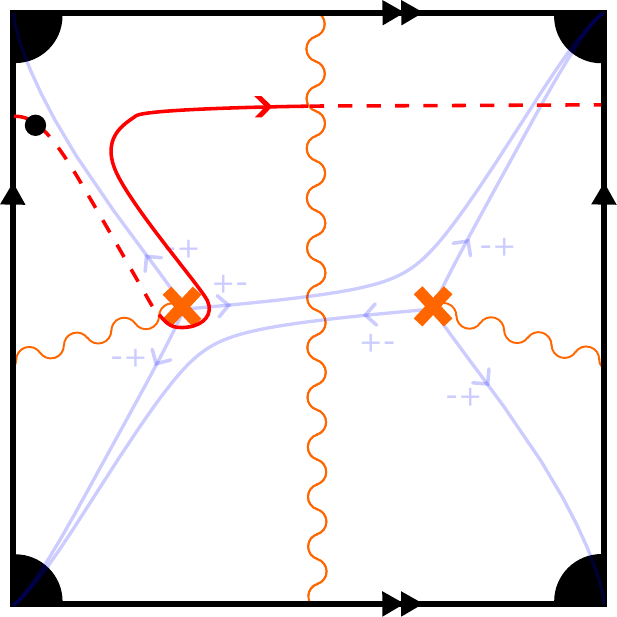}{.3}.
\end{equation*}
Their weights and writhe information are as follows.
\begin{center}
\begin{tabular}{|c|c|c|c|c|c|c|} 
 \hline
 $\tilde{\varrho}$ & $\alpha_f(\tilde{\varrho})$ & $\alpha_e(\tilde{\varrho})$ & $\alpha_d(\tilde{\varrho})$ & $\alpha_w(\tilde{\varrho})$ & $(-1)^{n'(\tilde{\varrho})} \mq^{\mathfrak{wr}(\tilde{\varrho})} \hat{X}_{\tilde{\varrho}}$ & Total normal ordered product \\
 \hline\hline
$\tilde{\varrho}_{l}^{(1)}$ & $\mq^{\frac12}$   & 1  & $\mq^{-\frac12}$  & $1$  &  $\hat{X}_{\gamma_{m,+}-\gamma_{m,-}+\gamma_{l,+}-\gamma_p}$  & $-\mq^{-1}\hat{X}_{\gamma_{m,+}-\gamma_{m,-}-\gamma_p}\hat{X}_{\gamma_{l,+}}$\\
$\tilde{\varrho}_{l}^{(2)}$ &  $\mq^{\frac12}$ &  1 &  $\mq^{-\frac12}$ & $ 1$  &  $\hat{X}_{\gamma_{l,-}}$  & $\hat{X}_{\gamma_{l,-}}$\\
$\tilde{\varrho}_{l}^{(3)}$ &  $\mq^{\frac12}$  & 1  &  $\mq^{-\frac12}$ &  $ 1$ &  $\hat{X}_{\gamma_{l,+}}$  & $\hat{X}_{\gamma_{l,+}}$\\
 \hline
\end{tabular}
\end{center}
Here the winding factor of each lift is trivial, because it receives two canceling contributions from tangency of $\tilde\varrho$ with the flow lines: one from the flow in upper part of the diagram, and one (with opposite sign) from the detours.
This gives us
\begin{equation}
    \hat\CL^{\FG}_l = \hat{X}_{\gamma_{l,-}} + \left(1
    - \mq^{-1}\hat{X}_{\gamma_{m,+}-\gamma_{m,-}-\gamma_p}\right)\hat{X}_{\gamma_{l,+}}.
    \label{eq:FG-Ll}
\end{equation}

\paragraph{The operator $\hat\CL_{m^2}^{\FG}$.} 
The lifts are
\allowdisplaybreaks   
\begin{align*}
    \tilde{\varrho}_{m^2}^{(1)} &= \pic{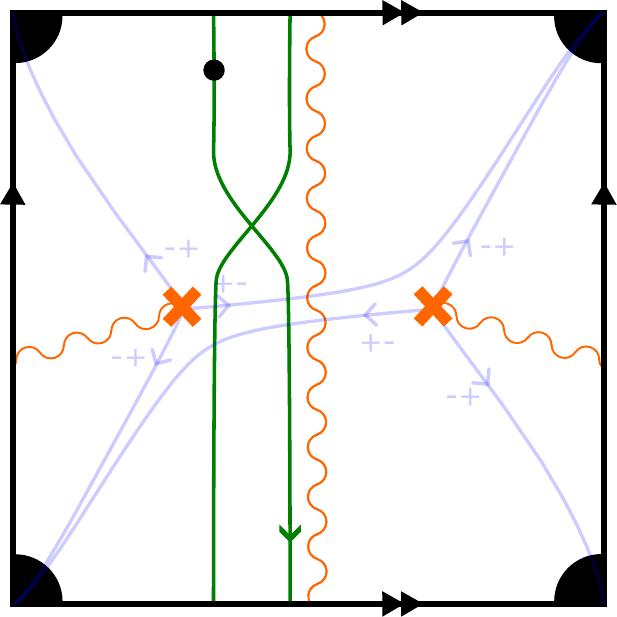}{.3}, &\qquad
    \tilde{\varrho}_{m^2}^{(2)} &= \pic{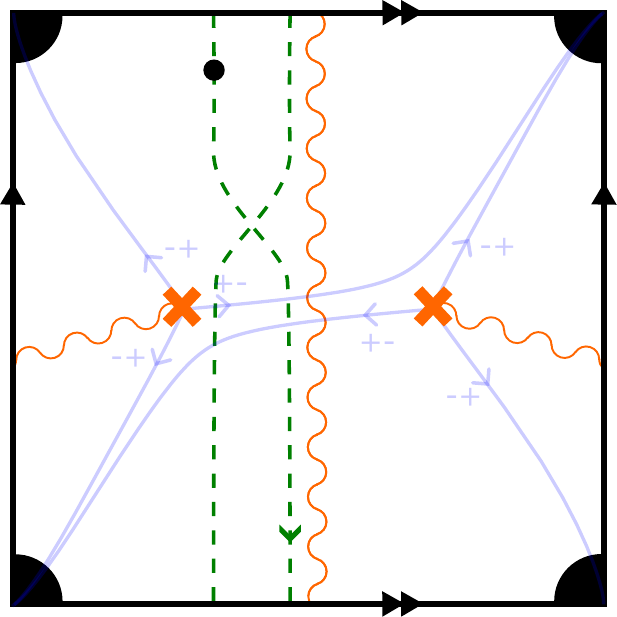}{.3}, &\qquad
    \tilde{\varrho}_{m^2}^{(3)} &= \pic{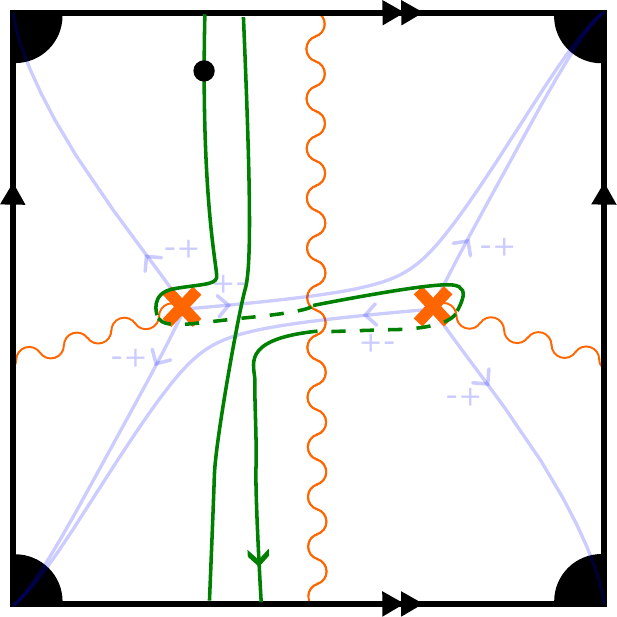}{.3}, \\[1ex]
    \tilde{\varrho}_{m^2}^{(4)} &= \pic{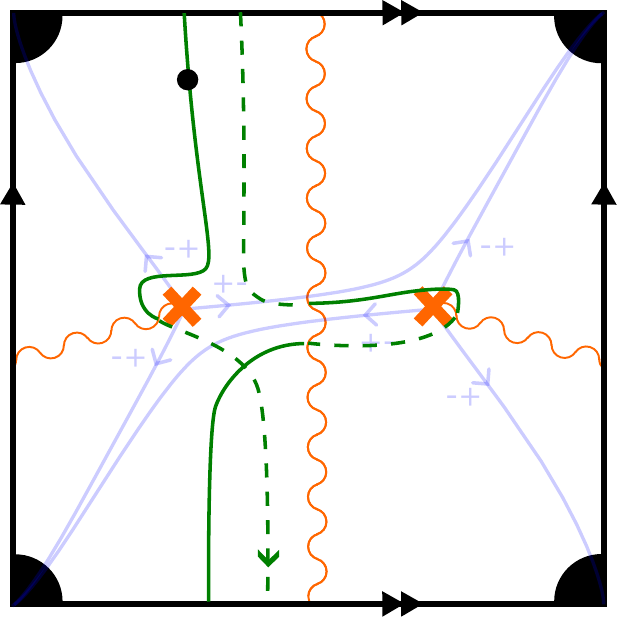}{.3}, &\qquad
    \tilde{\varrho}_{m^2}^{(5)} &= \pic{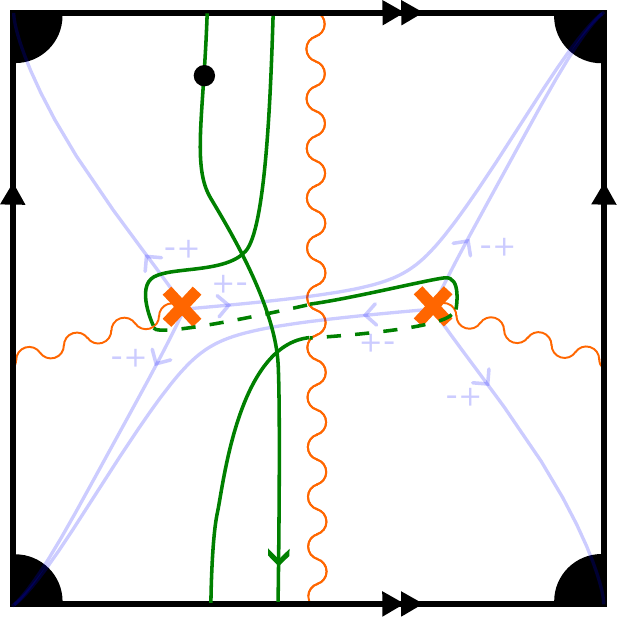}{.3}, &\qquad
    \tilde{\varrho}_{m^2}^{(6)} &= \pic{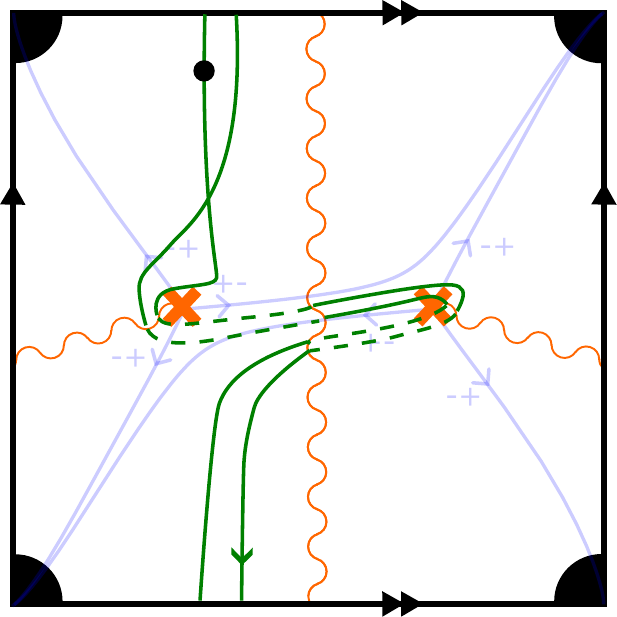}{.3}, \\[1ex]
    \tilde{\varrho}_{m^2}^{(7)} &= \pic{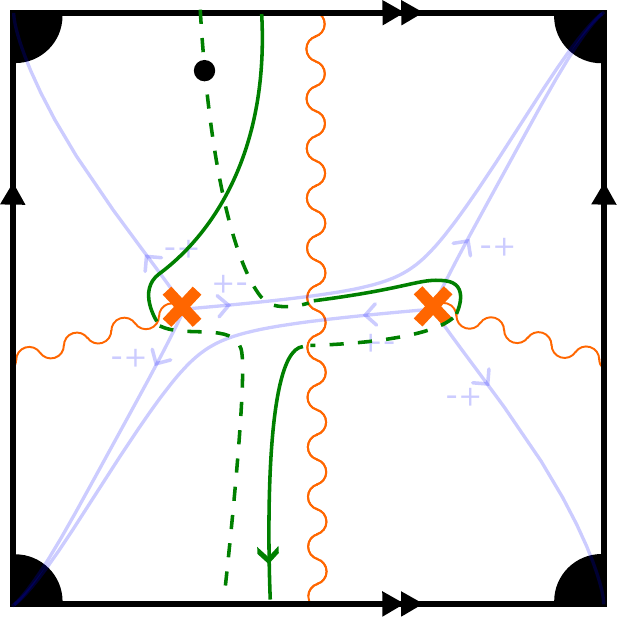}{.3}, &\qquad
    \tilde{\varrho}_{m^2}^{(8)} &= \pic{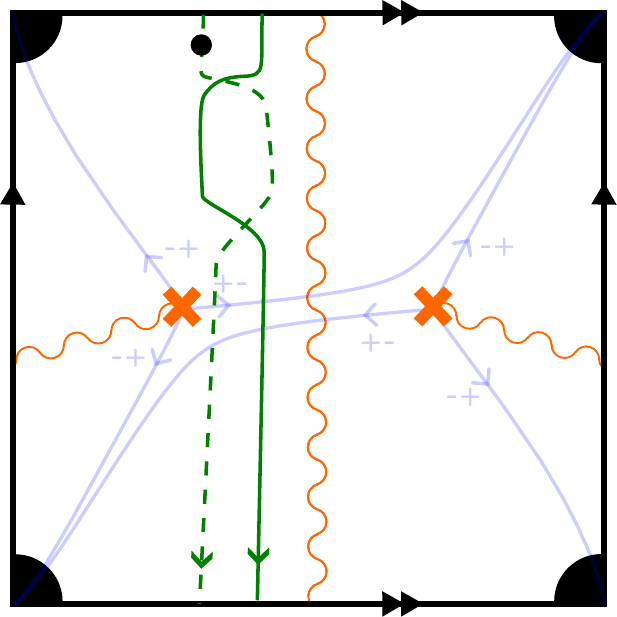}{.3}, &\qquad
    \tilde{\varrho}_{m^2}^{(9)} &= \pic{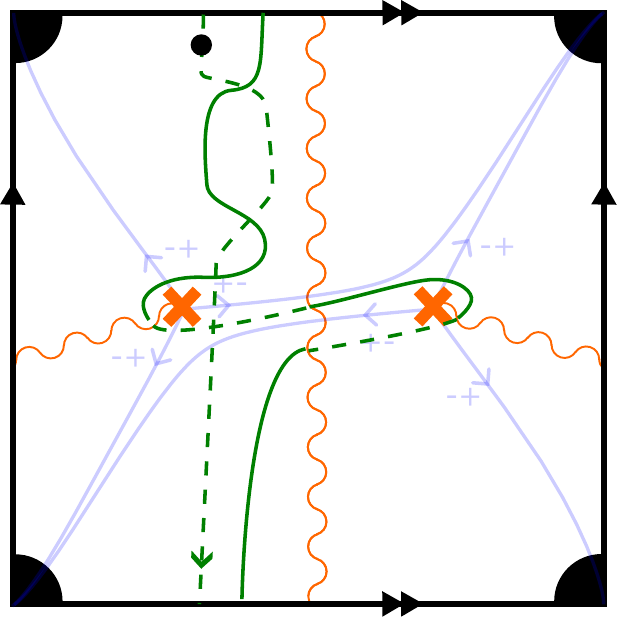}{.3}.
\end{align*}
We remark that lifts $\tilde{\varrho}_{m^2}^{(8)}$ and $\tilde{\varrho}_{m^2}^{(9)}$ include contributions from exchange paths. The weights and writhe data of these lifts is listed below
\begin{center}
\begin{tabular}{|c|c|c|c|c|c|c|} 
 \hline
 $\tilde{\varrho}$ & $\alpha_f(\tilde{\varrho})$ & $\alpha_e(\tilde{\varrho})$ & $\alpha_d(\tilde{\varrho})$ & $\alpha_w(\tilde{\varrho})$ & $(-1)^{n'(\tilde{\varrho})} \mq^{\mathfrak{wr}(\tilde{\varrho})} \hat{X}_{\tilde{\varrho}}$ & Total normal ordered product \\
 \hline\hline
 $\tilde{\varrho}_{m^2}^{(1)}$ & 1 & 1 & 1 & 1 & $\mq^{-1} \hat{X}_{2\gamma_{m,+}}$ &  $\mq^{-1} \hat{X}_{2\gamma_{m,+}}$ \\ 
 
 $\tilde{\varrho}_{m^2}^{(2)}$ & 1 & 1 & 1 & 1 & $\mq^{-1} \hat{X}_{2\gamma_{m,-}}$ & $\mq^{-1} \hat{X}_{2\gamma_{m,-}}$  \\
 
 $\tilde{\varrho}_{m^2}^{(3)}$ & 1 & 1 & 1 & 1 & $-\hat{X}_{\gamma_{m,+}+\gamma_{m,-}-\gamma_{l,+}+\gamma_{l,-}+\gamma_p}$ &  $-\hat{X}_{\gamma_{m,+}+\gamma_{m,-}}\hat{X}_{\gamma_{l,-}-\gamma_{l,+}+\gamma_p}$\\
 
 $\tilde{\varrho}_{m^2}^{(4)}$ & 1 & 1 & 1 & 1 & $-\hat{X}_{2\gamma_{m,-}+\gamma_p-\gamma_{l,+}+\gamma_{l,-}}$ & $-\mq^{-2}\hat{X}_{2\gamma_{m,-}}\hat{X}_{\gamma_{l,-}-\gamma_{l,+}+\gamma_p}$ \\
 
 $\tilde{\varrho}_{m^2}^{(5)}$ & 1 & 1 & 1 & 1 & $-\mq^{-2}\hat{X}_{\gamma_{m,+}+\gamma_{m,-}-\gamma_{l,+}+\gamma_{l,-}+\gamma_p}$ & $-\mq^{-2}\hat{X}_{\gamma_{m,+}+\gamma_{m,-}}\hat{X}_{\gamma_{l,-}-\gamma_{l,+}+\gamma_p}$ \\
 
 $\tilde{\varrho}_{m^2}^{(6)}$ & 1 & 1 & 1 & 1 & $\mq^{-1}\hat{X}_{2\gamma_p+2\gamma_{m,-}-2\gamma_{l,+}+2\gamma_{l,-}}$ & $\mq^{-5}\hat{X}_{2\gamma_{m,-}}\hat{X}_{2\gamma_{l,-}-2\gamma_{l,+}+2\gamma_p}$ \\
 
 $\tilde{\varrho}_{m^2}^{(7)}$ & 1 & 1 & 1 & 1 & $-\hat{X}_{2\gamma_{m,-}+\gamma_p-\gamma_{l,+}+\gamma_{l,-}}$ & $-\mq^{-2}\hat{X}_{2\gamma_{m,-}}\hat{X}_{\gamma_{l,-}-\gamma_{l,+}+\gamma_p}$ \\
 
 $\tilde{\varrho}_{m^2}^{(8)}$ & 1 & $\mq^{-1}-\mq$ & 1 & 1 & $\hat{X}_{\gamma_{m,+}+\gamma_{m,-}}$ & $(\mq^{-1}-\mq)\hat{X}_{\gamma_{m,+}+\gamma_{m,-}}$ \\
 
 $\tilde{\varrho}_{m^2}^{(9)}$ & 1 & $\mq^{-1}-\mq$ & 1 & 1 & $-\mq^{-1}\hat{X}_{2\gamma_{m,-}+\gamma_{l,-}-\gamma_{l,+}+\gamma_p}$ & $(\mq^{-2}-\mq^{-4})\hat{X}_{2\gamma_{m,-}}\hat{X}_{\gamma_{l,-}-\gamma_{l,+}+\gamma_p}$\\
 
 \hline
\end{tabular}
\end{center}
This gives us
\begin{equation}
\begin{aligned}
    \hat{\CL}^{\FG}_{m^2} ={}& \mq^{-1}\,\hat{X}_{2\gamma_{m,+}}
    + \hat{X}_{2\gamma_{m,-}}\left(\mq^{-1} - \left(\mq^{-2}+\mq^{-4}\right)\hat{X}_{\gamma_{l,-}-\gamma_{l,+}+\gamma_p} + \mq^{-5}\hat{X}_{\gamma_{l,-}-\gamma_{l,+}+\gamma_p}^{\,2}\right) \\[.5ex]
    & + \hat{X}_{\gamma_{m,+}+\gamma_{m,-}}\left(\mq^{-1}-\mq - \left(1+\mq^{-2}\right)\hat{X}_{\gamma_{l,-}-\gamma_{l,+}+\gamma_p}\right).
\end{aligned}
\label{eq:FG-Lm2}
\end{equation}
This operator commutes with the once-around counterpart 
\be\label{eq:FG-Commutation-m}
	[\hat\CL^{\FG}_{m^2},\hat\CL^{\FG}_m]=0\,.
\ee

\paragraph{The operator $\hat\CL_{l^2}^{\FG}$.} 
The lifts of $\varrho_{l^2}$ are given by
\allowdisplaybreaks   
\begin{align*}
    \tilde{\varrho}_{l^2}^{(1)} &= \pic{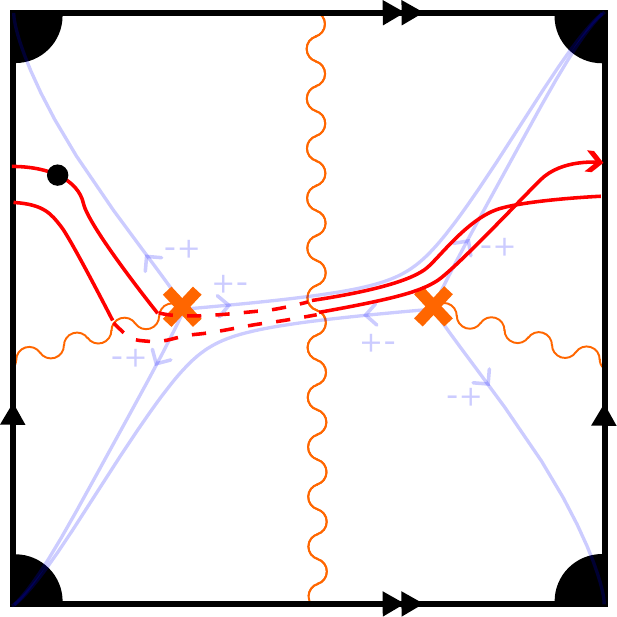}{.3}, &\qquad
    \tilde{\varrho}_{l^2}^{(2)} &= \pic{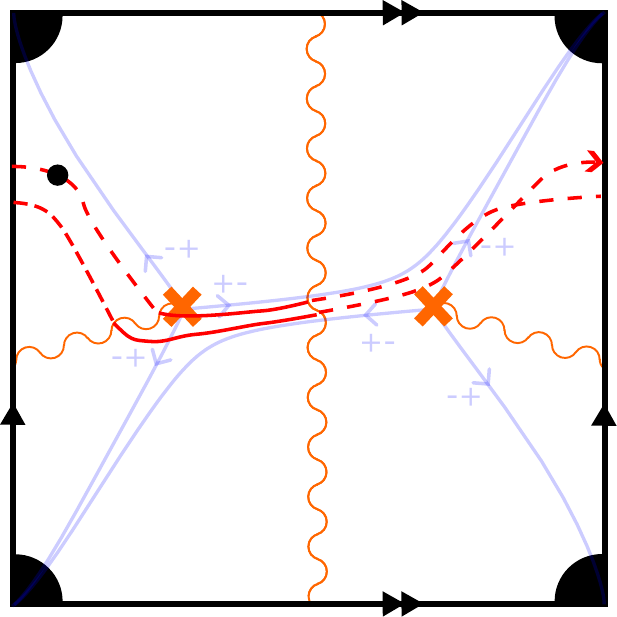}{.3}, &\qquad
    \tilde{\varrho}_{l^2}^{(3)} &= \pic{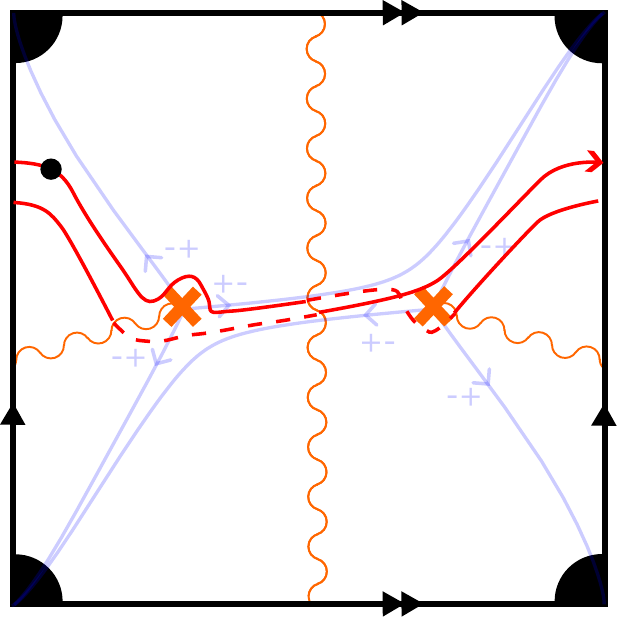}{.3}, \\[1ex]
    \tilde{\varrho}_{l^2}^{(4)} &= \pic{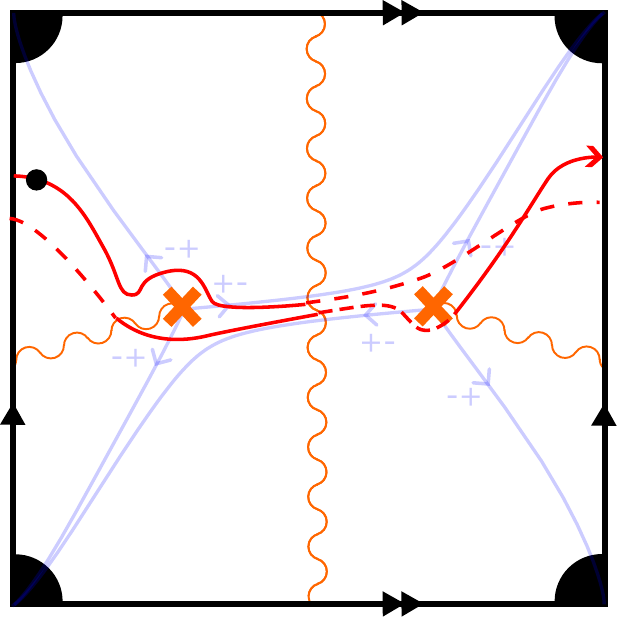}{.3}, &\qquad
    \tilde{\varrho}_{l^2}^{(5)} &= \pic{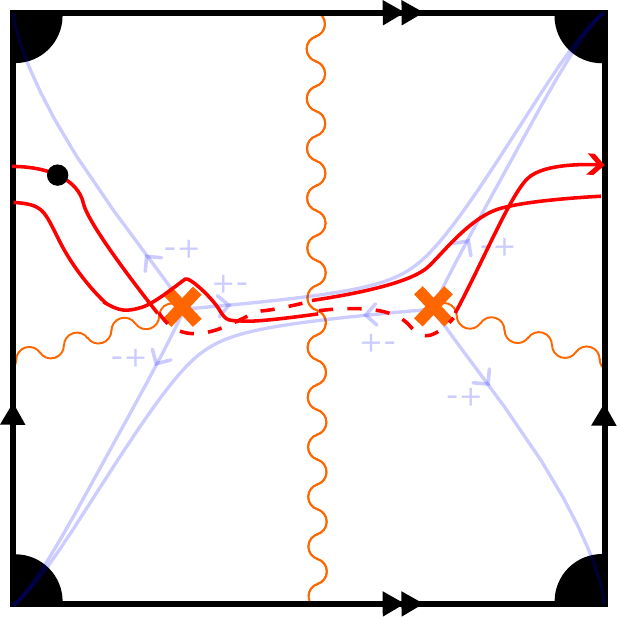}{.3}, &\qquad
    \tilde{\varrho}_{l^2}^{(6)} &= \pic{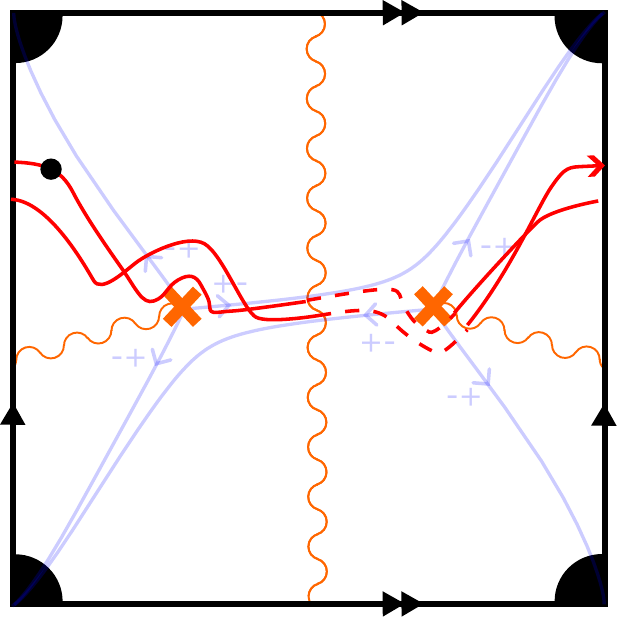}{.3}, \\[1ex]
    \tilde{\varrho}_{l^2}^{(7)} &= \pic{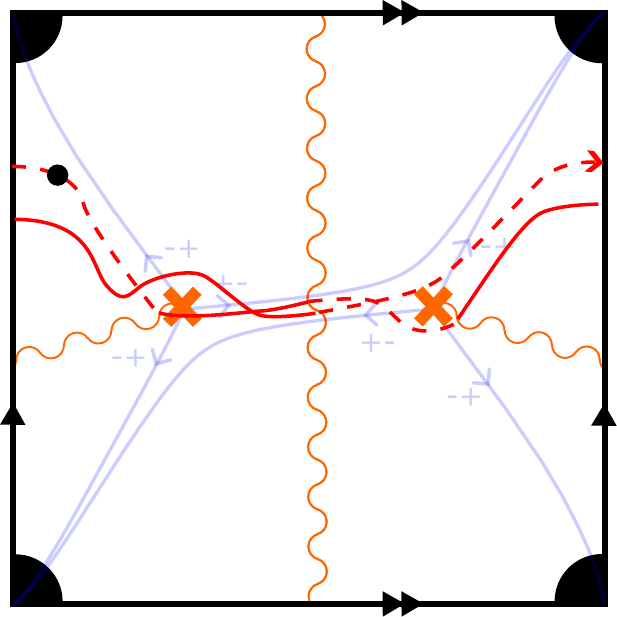}{.3}, &\qquad
    \tilde{\varrho}_{l^2}^{(8)} &= \pic{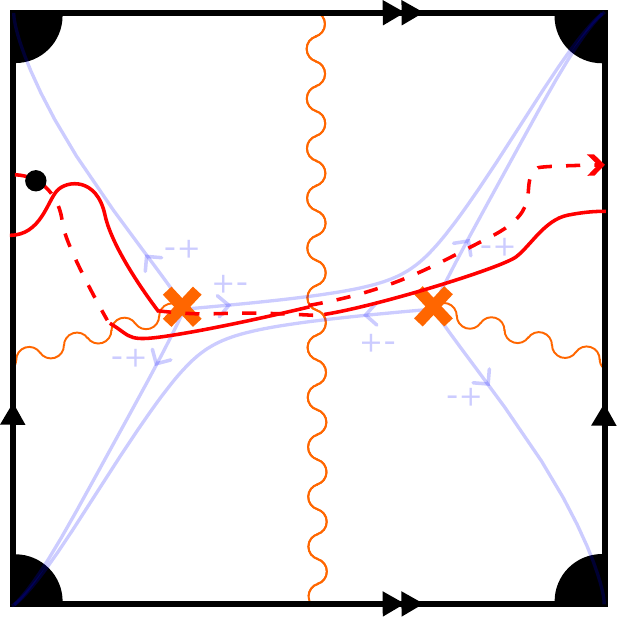}{.3}, &\qquad
    \tilde{\varrho}_{l^2}^{(9)} &= \pic{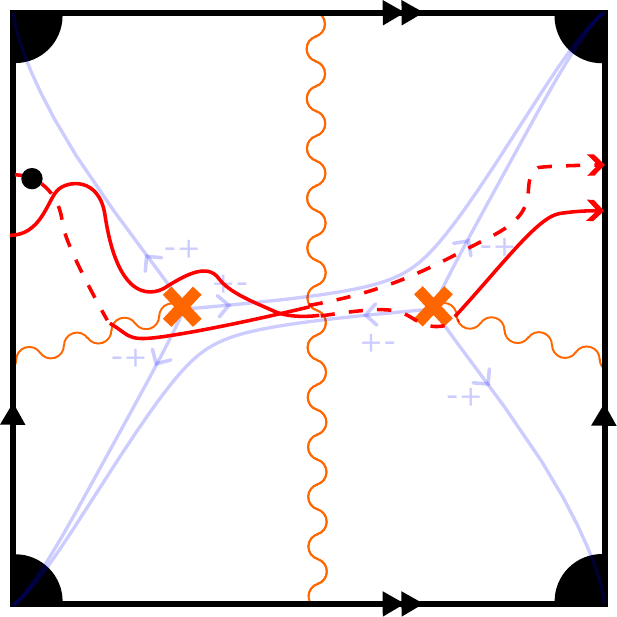}{.3}.
\end{align*}
Their weight and writhe data is collected in the table below.
\begin{center}
\begin{tabular}{|c|c|c|c|c|c|c|} 
 \hline
 $\tilde{\varrho}$ & $\alpha_f(\tilde{\varrho})$ & $\alpha_e(\tilde{\varrho})$ & $\alpha_d(\tilde{\varrho})$ & $\alpha_w(\tilde{\varrho})$ & $(-1)^{n'(\tilde{\varrho})} \mq^{\mathfrak{wr}(\tilde{\varrho})} \hat{X}_{\tilde{\varrho}}$ & Total normal ordered product \\
 \hline\hline
 $\tilde{\varrho}_{l^2}^{(1)}$ & 1 & 1 & 1 & 1 & $\mq \hat{X}_{2\gamma_{l,+}}$ & $\mq \hat{X}_{2\gamma_{l,+}}$ \\ 
 
 $\tilde{\varrho}_{l^2}^{(2)}$ & 1 & 1 & 1 & 1 & $\mq \hat{X}_{2\gamma_{l,-}}$ & $\mq \hat{X}_{2\gamma_{l,-}}$ \\
 
 $\tilde{\varrho}_{l^2}^{(3)}$ & 1 & 1 & 1 & 1 & $-\hat{X}_{\gamma_{l,+}+\gamma_{l,-}+\gamma_{m,+}-\gamma_{m,-}-\gamma_p}$ & $-\hat{X}_{\gamma_{m,+}-\gamma_{m,-}-\gamma_p}\hat{X}_{\gamma_{l,+}+\gamma_{l,-}}$ \\
 
 $\tilde{\varrho}_{l^2}^{(4)}$ & 1 & 1 & 1 & 1 & $-\hat{X}_{2\gamma_{l,+}+\gamma_{m,+}-\gamma_{m,-}-\gamma_p}$ & $-\mq^{-2}\hat{X}_{\gamma_{m,+}-\gamma_{m,-}-\gamma_p}\hat{X}_{2\gamma_{l,+}}$ \\
 
 $\tilde{\varrho}_{l^2}^{(5)}$ & 1 & 1 & 1 & 1 & $-\mq^2\hat{X}_{\gamma_{l,+}+\gamma_{l,-}+\gamma_{m,+}-\gamma_{m,-}-\gamma_p}$ & $-\mq^{2}\hat{X}_{\gamma_{m,+}-\gamma_{m,-}-\gamma_p}\hat{X}_{\gamma_{l,+}+\gamma_{l,-}}$ \\
 
 $\tilde{\varrho}_{l^2}^{(6)}$ & 1 & 1 & 1 & 1 & $\mq\hat{X}_{2\gamma_{l,+}+2\gamma_{m,+}-2\gamma_{m,-}-2\gamma_p}$ & $\mq^{-3}\hat{X}_{2\gamma_{m,+}-2\gamma_{m,-}-2\gamma_p}\hat{X}_{2\gamma_{l,+}}$ \\
 
 $\tilde{\varrho}_{l^2}^{(7)}$ & 1 & 1 & 1 & 1 & $-\hat{X}_{2\gamma_{l,+}+\gamma_{m,+}-\gamma_{m,-}-\gamma_p}$ & $-\mq^{-2}\hat{X}_{\gamma_{m,+}-\gamma_{m,-}-\gamma_p}\hat{X}_{2\gamma_{l,+}}$ \\
 
 $\tilde{\varrho}_{l^2}^{(8)}$ & 1 & $\mq-\mq^{-1}$ & 1 & 1 & $\hat{X}_{\gamma_{l,+}+\gamma_{l,-}}$ & $(\mq-\mq^{-1})\hat{X}_{\gamma_{l,+}+\gamma_{l,-}}$ \\
 
 $\tilde{\varrho}_{l^2}^{(9)}$ & 1 & $\mq-\mq^{-1}$ & 1 & 1 & $-\mq\hat{X}_{2\gamma_{l,+}+\gamma_{m,+}-\gamma_{m,-}-\gamma_p}$ & $(\mq^{-2}-1)\hat{X}_{\gamma_{m,+}-\gamma_{m,-}-\gamma_p}\hat{X}_{2\gamma_{l,+}}$ \\
 
 \hline
\end{tabular}
\end{center}
This gives us
\begin{equation}
\begin{aligned}
    \hat{\CL}^{\FG}_{l^2} ={}& \mq\,\hat{X}_{2\gamma_{l,-}}
    + \left(\mq - \left(1+\mq^{-2}\right)\hat{X}_{\gamma_{m,+}-\gamma_{m,-}-\gamma_p} + \mq^{-3}\hat{X}_{\gamma_{m,+}-\gamma_{m,-}-\gamma_p}^{\,2}\right)\hat{X}_{2\gamma_{l,+}} \\[.5ex]
    & + \left(\mq-\mq^{-1} - \left(1+\mq^{2}\right)\hat{X}_{\gamma_{m,+}-\gamma_{m,-}-\gamma_p}\right)\hat{X}_{\gamma_{l,+}+\gamma_{l,-}}.
\end{aligned}
\label{eq:FG-Ll2}
\end{equation}
This operator commutes with the once-around counterpart 
\be\label{eq:FG-Commutation-l}
	[\hat\CL^{\FG}_{l^2},\hat\CL^{\FG}_l]=0\,.
\ee

\subsection{Fenchel-Nielsen charts}
We next compute the $\mq$-nonabelianization map for the Fenchel-Nielsen network of Figure \ref{fig:n2starFNA} in American resolution. 
Here we will focus on the computation of images $\hat\CL_m^{\FN_-}, \hat\CL_l^{\FN_-},\hat\CL_{m^2}^{\FN_-} \in \Sk(\Sigma\times I, GL_1)$ corresponding to generators $m,l,m^2$ of $\Sk(C\times I, GL_2)$. We choose path representatives as shown below
\begin{align*}
    \varrho_{m}   = \pic{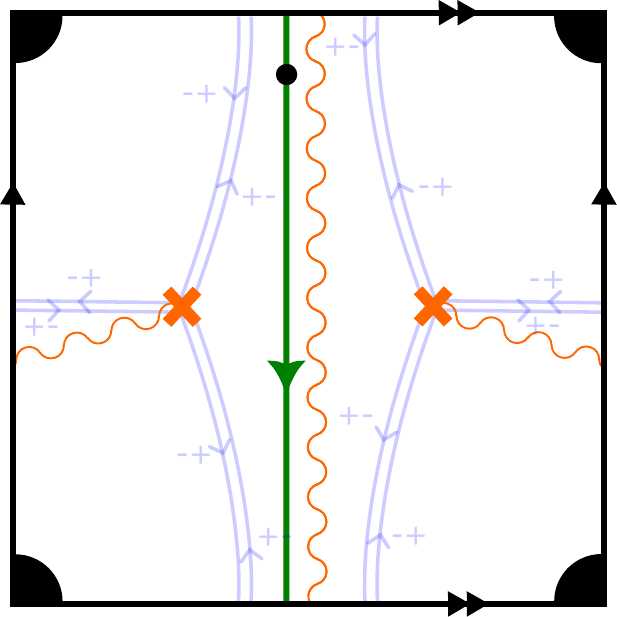}{.4}, \quad 
    \varrho_{l}   = \pic{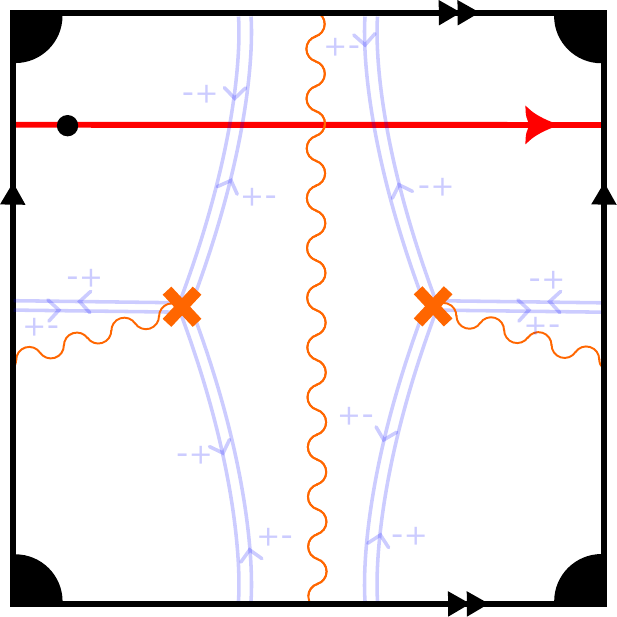}{.4},  \quad
    \varrho_{m^2} = \pic{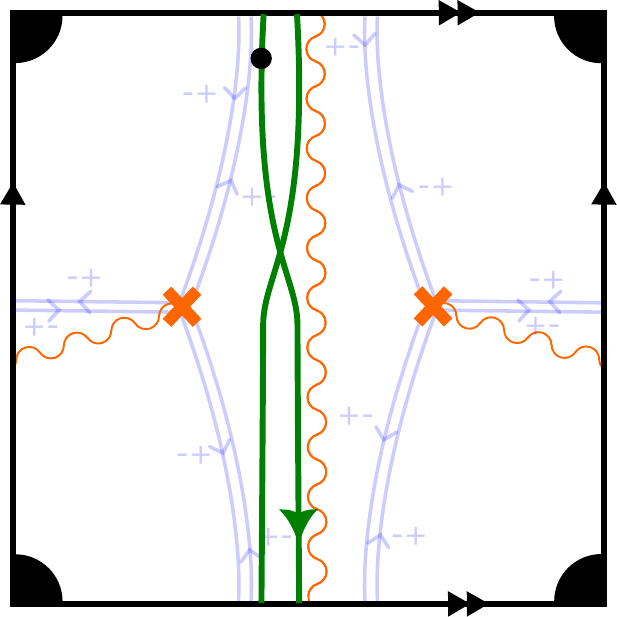}{.4}.
\end{align*}
The analysis of $\hat\CL_{l^2}^{\FN_-}$ turns out to be significantly more involved. We will omit this and obtain the desired result instead using framed wall-crossing relations later in Section \ref{eq:quantum-chart-relations}.
The analogous computation for British resolution shown in Figure \ref{fig:n2starFNB} is very similar and we omit it, but provide its results in Section \ref{sec:FG-FN-Results}.

\paragraph{The operator $\hat\CL_m^{\FN_-}$.}
There are just two lifts of $\varrho_m$ to $\Sigma\times I$
    \begin{align*}
    \tilde{\varrho}_{m}^{(1)} = \pic{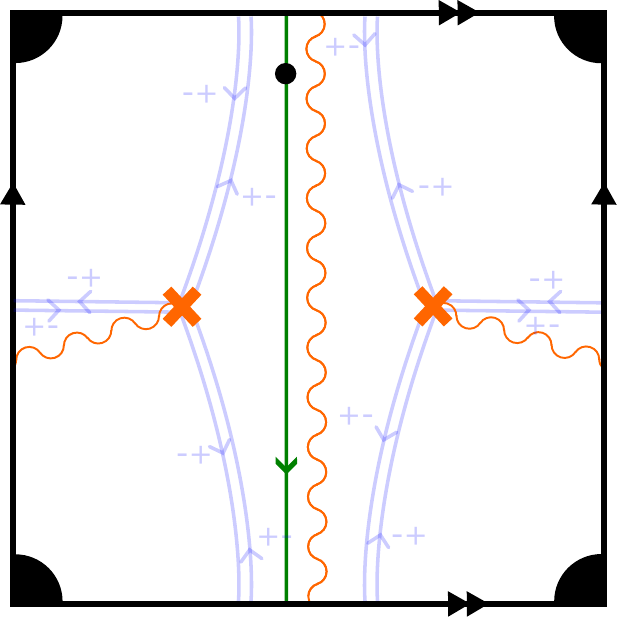}{.3}, \qquad
    \tilde{\varrho}_{m}^{(2)} = \pic{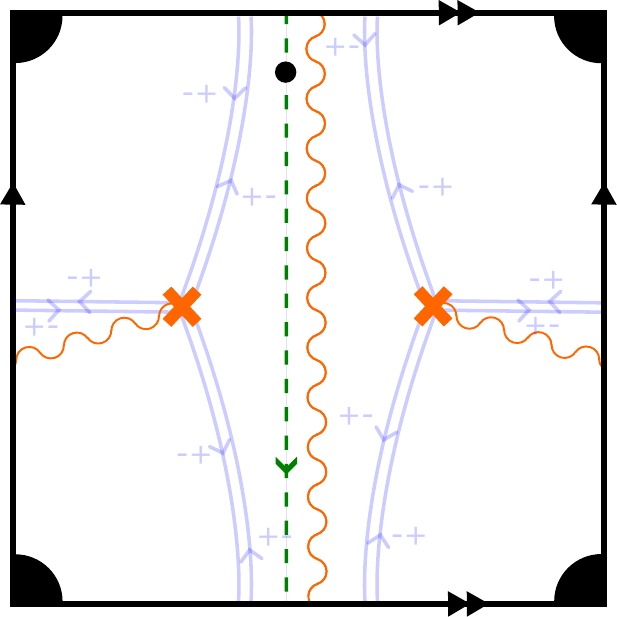}{.3}.
\end{align*}
The corresponding table of weights and writhe is as follows.
\begin{center}
\begin{tabular}{|c|c|c|c|c|c|c|}
 \hline
 $\tilde{\varrho}$ & $\alpha_f(\tilde{\varrho})$ & $\alpha_e(\tilde{\varrho})$ & $\alpha_d(\tilde{\varrho})$ & $\alpha_w(\tilde{\varrho})$ & $(-1)^{n'(\tilde{\varrho})} \mq^{\mathfrak{wr}(\tilde{\varrho})} \hat{X}_{\tilde{\varrho}}$ & Total normal ordered product \\
 \hline\hline
 $\tilde{\varrho}_{m}^{(1)}$ & 1 & 1  &  1 & 1  & $\hat{X}_{\gamma_{m,+}}$  & $\hat{X}_{\gamma_{m,+}}$   \\
 $\tilde{\varrho}_{m}^{(2)}$ & 1  & 1   & 1 & 1  & $\hat{X}_{\gamma_{m,-}}$  & $\hat{X}_{\gamma_{m,-}}$    \\
 \hline
\end{tabular}
\end{center}
Summimng them up gives the following element of $\Sk(\Sigma\times I, GL_1)$
\begin{equation}
    \hat\CL^{\FN_-}_m = \hat{X}_{\gamma_{m,+}} + \hat{X}_{\gamma_{m,-}}.
    \label{eq:FN-Lm}
\end{equation}

\paragraph{The operator $\hat\CL_l^{\FN_-}$.} 
There is an infinite number of lifts of $\varrho_l$ to $\Sigma\times I$, because of the infinite number of detours possible. 
However, all detours are generated by intersections with only three trajectories of the network, and the source of infinity is that these trajectories wind around annular regions on $C$ multiple times.
This motivates a classification of lifts into two broad classes: those that involve detours from just one of the three trajectories, and those that involve detours from all three.\footnote{An odd number of detours is necessary to obtain a closed lift because $\varrho_l$ crosses a square root branch cut.}
We refer to these as the class of `single-detour' lifts and `triple-detour' lifts. We remark that the former will give us a $\mathbb{Z}_{\geq0}$-graded geometric series while the latter will give us $\mathbb{Z}_{\geq 0}^3$-graded geometric series. 
Additionally there are contributions from exchange flow lines. As we explain below, these turn out to contribute overall multiplicative constant to certain classes of detour paths. 
Therefore our strategy will be to first analyse the single-detours and triple-detours without the exchanges, and then include the effect of exchanges as appropriate.

\bigskip

There are ten families of single-detour paths, each giving us a $\mathbb{Z}_{\geq 0}$-graded geometric series.

\begin{enumerate}
    \item The first family of single-detours is given by
\begin{align*}
    \tilde{\varrho}_{l}^{(1,0)} = \pic{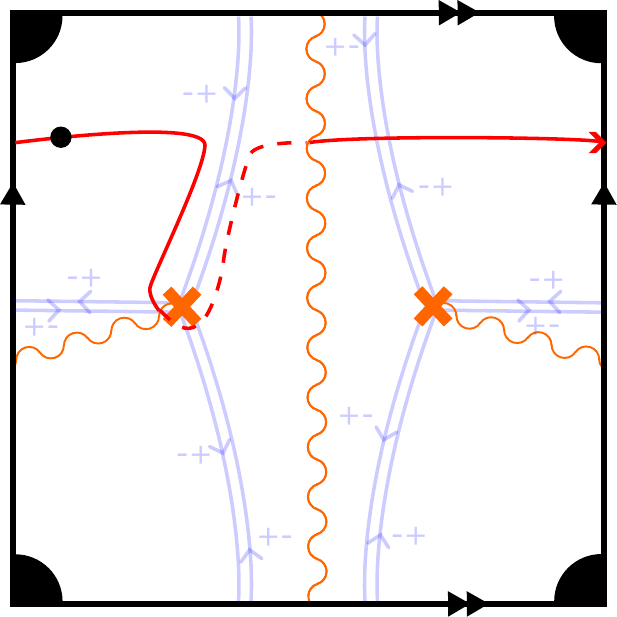}{.3}, \qquad
    \tilde{\varrho}_{l}^{(1,1)} = \pic{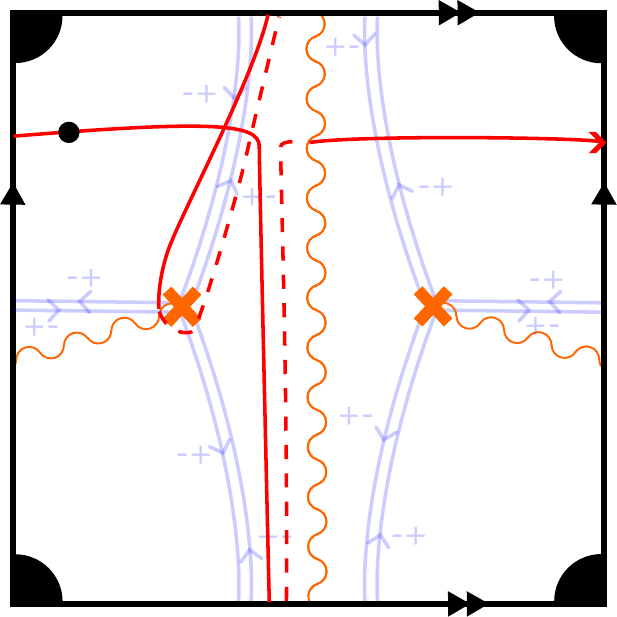}{.3}, \qquad \dots
\end{align*}
where ellipses denote additional winding around the meridian cycle. The weights of each lifted path are summarized below
\begin{center}
\begin{tabular}{|c|c|c|c|c|c|c|}
 \hline
 $\tilde{\varrho}$ & $\alpha_f(\tilde{\varrho})$ & $\alpha_e(\tilde{\varrho})$ & $\alpha_d(\tilde{\varrho})$ & $\alpha_w(\tilde{\varrho})$ & $(-1)^{n'(\tilde{\varrho})} \mq^{\mathfrak{wr}(\tilde{\varrho})} \hat{X}_{\tilde{\varrho}}$ & Total normal ordered product \\
 \hline\hline
 $\tilde{\varrho}_{l}^{(1,0)}$ & $\mq^\frac12$ & 1  &  $\mq^{-\frac12}$ & $1$ & $\hat{X}_{\gamma_{l,-}}$  & $\hat{X}_{\gamma_{l,-}}$   \\
 $\tilde{\varrho}_{l}^{(1,1)}$ & $\mq^\frac12$  & 1   & $\mq^{-\frac12}$ & $1$  &
 $(-\mq)^{-1}\hat{X}_{\gamma_{l,-}+\gamma_{m,+}-\gamma_{m,-}}$  & $\hat{X}_{\gamma_{m,+}-\gamma_{m,-}} \hat{X}_{\gamma_{l,-}}$    \\
 \vdots &  \vdots & \vdots & \vdots & \vdots & \vdots & \vdots     \\
 $\tilde{\varrho}_{l}^{(1,n)}$ & $\mq^\frac12$  & 1   & $\mq^{-\frac12}$ & $1$  &
 $(-\mq)^{-n}\hat{X}_{\gamma_{l,-}+n(\gamma_{m,+}-\gamma_{m,-})}$  & $\hat{X}_{n(\gamma_{m,+}-\gamma_{m,-})}\hat{X}_{\gamma_{l,-}}$    \\
 \vdots &  \vdots & \vdots & \vdots & \vdots & \vdots & \vdots    \\
 \hline
\end{tabular}
\end{center}
Summing over all contributions gives the following element of $\Sk(\Sigma\times I, GL_1)$
\begin{equation}
    \hat\CL_{l}^{(1)} = \frac{1}{1- \hat{X}_{\gamma_{m,+}-\gamma_{m,-}}}\hat{X}_{\gamma_{l,-}} \;.
\end{equation}

\item The second family of single-detours is given by
\begin{align*}
    \tilde{\varrho}_{l}^{(2,0)} = \pic{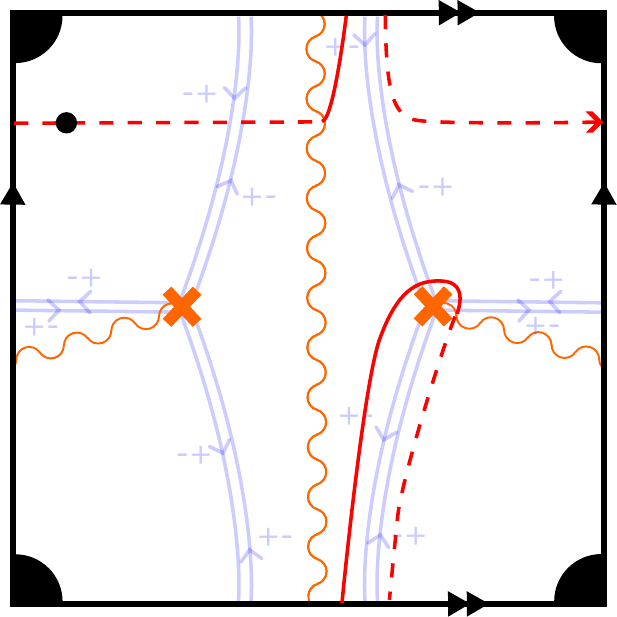}{.3}, \qquad
    \tilde{\varrho}_{l}^{(2,1)} = \pic{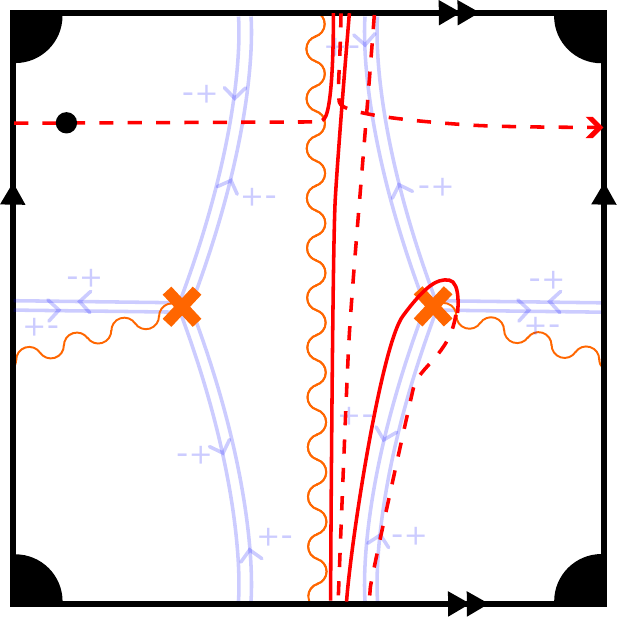}{.3}, \qquad \dots
\end{align*}
where ellipses denote paths with adiditional winding along the meridian cycle. The weights of these detours are given in the table below
\begin{center}
\begin{tabular}{|c|c|c|c|c|c|c|}
 \hline
 $\tilde{\varrho}$ & $\alpha_f(\tilde{\varrho})$ & $\alpha_e(\tilde{\varrho})$ & $\alpha_d(\tilde{\varrho})$ & $\alpha_w(\tilde{\varrho})$ & $(-1)^{n'(\tilde{\varrho})} \mq^{\mathfrak{wr}(\tilde{\varrho})} \hat{X}_{\tilde{\varrho}}$ & Total normal ordered product \\
 \hline\hline
 $\tilde{\varrho}_{l}^{(2,0)}$ & $\mq^\frac12$ & 1  &  $\mq^{\frac12}$ & $1$ & $\hat{X}_{\gamma_p+\gamma_{l,-}}$  & $\mq\hat{X}_{\gamma_p+\gamma_{l,-}}$   \\
 $\tilde{\varrho}_{l}^{(2,1)}$ & $\mq^\frac12$  & 1   & $\mq^{\frac12}$ & $1$  &
 $-\mq\hat{X}_{\gamma_p+\gamma_{l,-}+\gamma_{m,+}-\gamma_{m,-}}$  & $\mq^3 \hat{X}_{\gamma_p+\gamma_{m,+}-\gamma_{m,-}} \hat{X}_{\gamma_{l,-}}$    \\
 \vdots &  \vdots & \vdots & \vdots & \vdots & \vdots & \vdots     \\
 $\tilde{\varrho}_{l}^{(2,n)}$ & $\mq^\frac12$  & 1   & $\mq^{\frac12}$ & $1$  &
 $(-\mq)^{n}\hat{X}_{\gamma_p+\gamma_{l,-}+n(\gamma_{m,+}-\gamma_{m,-})}$  & $\mq^{2n+1}\hat{X}_{\gamma_p+n(\gamma_{m,+}-\gamma_{m,-})}\hat{X}_{\gamma_{l,-}}$    \\
 \vdots &  \vdots & \vdots & \vdots & \vdots & \vdots & \vdots    \\
 \hline
\end{tabular}
\end{center}
Summing up all contributions gives
\begin{equation}
    \hat\CL_l^{(2)} = \frac{\mq\hat{X}_{\gamma_p}}{1-\mq^2 \hat{X}_{\gamma_{m,+}-\gamma_{m,-}}}\hat{X}_{\gamma_{l,-}}.
\end{equation}

\item The third family of single-detours is given by
\begin{align*}
    \tilde{\varrho}_{l}^{(3,0)} = \pic{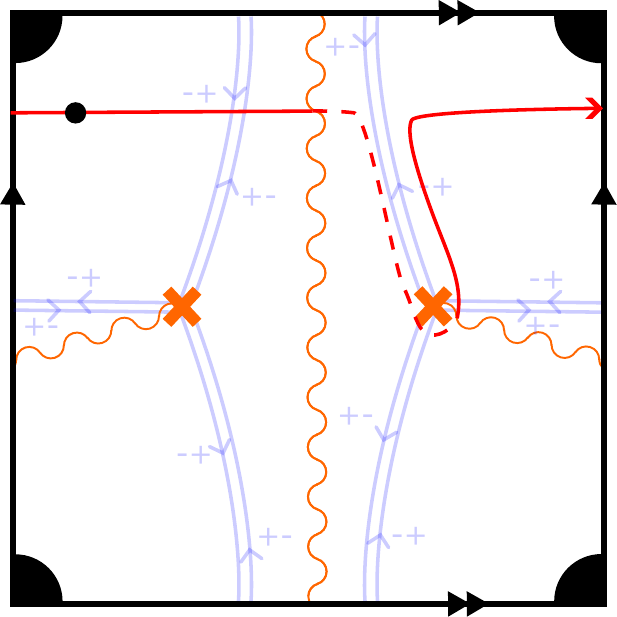}{.3}, \qquad
    \tilde{\varrho}_{l}^{(3,1)} = \pic{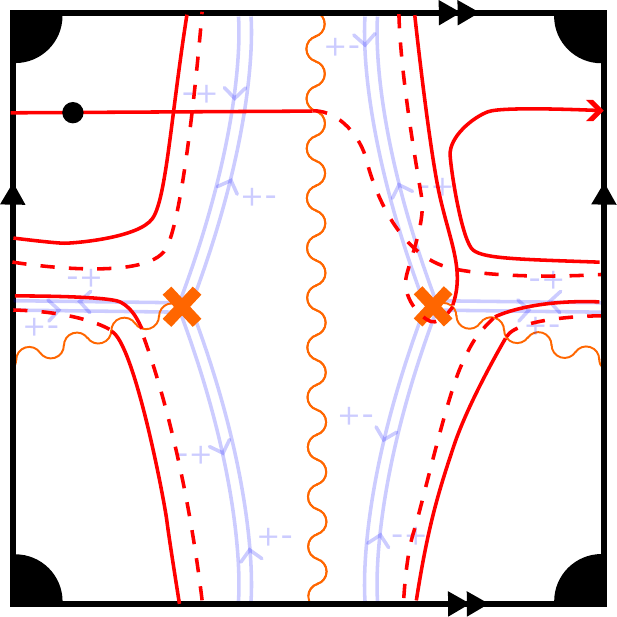}{.3}, \qquad \dots
\end{align*}
with weights
\begin{center}
\resizebox{\textwidth}{!}{%
\begin{tabular}{|c|c|c|c|c|c|c|}
 \hline
 $\tilde{\varrho}$ & $\alpha_f(\tilde{\varrho})$ & $\alpha_e(\tilde{\varrho})$ & $\alpha_d(\tilde{\varrho})$ & $\alpha_w(\tilde{\varrho})$ & $(-1)^{n'(\tilde{\varrho})} \mq^{\mathfrak{wr}(\tilde{\varrho})} \hat{X}_{\tilde{\varrho}}$ & Total normal ordered product \\
 \hline\hline
 $\tilde{\varrho}_{l}^{(3,0)}$ & $\mq^\frac12$ & 1  &  $\mq^{-\frac12}$ & $1$ & $\hat{X}_{\gamma_{l,+}+\gamma_{m,+}-\gamma_{m,-}-\gamma_p}$  & $-\mq^{-1}\hat{X}_{\gamma_{m,+}-\gamma_{m,-}-\gamma_p}\hat{X}_{\gamma_{l,+}}$   \\
 $\tilde{\varrho}_{l}^{(3,1)}$ & $\mq^\frac12$  & 1   & $\mq^{-\frac12}$ & $1$  &
 $\mq^{-2}\hat{X}_{\gamma_{l,+}+\gamma_{m,+}-\gamma_{m,-}+\gamma_p}$  & $-\mq^{-3} \hat{X}_{\gamma_{m,+}-\gamma_{m,-}+\gamma_p} \hat{X}_{\gamma_{l,+}}$    \\
 \vdots &  \vdots & \vdots & \vdots & \vdots & \vdots & \vdots     \\
 $\tilde{\varrho}_{l}^{(3,n)}$ & $\mq^\frac12$  & 1   & $\mq^{-\frac12}$ & $1$  &
 $\mq^{-2n}\hat{X}_{\gamma_{l,+}+\gamma_{m,+}-\gamma_{m,-}+(2n-1)\gamma_p}$  & $-\mq^{-2n-1}\hat{X}_{\gamma_{m,+}-\gamma_{m,-}+(2n-1)\gamma_p}\hat{X}_{\gamma_{l,+}}$    \\
 \vdots &  \vdots & \vdots & \vdots & \vdots & \vdots & \vdots    \\
 \hline
\end{tabular}}
\end{center}
The sum of these lifts gives
\begin{equation}
    \hat\CL_l^{(3)}=-\frac{\mq^{-1}\hat{X}_{\gamma_{m,+}-\gamma_{m,-}-\gamma_p}}{1-\mq^{-2}\hat{X}_{2\gamma_p}} \hat{X}_{\gamma_{l,+}}\;.
\end{equation}

\item
The fourth, fifth and sixth family of single-detours has the same geometric series (except for the initial element) as the $\tilde{\varrho}_l^{(3,n)}$ family (with $n\in\mathbb{Z}_{\geq 0}$), because the detour is taken at exactly the same point, and the full loop contribution is also the same. Hence, knowing the initial elements of each of those families,
\begin{align*}
    \tilde{\varrho}_{l}^{(4,0)} = \pic{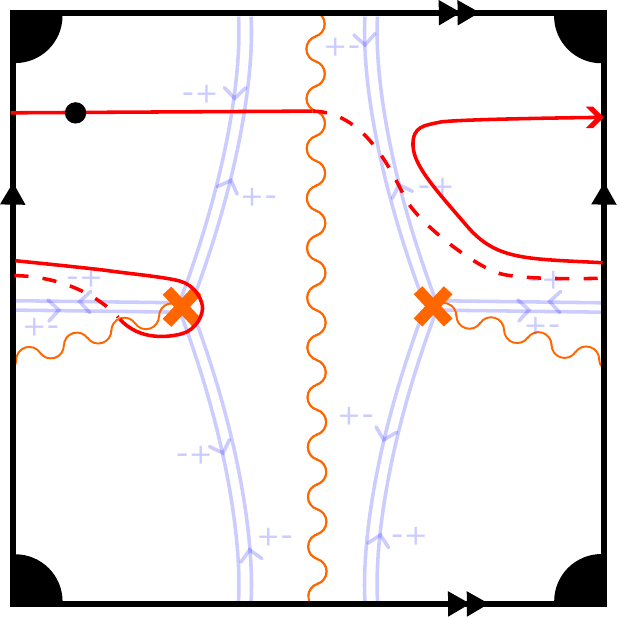}{.3}, \qquad
    \tilde{\varrho}_{l}^{(5,0)} = \pic{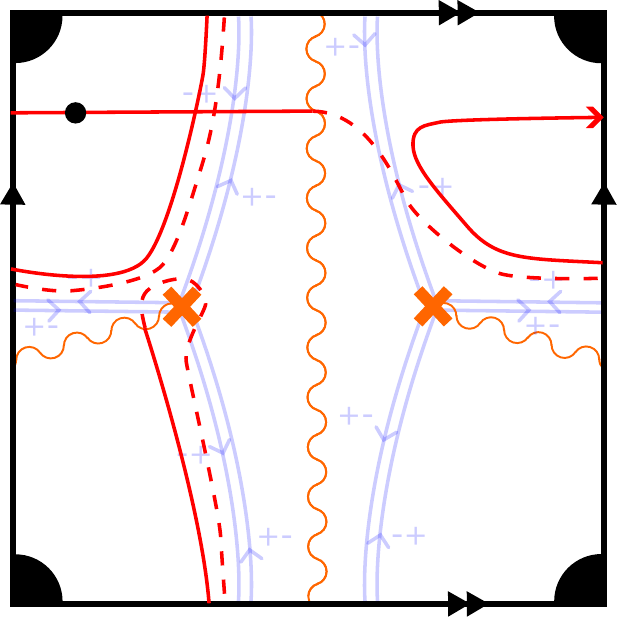}{.3}, \qquad
    \tilde{\varrho}_{l}^{(6,0)} = \pic{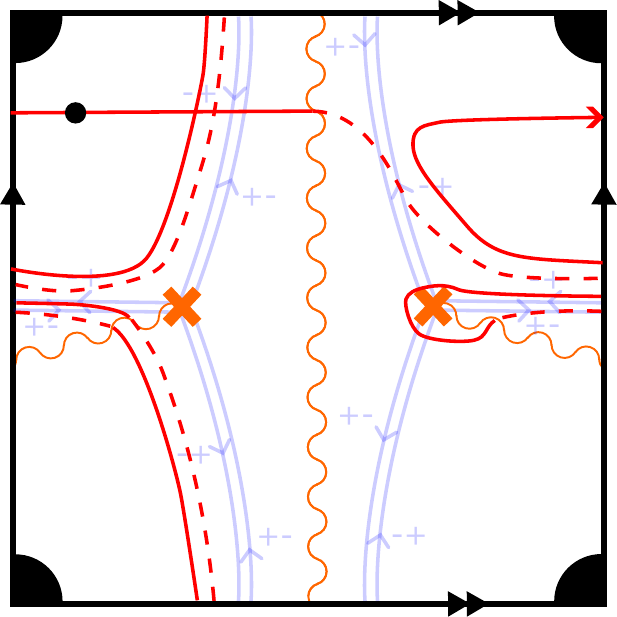}{.3}, \qquad
\end{align*}
gives us
\begin{equation}
    \hat\CL_l^{(4)} = \frac{1}{1-\mq^{-2}\hat{X}_{2\gamma_p}} \hat{X}_{\gamma_{l,+}}\;,
\end{equation}
\begin{equation}
    \hat\CL_l^{(5)} = \frac{\mq^{-2}\hat{X}_{\gamma_{m,+}-\gamma_{m,-}}}{1-\mq^{-2}\hat{X}_{2\gamma_p}}\hat{X}_{\gamma_{l,+}} \;,
\end{equation}
\begin{equation}
    \hat\CL_l^{(6)}=-\frac{\mq^{-1}\hat{X}_{\gamma_p}}{1-\mq^{-2}\hat{X}_{2\gamma_p}}\hat{X}_{\gamma_{l,+}} \;.
\end{equation}

\item Just like the previous point, the seventh till tenth families also have the same geometric series modulo the initial element. To that end, we analyse the seventh family of detours using the lifts,
\begin{align*}
    \tilde{\varrho}_{l}^{(7,0)} = \pic{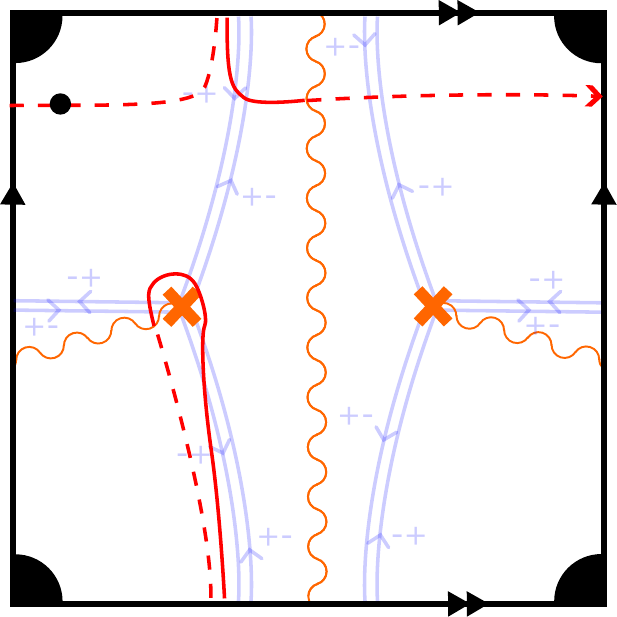}{.3}, \qquad
    \tilde{\varrho}_{l}^{(7,1)} = \pic{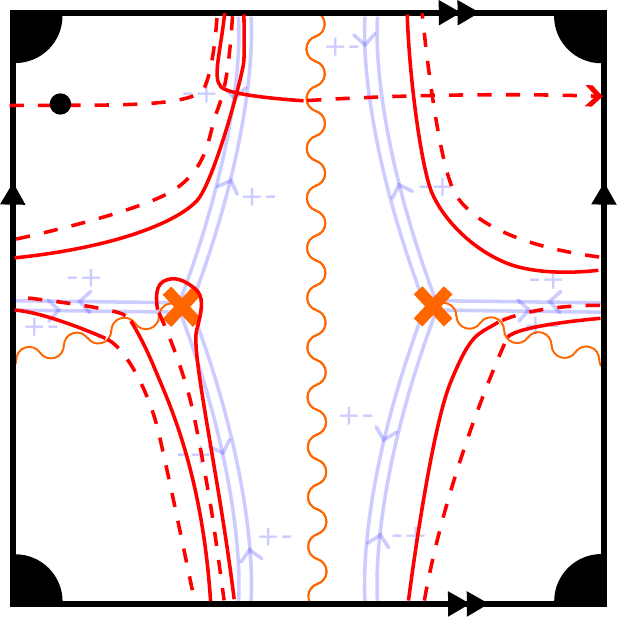}{.3}, \qquad \dots
\end{align*}
These have the following weights
\begin{center}
\begin{tabular}{|c|c|c|c|c|c|c|}
 \hline
 $\tilde{\varrho}$ & $\alpha_f(\tilde{\varrho})$ & $\alpha_e(\tilde{\varrho})$ & $\alpha_d(\tilde{\varrho})$ & $\alpha_w(\tilde{\varrho})$ & $(-1)^{n'(\tilde{\varrho})} \mq^{\mathfrak{wr}(\tilde{\varrho})} \hat{X}_{\tilde{\varrho}}$ & Total normal ordered product \\
 \hline\hline
 $\tilde{\varrho}_{l}^{(7,0)}$ & $\mq^\frac12$ & 1  &  $\mq^{\frac12}$ & $1$ & $\hat{X}_{\gamma_{l,+}+\gamma_{m,+}-\gamma_{m,-}}$  & $-\hat{X}_{\gamma_{m,+}-\gamma_{m,-}}\hat{X}_{\gamma_{l,+}}$   \\
 $\tilde{\varrho}_{l}^{(7,1)}$ & $\mq^\frac12$  & 1   & $\mq^{\frac12}$ & $1$  &
 $\mq^{2}\hat{X}_{\gamma_{l,+}+\gamma_{m,+}-\gamma_{m,-}+\gamma_p}$  & $-\mq^2 \hat{X}_{\gamma_{m,+}-\gamma_{m,-}+2\gamma_p} \hat{X}_{\gamma_{l,+}}$    \\
 \vdots &  \vdots & \vdots & \vdots & \vdots & \vdots & \vdots     \\
 $\tilde{\varrho}_{l}^{(7,n)}$ & $\mq^\frac12$  & 1   & $\mq^{\frac12}$ & $1$  &
 $\mq^{2n}\hat{X}_{\gamma_{l,+}+\gamma_{m,+}-\gamma_{m,-}+n\gamma_p}$  & $-\mq^{2n}\hat{X}_{\gamma_{m,+}-\gamma_{m,-}+2n\gamma_p}\hat{X}_{\gamma_{l,+}}$    \\
 \vdots &  \vdots & \vdots & \vdots & \vdots & \vdots & \vdots    \\
 \hline
\end{tabular}
\end{center}
Summing up their contributions we obtain the following element of $\Sk(\Sigma\times I, GL_1)$
\begin{equation}
    \hat\CL_{l}^{(7)} = -\frac{ \hat{X}_{\gamma_{m,+}-\gamma_{m,-}}}{1-\mq^2 \hat{X}_{2\gamma_p}}\hat{X}_{\gamma_{l,+}}\;.
\end{equation}
Now, using the same geometric series but with initial elements 
\begin{align*}
    \tilde{\varrho}_{l}^{(8,0)} = \pic{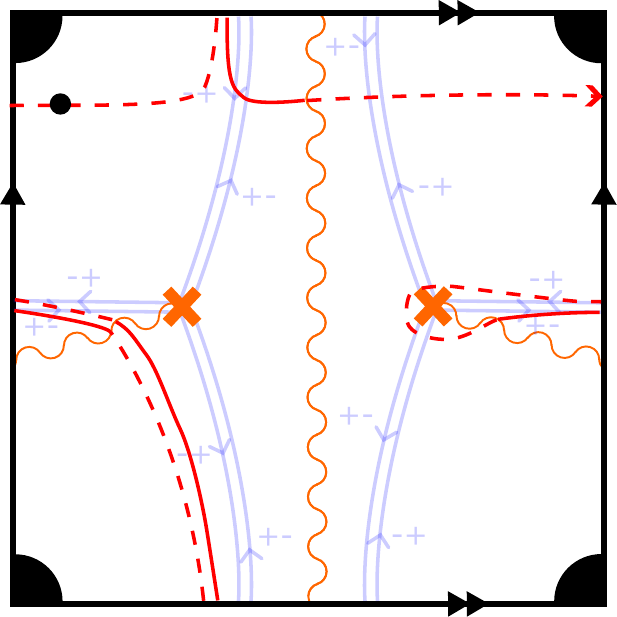}{.3}, \qquad
    \tilde{\varrho}_{l}^{(9,0)} = \pic{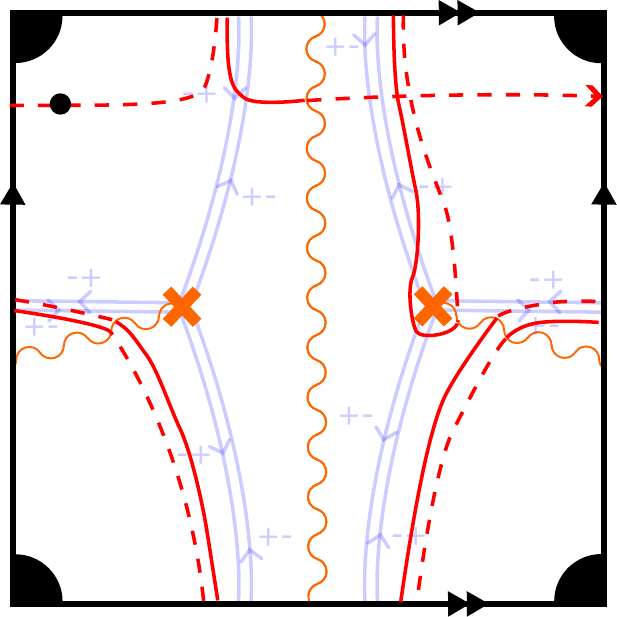}{.3}, \qquad
    \tilde{\varrho}_{l}^{(10,0)} = \pic{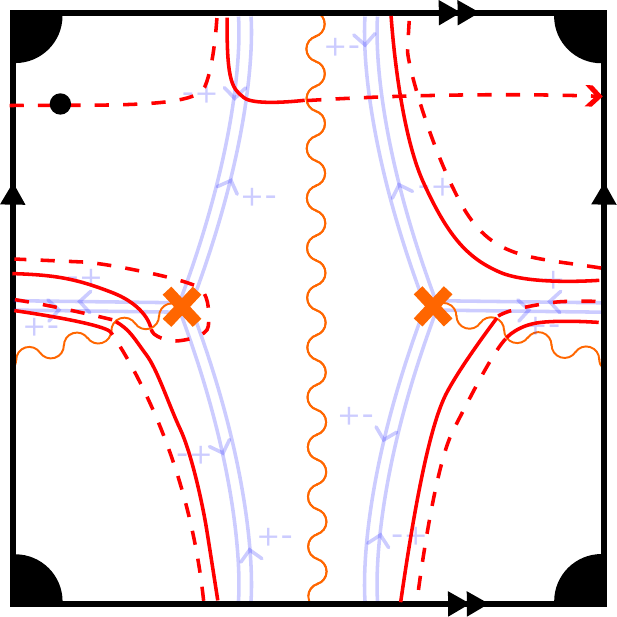}{.3},
\end{align*}
we get
\begin{equation}
    \hat\CL_{l}^{(8)} =\frac{\mq \hat{X}_{\gamma_p}}{1-\mq^2 \hat{X}_{2\gamma_p}}\hat{X}_{\gamma_{l,+}}\;,
\end{equation}
\begin{equation}
    \hat\CL_{l}^{(9)} =\frac{\mq \hat{X}_{\gamma_{m,+}-\gamma_{m,-}+\gamma_p}}{1-\mq^2 \hat{X}_{2\gamma_p}}\hat{X}_{\gamma_{l,+}}\;,
\end{equation}
\begin{equation}
    \hat\CL_{l}^{(10)} =-\frac{\mq^2 \hat{X}_{2\gamma_p}}{1-\mq^2 \hat{X}_{2\gamma_p}}\hat{X}_{\gamma_{l,+}}\;.
\end{equation}
\end{enumerate}

Next we consider lifts involving three detours. These can be organized into eight families. Each comes with a natural $\mathbb{Z}_{\geq 0}^3$-grading.

\begin{enumerate}
    \item The first four families of triple-detour lifts have the same $\mathbb{Z}^3_{\geq0}$ graded geometric series, but with different initial elements. To that end, let us start with the first triple-detour family $\tilde\varrho_l^{(11;a,b,c)}$ with $a,b,c\in\mathbb{Z}_{\geq 0}$ representing the three gradings. Then, we have the following characteristic lifts
\allowdisplaybreaks 
\begin{align*}
    \tilde{\varrho}_{l}^{(11;0,0,0)} &= \pic{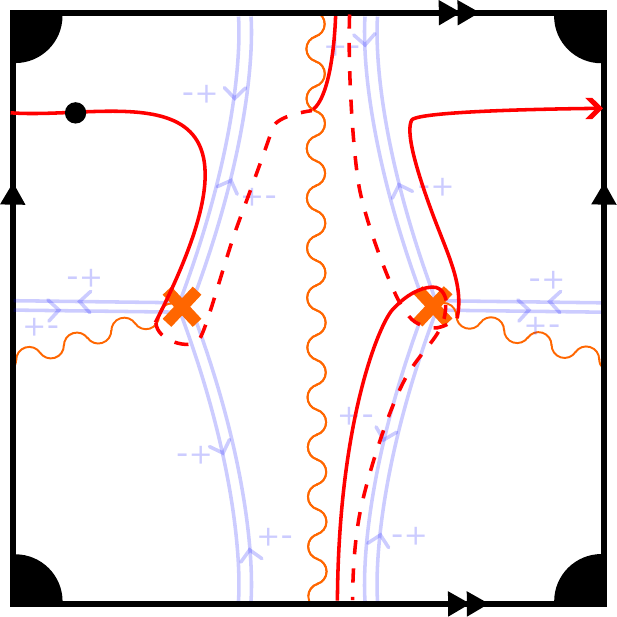}{.3}, &\qquad
    \tilde{\varrho}_{l}^{(11;1,0,0)} &= \pic{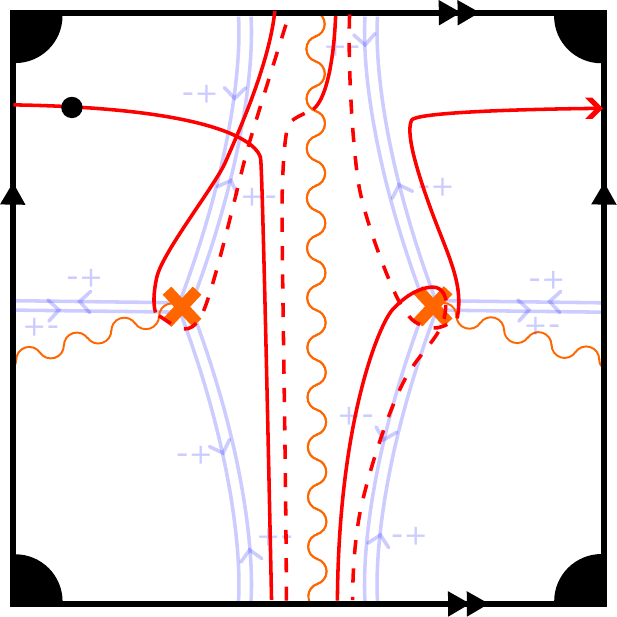}{.3}, \\[1ex]
    \tilde{\varrho}_{l}^{(11;0,1,0)} &= \pic{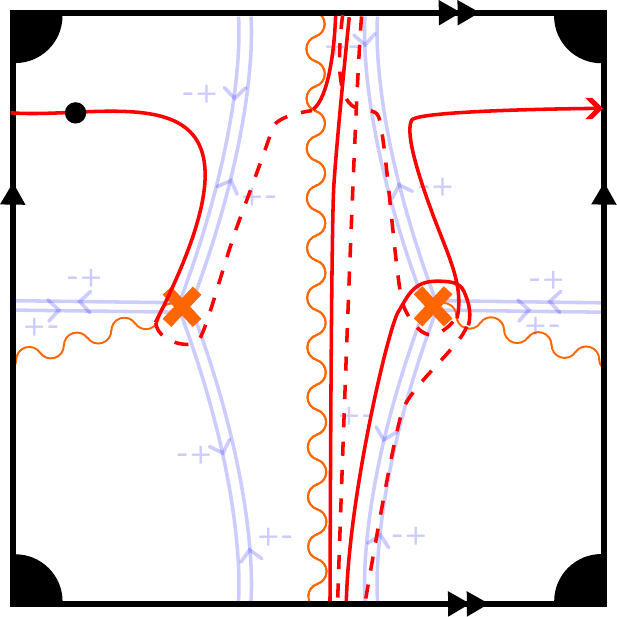}{.3}, &\qquad
    \tilde{\varrho}_{l}^{(11;0,0,1)} &= \pic{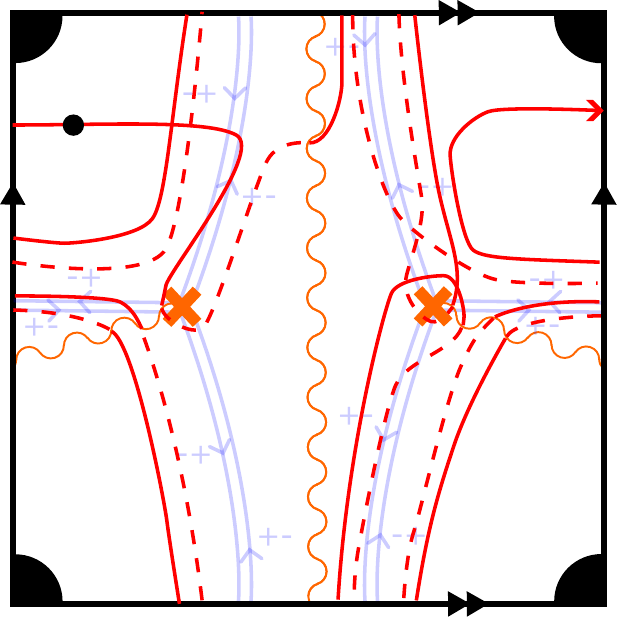}{.3},
\end{align*}
with corresponding weights
\begin{center}
\footnotesize
\setlength{\tabcolsep}{4pt}
\renewcommand{\arraystretch}{1.25}
\makebox[\textwidth][c]{%
\begin{tabular}{|c|c|c|c|c|c|c|}
\hline
$\tilde{\varrho}$ & $\alpha_f(\tilde{\varrho})$ & $\alpha_e(\tilde{\varrho})$ & $\alpha_d(\tilde{\varrho})$ & $\alpha_w(\tilde{\varrho})$ & $(-1)^{n'(\tilde{\varrho})} \mq^{\mathfrak{wr}(\tilde{\varrho})} \hat{X}_{\tilde{\varrho}}$
 & Total normal ordered product \\
\hline\hline
$\tilde{\varrho}_{l}^{(11;0,0,0)}$ & $\mq^{\frac12}$ & 1
 & $\mq^{-\frac12}$
 & 1
 & $-\mq\hat{X}_{\gamma_{l,-}+\gamma_{m,+}-\gamma_{m,-}}$
 & $\mq^{2}\hat{X}_{\gamma_{m,+}-\gamma_{m,-}}\hat{X}_{\gamma_{l,-}}$ \\
$\tilde{\varrho}_{l}^{(11;1,0,0)}$ & $\mq^{\frac12}$ & 1
 & $\mq^{-\frac12}$
 & 1
 & $\hat{X}_{\gamma_{l,-}+2(\gamma_{m,+}-\gamma_{m,-})}$
 & $\mq^{2}\hat{X}_{2(\gamma_{m,+}-\gamma_{m,-})}\hat{X}_{\gamma_{l,-}}$ \\
$\tilde{\varrho}_{l}^{(11;0,1,0)}$ & $\mq^{\frac12}$ & 1
 & $\mq^{-\frac12}$
 & 1
 & $\mq^{2}\hat{X}_{\gamma_{l,-}+2(\gamma_{m,+}-\gamma_{m,-})}$
 & $\mq^{4}\hat{X}_{2(\gamma_{m,+}-\gamma_{m,-})}\hat{X}_{\gamma_{l,-}}$ \\
$\tilde{\varrho}_{l}^{(11;0,0,1)}$ & $\mq^{\frac12}$ & 1
 & $\mq^{-\frac12}$
 & 1
 & $-\mq^{-1}\hat{X}_{\gamma_{l,-}+\gamma_{m,+}-\gamma_{m,-}+2\gamma_p}$
 & $\hat{X}_{\gamma_{m,+}-\gamma_{m,-}+2\gamma_p}\hat{X}_{\gamma_{l,-}}$ \\
\vdots & \vdots & \vdots & \vdots & \vdots & \vdots & \vdots \\
$\tilde{\varrho}_{l}^{(11;a,b,c)}$ & $\mq^{\frac12}$ & 1 & $\mq^{-\frac12}$ & 1
 & \begin{tabular}[c]{@{}c@{}}$(-\mq)^{1-a+b-2c}$\\[2pt]$\times\,\hat{X}_{\gamma_{l,-}+(1+a+b)(\gamma_{m,+}-\gamma_{m,-})+2c\gamma_p}$\end{tabular}
 & \begin{tabular}[c]{@{}c@{}}$\mq^{2+2b-2c}$\\[2pt]$\times\,\hat{X}_{(1+a+b)(\gamma_{m,+}-\gamma_{m,-})+2c\gamma_p}\hat{X}_{\gamma_{l,-}}$\end{tabular} \\
\vdots & \vdots & \vdots & \vdots & \vdots & \vdots & \vdots \\
\hline
\end{tabular}}
\end{center}
Taken all together, these sum up to
\begin{equation}
    \hat\CL_l^{(11)} = \frac{\mq^2\hat{X}_{\gamma_{m,+}-\gamma_{m,-}}}{(1-\hat{X}_{\gamma_{m,+}-\gamma_{m,-}})(1-\mq^2\hat{X}_{\gamma_{m,+}-\gamma_{m,-}})(1-\mq^{-2}\hat{X}_{2\gamma_p})}\hat{X}_{\gamma_{l,-}}\;.
\end{equation}

The next three families feature the same type of geometric series, but with different initial detours
\begin{align*}
    \tilde{\varrho}_{l}^{(12;0,0,0)} = \pic{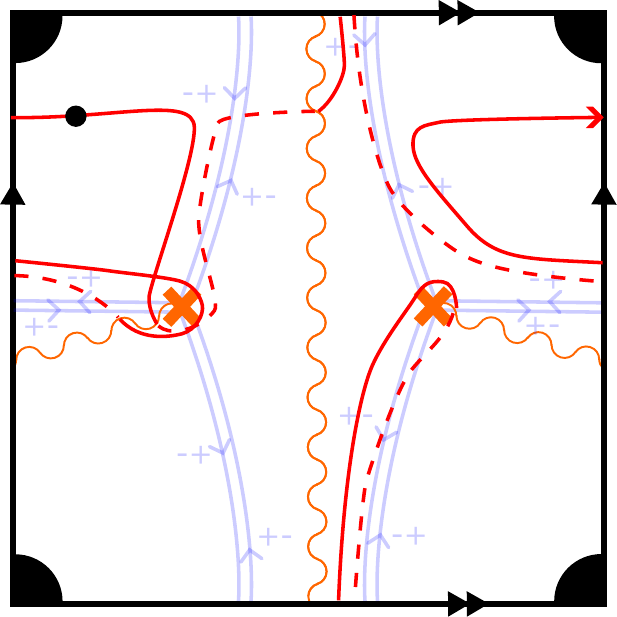}{.3}, 
    &\qquad
    \tilde{\varrho}_{l}^{(13;0,0,0)} = \pic{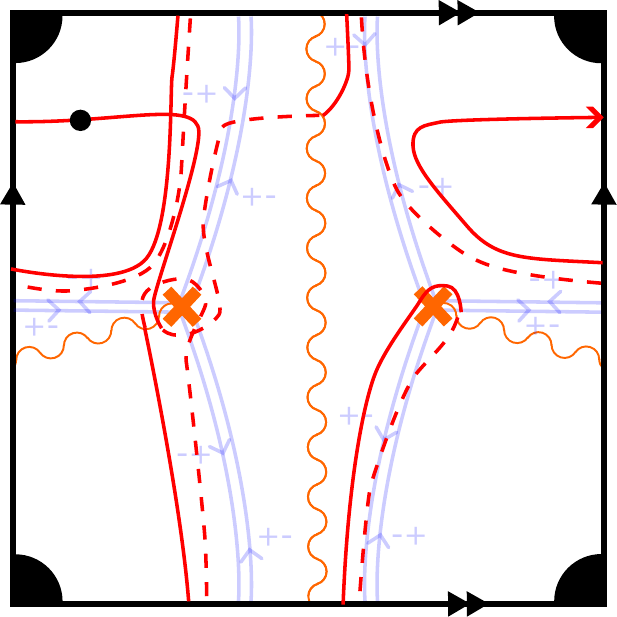}{.3}, \\
    \tilde{\varrho}_{l}^{(14;0,0,0)} = 
    &\pic{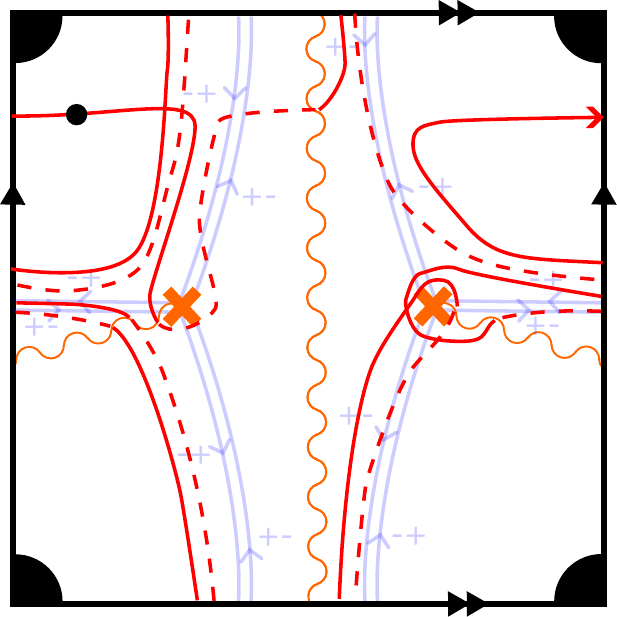}{.3},
\end{align*}
Summing up all the contributions of each family, as we did the previous triple-detour family we obtain
\begin{equation}
    \hat\CL_l^{(12)} = -\frac{\mq^{-1}\hat{X}_{\gamma_p}}{(1-\hat{X}_{\gamma_{m,+}-\gamma_{m,-}})(1-\mq^2\hat{X}_{\gamma_{m,+}-\gamma_{m,-}})(1-\mq^{-2}\hat{X}_{2\gamma_p})}\hat{X}_{\gamma_{l,-}}\;,
\end{equation}
\begin{equation}
    \hat\CL_l^{(13)} = -\frac{\mq\hat{X}_{\gamma_{m,+}-\gamma_{m,-}+\gamma_p}}{(1-\hat{X}_{\gamma_{m,+}-\gamma_{m,-}})(1-\mq^2\hat{X}_{\gamma_{m,+}-\gamma_{m,-}})(1-\mq^{-2}\hat{X}_{2\gamma_p})}\hat{X}_{\gamma_{l,-}}\;,
\end{equation}
\begin{equation}
    \hat\CL_l^{(14)} = \frac{\mq^{-2}\hat{X}_{2\gamma_p}}{(1-\hat{X}_{\gamma_{m,+}-\gamma_{m,-}})(1-\mq^2\hat{X}_{\gamma_{m,+}-\gamma_{m,-}})(1-\mq^{-2}\hat{X}_{2\gamma_p})}\hat{X}_{\gamma_{l,-}}\;.
\end{equation}

\item The same can be done for the last four families of triple-detour lifts. Starting with the characteristic lifts of the $\tilde\varrho_l^{(15;,a,b,c)}$ family, we have
\allowdisplaybreaks 
\begin{align*}
    \tilde{\varrho}_{l}^{(15;0,0,0)} &= \pic{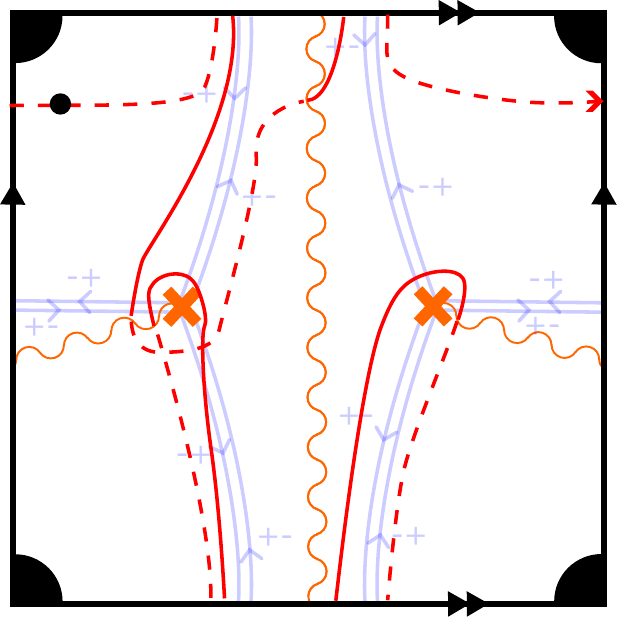}{.3}, &\qquad
    \tilde{\varrho}_{l}^{(15;1,0,0)} &= \pic{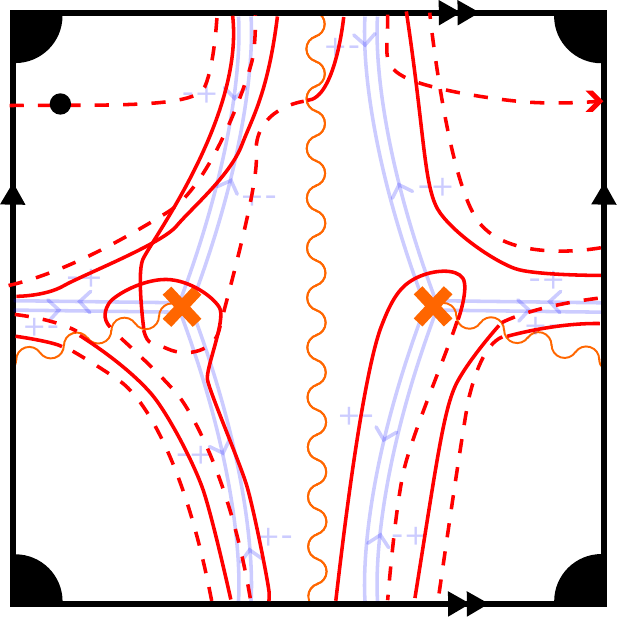}{.3}, \\[1ex]
    \tilde{\varrho}_{l}^{(15;0,1,0)} &= \pic{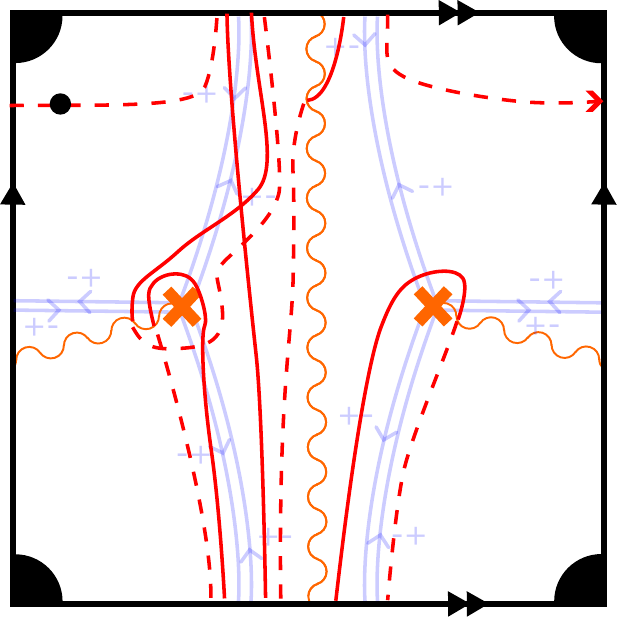}{.3}, &\qquad
    \tilde{\varrho}_{l}^{(15;0,0,1)} &= \pic{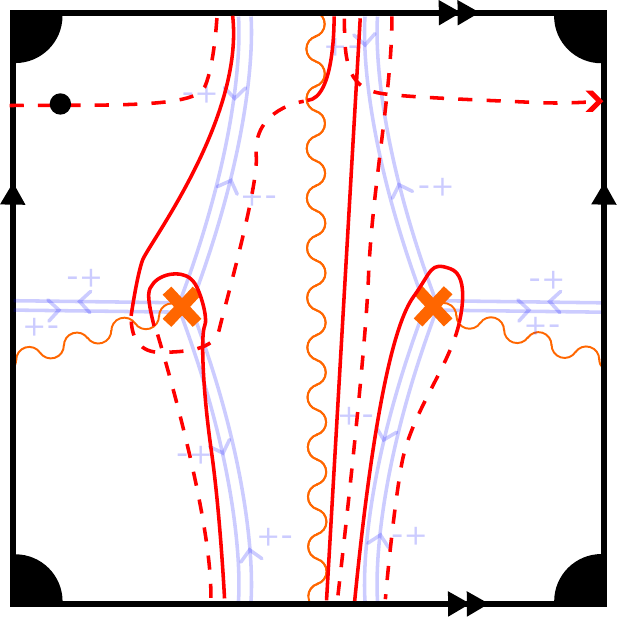}{.3},
\end{align*}
with respective weights
\begin{center}
\footnotesize
\setlength{\tabcolsep}{4pt}
\renewcommand{\arraystretch}{1.25}
\makebox[\textwidth][c]{%
\begin{tabular}{|c|c|c|c|c|c|c|}
\hline
$\tilde{\varrho}$ & $\alpha_f(\tilde{\varrho})$ & $\alpha_e(\tilde{\varrho})$ & $\alpha_d(\tilde{\varrho})$ & $\alpha_w(\tilde{\varrho})$ & $(-1)^{n'(\tilde{\varrho})} \mq^{\mathfrak{wr}(\tilde{\varrho})} \hat{X}_{\tilde{\varrho}}$
 & Total normal ordered product \\
\hline\hline

$\tilde{\varrho}_{l}^{(15;0,0,0)}$ & $\mq^{\frac12}$ & 1
 & $\mq^{\frac12}$
 & 1
 & $-\mq^{-1}\hat{X}_{\gamma_{l,-}+\gamma_{m,+}-\gamma_{m,-}+\gamma_p}$
 & $\mq\hat{X}_{\gamma_{m,+}-\gamma_{m,-}+\gamma_p}\hat{X}_{\gamma_{l,-}}$ \\

$\tilde{\varrho}_{l}^{(15;1,0,0)}$ & $\mq^{\frac12}$ & 1
 & $\mq^{\frac12}$
 & 1
 & $-\mq\hat{X}_{\gamma_{l,-}+\gamma_{m,+}-\gamma_{m,-}+3\gamma_p}$
 & $\mq^{3}\hat{X}_{\gamma_{m,+}-\gamma_{m,-}+3\gamma_p}\hat{X}_{\gamma_{l,-}}$ \\

$\tilde{\varrho}_{l}^{(15;0,1,0)}$ & $\mq^{\frac12}$ & 1
 & $\mq^{\frac12}$
 & 1
 & $\mq^{-2}\hat{X}_{\gamma_{l,-}+2(\gamma_{m,+}-\gamma_{m,-})+\gamma_p}$
 & $\mq\hat{X}_{2(\gamma_{m,+}-\gamma_{m,-})+\gamma_p}\hat{X}_{\gamma_{l,-}}$ \\

$\tilde{\varrho}_{l}^{(15;0,0,1)}$ & $\mq^{\frac12}$ & 1
 & $\mq^{\frac12}$
 & 1
 & $\hat{X}_{\gamma_{l,-}+2(\gamma_{m,+}-\gamma_{m,-})+2\gamma_p}$
 & $\mq^{3}\hat{X}_{2(\gamma_{m,+}-\gamma_{m,-})+\gamma_p}\hat{X}_{\gamma_{l,-}}$ \\

\vdots & \vdots & \vdots & \vdots & \vdots & \vdots & \vdots \\

$\tilde{\varrho}_{l}^{(15;a,b,c)}$ & $\mq^{\frac12}$ & 1 & $\mq^{\frac12}$ & 1
 & \begin{tabular}[c]{@{}c@{}}$(-\mq)^{-1+2a-b+c}$\\[2pt]$\times\,\hat{X}_{\gamma_{l,-}+(1+b+c)(\gamma_{m,+}-\gamma_{m,-})+(2a+1)\gamma_p}$\end{tabular}
 & \begin{tabular}[c]{@{}c@{}}$\mq^{2a+2c+1}$\\[2pt]$\times\,\hat{X}_{(1+b+c)(\gamma_{m,+}-\gamma_{m,-})+(2a+1)\gamma_p}\hat{X}_{\gamma_{l,-}}$\end{tabular} \\

\vdots & \vdots & \vdots & \vdots & \vdots & \vdots & \vdots \\
\hline
\end{tabular}}
\end{center}
The lifts within this family sum up to
\begin{equation}
    \hat\CL_l^{(15)} = \frac{\mq \hat{X}_{\gamma_{m,+}-\gamma_{m,-}+\gamma_p}}{(1-\mq^2\hat{X}_{2\gamma_p})(1-\hat{X}_{\gamma_{m,+}-\gamma_{m,-}})(1-\mq^2\hat{X}_{\gamma_{m,+}-\gamma_{m,-}})}\hat{X}_{\gamma_{l,-}}\;.
\end{equation}
The last three families feature again the same type of geometric series, but with different initial detours
\begin{align*}
    \tilde{\varrho}_{l}^{(16;0,0,0)} = \pic{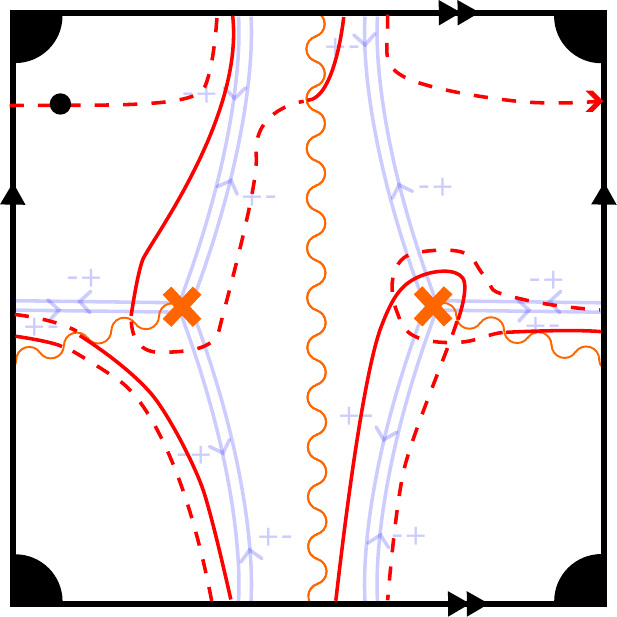}{.3}, 
    &\qquad
    \tilde{\varrho}_{l}^{(17;0,0,0)} = \pic{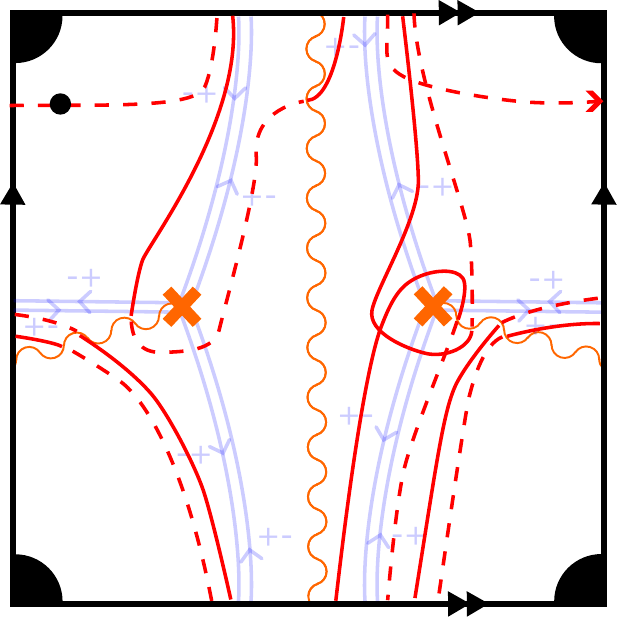}{.3}, \\
    \tilde{\varrho}_{l}^{(18;0,0,0)} = &\pic{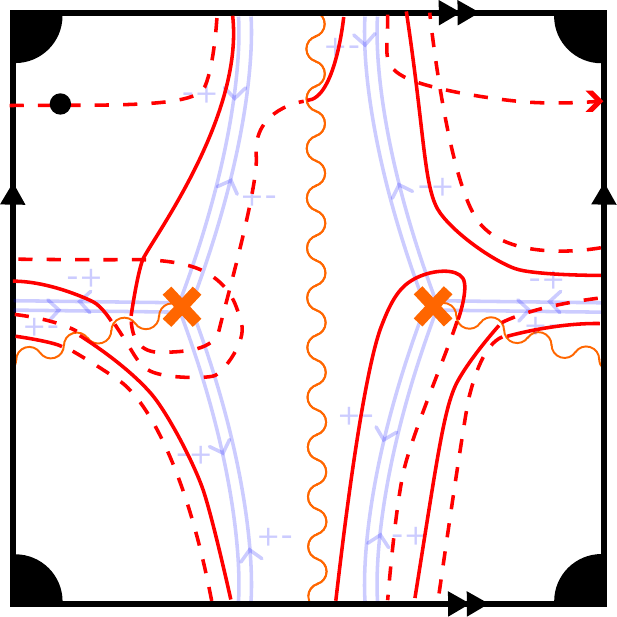}{.3},
\end{align*}
and summing up over all contributions we obtain, for each family respectively
\begin{equation}
    \hat\CL_l^{(16)} = -\frac{\mq^2\hat{X}_{2\gamma_p}}{(1-\mq^2\hat{X}_{2\gamma_p})(1-\hat{X}_{\gamma_{m,+}-\gamma_{m,-}})(1-\mq^2\hat{X}_{\gamma_{m,+}-\gamma_{m,-}})}\hat{X}_{\gamma_{l,-}}\;,
\end{equation}
\begin{equation}
    \hat\CL_l^{(17)} = -\frac{\mq^2\hat{X}_{\gamma_{m,+}-\gamma_{m,-}+2\gamma_p}}{(1-\mq^2\hat{X}_{2\gamma_p})(1-\hat{X}_{\gamma_{m,+}-\gamma_{m,-}})(1-\mq^2\hat{X}_{\gamma_{m,+}-\gamma_{m,-}})}\hat{X}_{\gamma_{l,-}}\;,
\end{equation}
\begin{equation}
    \hat\CL_l^{(18)} = \frac{\mq^3\hat{X}_{3\gamma_p}}{(1-\mq^2\hat{X}_{2\gamma_p})(1-\hat{X}_{\gamma_{m,+}-\gamma_{m,-}})(1-\mq^2\hat{X}_{\gamma_{m,+}-\gamma_{m,-}})}\hat{X}_{\gamma_{l,-}}\;.
\end{equation}

\end{enumerate}

\begin{figure}[h!]
    \centering
    \includegraphics[width=0.3\linewidth]{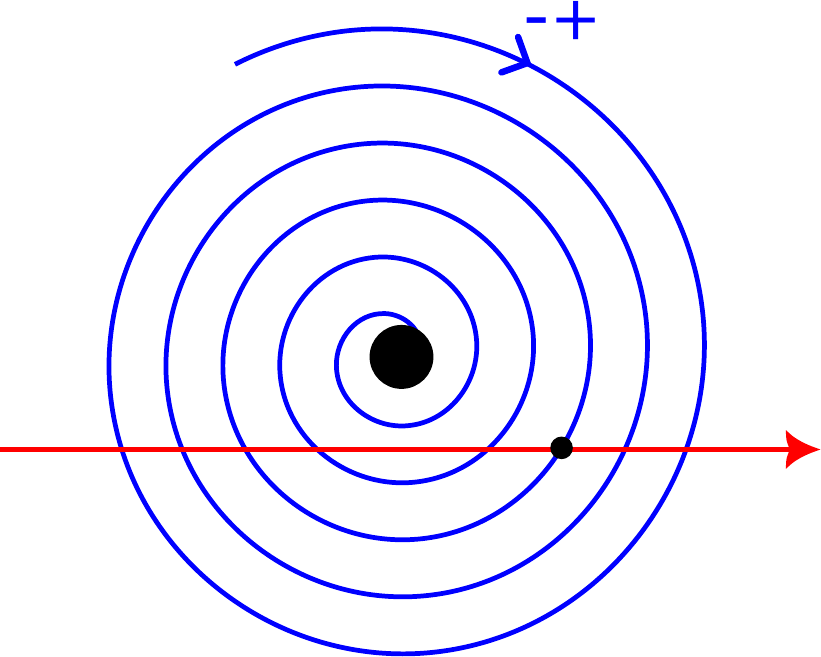}
    \caption{American resolved Fenchel-Nielsen network near the puncture with $\varrho_l$.}
    \label{fig:puncture-zoom-L}
\end{figure}

This completes the enumeration of all possible lifts involving detours alone. Next we consider contributions involving exchanges. 
Recall that, owing to our choice of conventions any exchange contributions must be generated by flow lines that end or begin on the basepoint of $\varrho$.
In the case of Fenchel-Nielsen networks these correspond to flow lines that wind around the puncture, see Figure \ref{fig:puncture-zoom-L}.

It is useful to observe that these contributions are `local', in the sense that they do not interfere with the contributions to lifts of $\varrho_l$ arising from detours.
For this reason, the corrections introduced by exchanges to each of the generating series of detour lifts computed above consists just of multiplication by overall factors.
Exchange contributions come in four distinct families.

\begin{enumerate}
    \item The first family of exchange contributions is given by
    \begin{equation*}
        e_{1,0}=\pic{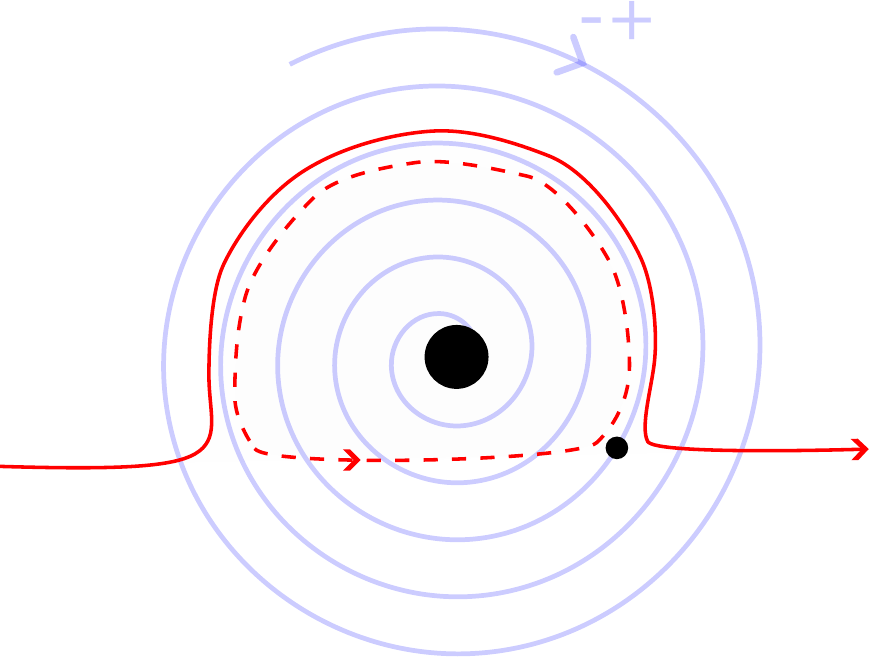}{.3}, \qquad e_{1,1} = \pic{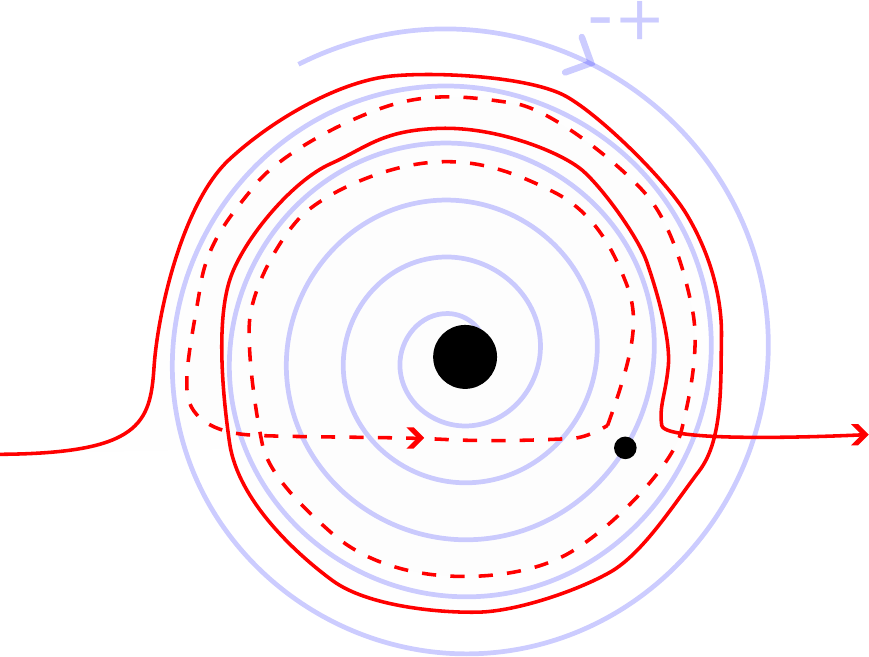}{.3}, \qquad \dots
    \end{equation*}
    These introduce the following \textit{multiplicative} shifts in the weights and in the write of each lifted path
    \begin{center}
\footnotesize
\setlength{\tabcolsep}{4pt}
\renewcommand{\arraystretch}{1.25}
\makebox[\textwidth][c]{%
\begin{tabular}{|c|c|c|c|c|c|c|}
\hline
$\Delta\tilde{\varrho}$ & $\Delta\alpha_f(\tilde{\varrho})$ & $\Delta\alpha_e(\tilde{\varrho})$ & $\Delta\alpha_d(\tilde{\varrho})$ & $\Delta\alpha_w(\tilde{\varrho})$ & $\Delta\left((-1)^{n'(\tilde{\varrho})} \mq^{\mathfrak{wr}(\tilde{\varrho})} \hat{X}_{\tilde{\varrho}}\right)$
 & $\Delta\left(\text{Total normal ordered product}\right)$ \\
\hline\hline
 $e_{1,0}$ & 1 & $1-\mq^{-2}$ & 1 & 1 & $X_{2\gamma_p}$ & $(1-\mq^{-2})X_{2\gamma_p}$ \\
 $e_{1,1}$ & 1 & $1-\mq^{-2}$ & 1 & 1 & $X_{4\gamma_p}$ & $(1-\mq^{-2})X_{4\gamma_p}$ \\
\vdots & \vdots & \vdots & \vdots & \vdots & \vdots & \vdots \\
 $e_{1,n}$ & 1 & $1-\mq^{-2}$ & 1 & 1 & $X_{2(1+n)\gamma_p}$ & $(1-\mq^{-2})X_{2(1+n)\gamma_p}$ \\
\vdots & \vdots & \vdots & \vdots & \vdots & \vdots & \vdots \\
\hline
\end{tabular}}
\end{center}
From the explicit diagrams of the exchanges shown above, it is clear that these exchanges contribute to those lifts that have the base point on the $+$ sheet of $\Sigma$. 
This implies that we should introduce the following multiplicative factors to the generating series of detour paths computed earlier
\begin{equation}
    \left\{\frac{\left(1-\mq^{-2}\right)X_{2\gamma_p}}{1-X_{2\gamma_p}}\times \hat\CL_l^{(i)}\right\}_{i \in \{1,3,4,5,6,11,12,13,14\}}.
\end{equation}

\item The family of second exchange contributions is
\begin{equation}
    e_{2,0} = \pic{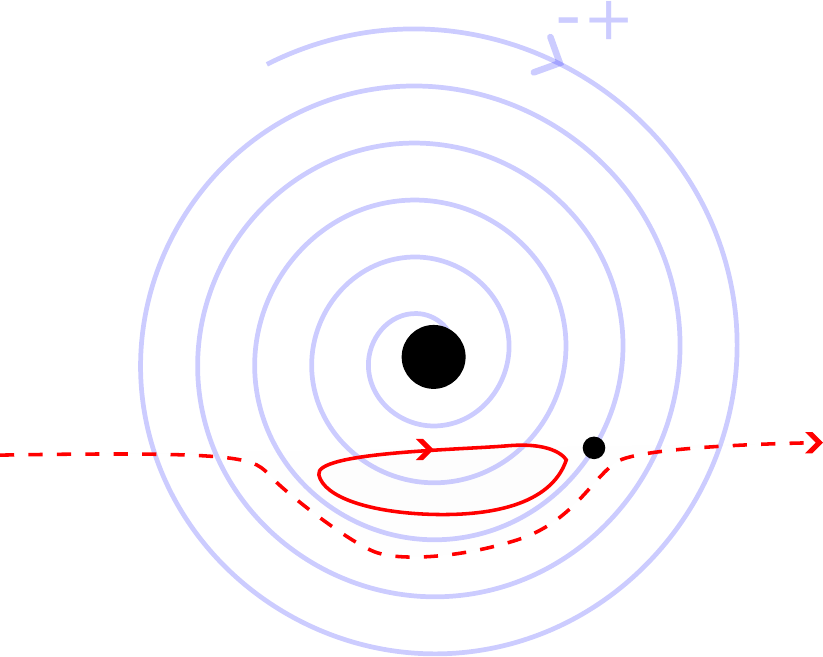}{.3}, \qquad e_{2,1} = \pic{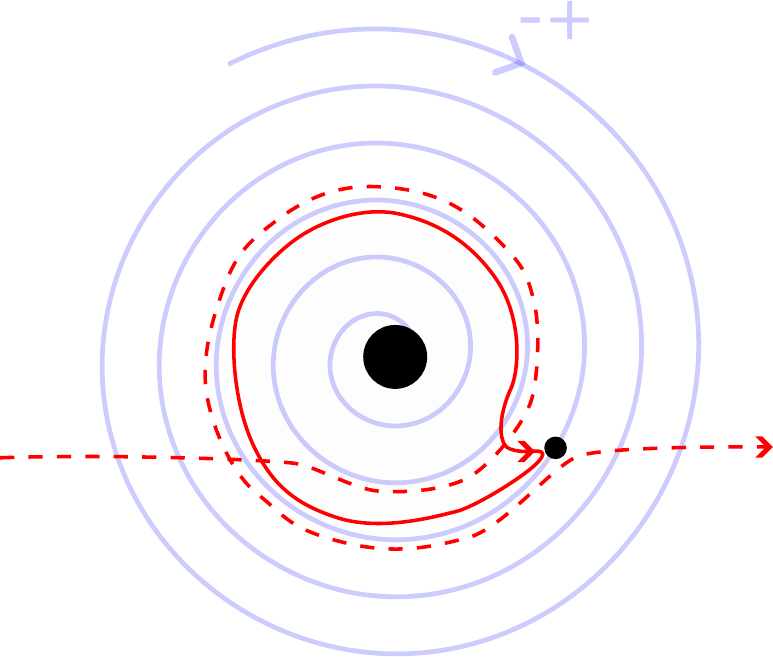}{.3}, \qquad \dots,
\end{equation}
These introduce multiplicative shifts of weights and writhe as follows
\begin{center}
\footnotesize
\setlength{\tabcolsep}{4pt}
\renewcommand{\arraystretch}{1.25}
\makebox[\textwidth][c]{%
\begin{tabular}{|c|c|c|c|c|c|c|}
\hline
$\Delta\tilde{\varrho}$ & $\Delta\alpha_f(\tilde{\varrho})$ & $\Delta\alpha_e(\tilde{\varrho})$ & $\Delta\alpha_d(\tilde{\varrho})$ & $\Delta\alpha_w(\tilde{\varrho})$ & $\Delta\left((-1)^{n'(\tilde{\varrho})} \mq^{\mathfrak{wr}(\tilde{\varrho})} \hat{X}_{\tilde{\varrho}}\right)$
 & $\Delta\left(\text{Total normal ordered product}\right)$ \\
\hline\hline
 $e_{2,0}$ & 1 & $1-\mq^{2}$ & 1 & $\mq^{-2}$ & $X_{2\gamma_p}$ & $\mq^{-2}(1-\mq^{2})$ \\
 $e_{2,1}$ & 1 & $1-\mq^{2}$ & 1 & $\mq^{-2}$ & $X_{4\gamma_p}$ & $\mq^{-2}(1-\mq^{2})X_{2\gamma_p}$ \\
\vdots & \vdots & \vdots & \vdots & \vdots & \vdots & \vdots \\
 $e_{2,n}$ & 1 & $1-\mq^{2}$ & 1 & $\mq^{-2}$ & $X_{2n\gamma_p}$ & $\mq^{-2}(1-\mq^{2})X_{2n\gamma_p}$  \\
\vdots & \vdots & \vdots & \vdots & \vdots & \vdots & \vdots \\
\hline
\end{tabular}}
\end{center}
These exchanges contribute to those lifts that have the base point on the $-$ sheet of $\Sigma$. 
Hence, they introduce the following multiplicative corrections to generating series of detour paths computed earlier
\begin{equation}
    \left\{\frac{\mq^{-2}\left(1-\mq^{2}\right)}{1-X_{2\gamma_p}}\times \hat\CL_l^{(i)}\right\}_{i \in \{2,7,8,9,10,15,16,17,18\}}.
\end{equation}

\item The remaining two families mutually cancel. To see this, we explicitly show the first element of the two remaining geometric series as follows
\begin{equation}
    e_{3,0} = \pic{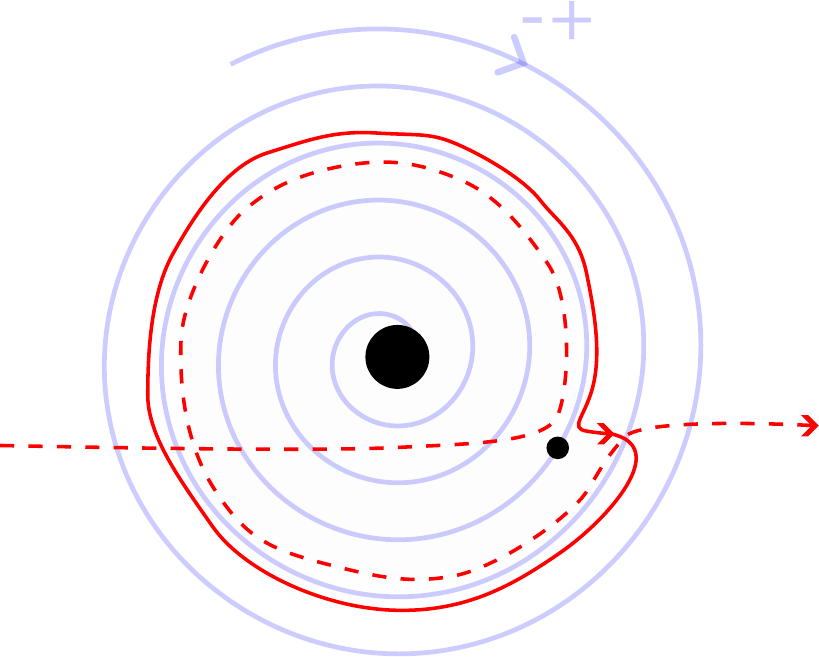}{.3}, \qquad e_{4,0} = \pic{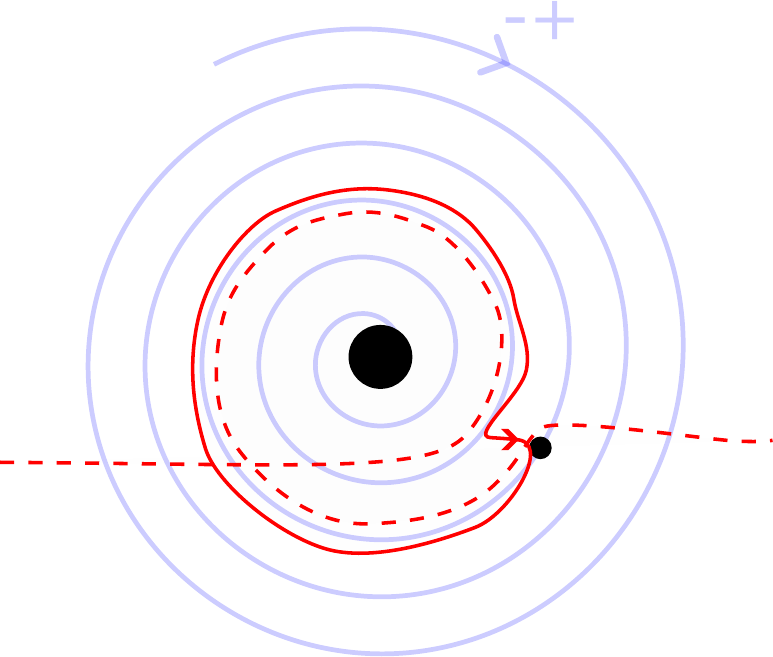}{.3}.
\end{equation}
We observe that both contributions above affect the lifts that have their base point on the $-$ branch of $\Sigma$, with the only difference among them being in the exchange factor, which is $(\mq-\mq^{-1})$ for $e_{3,0}$ and $(\mq^{-1}-\mq)$ for $e_{4,0}$. All the other weights and writhe factors are exactly the same owing to the diagramatic similarity of the above contributions, and hence they cancel, giving us no contribution.\footnote{It is conceivable that a suitable deformation of the representative lifted path to another choice of representative may get rid of these contributions altogether.}

\end{enumerate}

Summing up all contributions from detours and exchanges we find that $[\varrho_l]\in \Sk(C\times I,GL_2)$ maps to the following element of $\Sk(\Sigma\times I, GL_1)$
\begin{equation}
    \hat\CL^{\FN_-}_l = \left(1-\mq^{-1}\hat{X}_{\gamma_{m,+}-\gamma_{m,-}-\gamma_p}\right) \hat{X}_{\gamma_{l,+}} + \frac{1-\mq \hat{X}_{\gamma_{m,+}-\gamma_{m,-}+\gamma_p}}{(1-\hat{X}_{\gamma_{m,+}-\gamma_{m,-}})(1-\mq^2\hat{X}_{\gamma_{m,+}-\gamma_{m,-}})}\hat{X}_{\gamma_{l,-}}.
    \label{eq:FN-Ll}
\end{equation}

\paragraph{The operator $\hat\CL_{m^2}$.}
There are only three lifts of $\varrho_{m^2}$. Two of them correspond to direct lifts, and we show them below.
\begin{equation*}
    \tilde{\varrho}_{m^2}^{(1)} = \pic{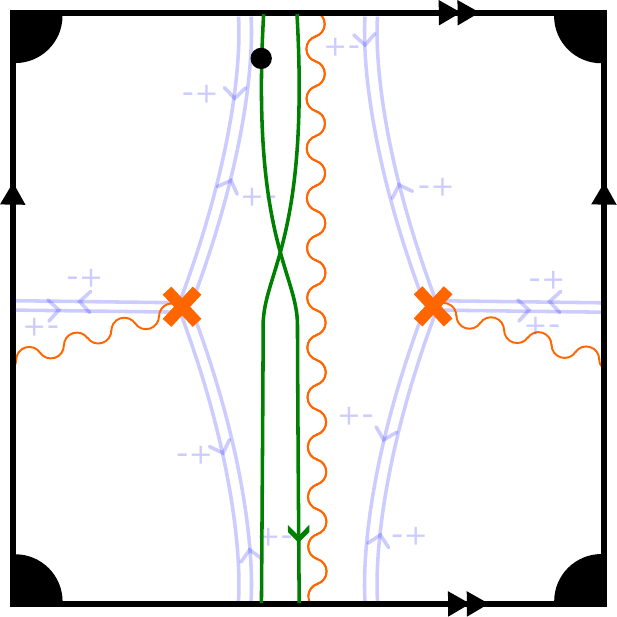}{.3}, \qquad
    \tilde{\varrho}_{m^2}^{(2)} = \pic{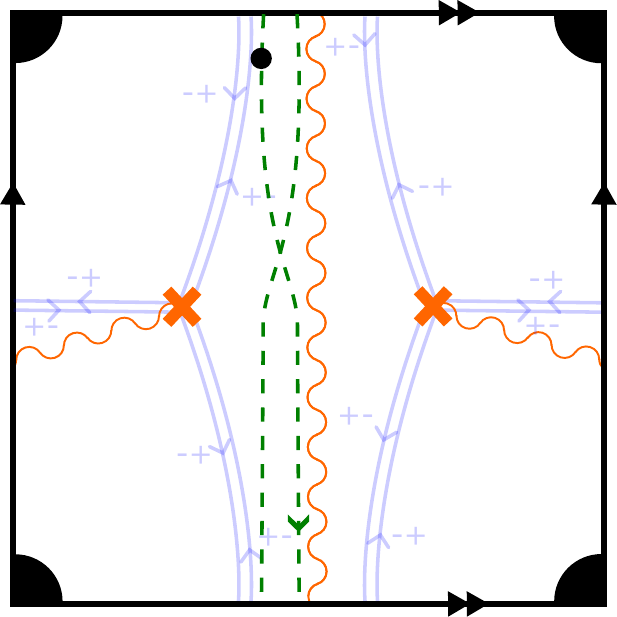}{.3},
\end{equation*}
Their weights and writhe are given in the following table.
\begin{center}
\begin{tabular}{|c|c|c|c|c|c|c|}
 \hline
 $\tilde{\varrho}$ & $\alpha_f(\tilde{\varrho})$ & $\alpha_e(\tilde{\varrho})$ & $\alpha_d(\tilde{\varrho})$ & $\alpha_w(\tilde{\varrho})$ & $(-1)^{n'(\tilde{\varrho})} \mq^{\mathfrak{wr}(\tilde{\varrho})} \hat{X}_{\tilde{\varrho}}$ & Total normal ordered product \\
 \hline\hline
 $\tilde{\varrho}_{m^2}^{(1)}$ & 1 & 1  &  1 & 1  & $\mq^{-1}\hat{X}_{2\gamma_{m,+}}$  & $\mq^{-1}\hat{X}_{2\gamma_{m,+}}$   \\
 $\tilde{\varrho}_{m^2}^{(2)}$ & 1 & 1  &  1 & 1  & $\mq^{-1}\hat{X}_{2\gamma_{m,-}}$  & $\mq^{-1}\hat{X}_{2\gamma_{m,-}}$     \\
 \hline
\end{tabular}
\end{center}
The third contribution involves exchanges, and a generic choice of $\varrho_{m^2}$ will lead to an infinite number of contributions. However, we observe that in the middle of the annular region defined by the Fenchel-Nielsen network (but still to the left of the branch cut), the flow lines look as follows
\begin{equation}
    \pic{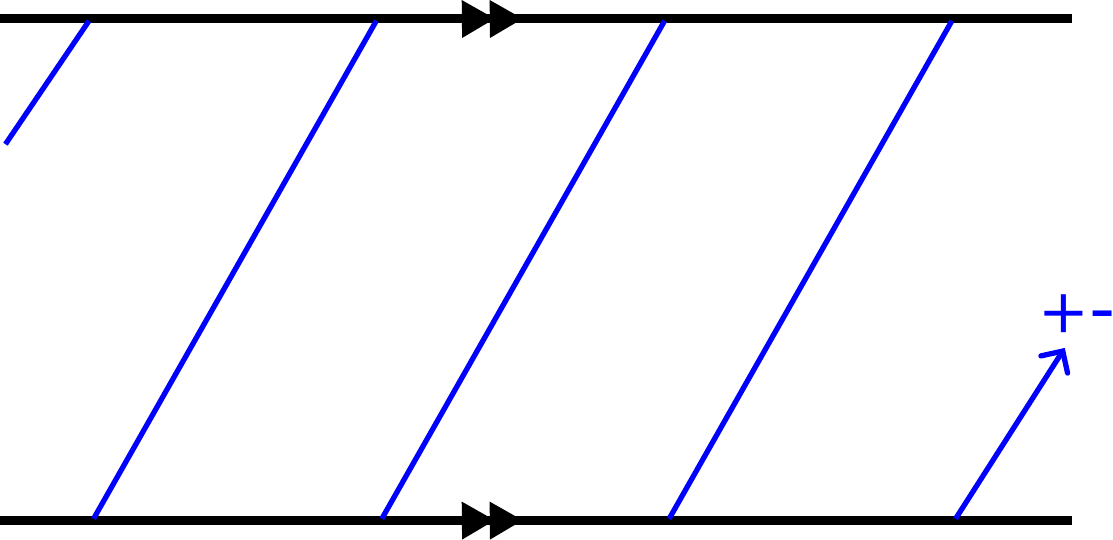}{.25}
\end{equation}
We remark that the above are diagonal because we work with the American resolution of the Fenchel-Nielsen network. To carry out the computation we take advantange of deformation invariance, and choose to arrange the loop $\rho_{m^2}$ relative to the flow of the foliation so that there is only a single exchange contribution, as follows
\begin{equation*}
    \varrho_{m^2} = \pic{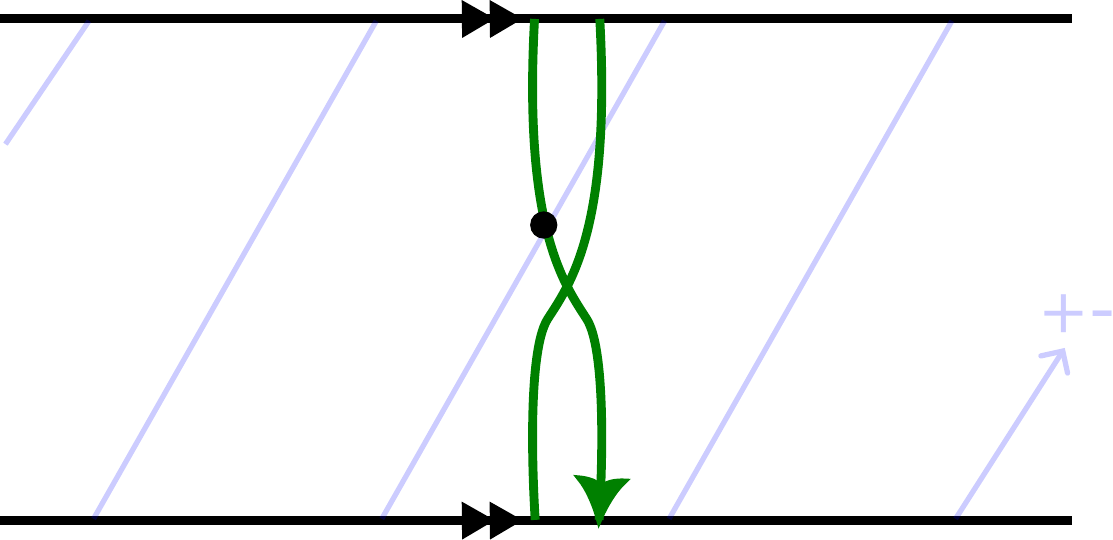}{.25} \xrightarrow[]{\text{Exchange Lift}} \pic{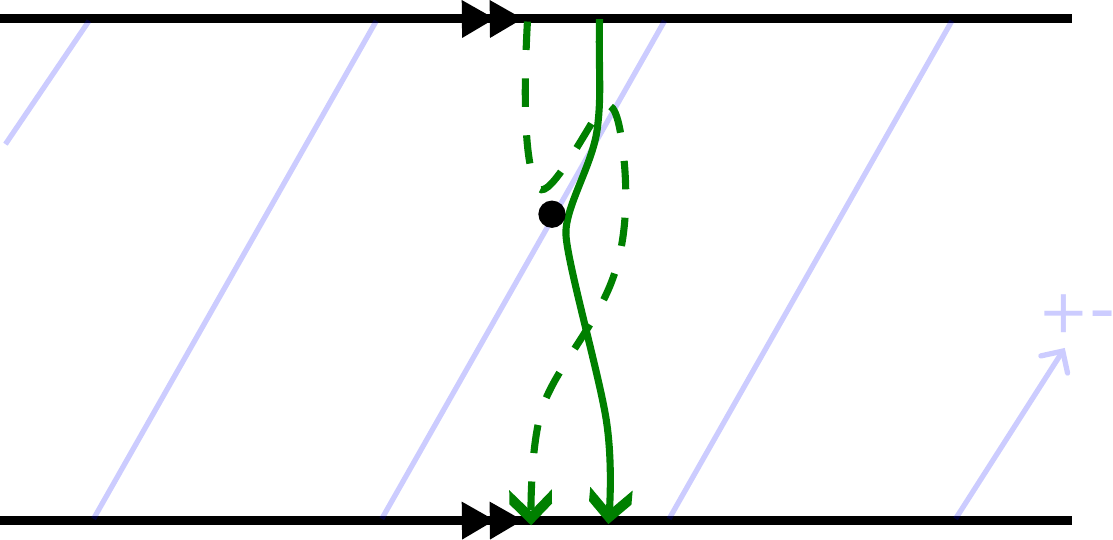}{.25} = \tilde\varrho_{m^2}^{(3)}
\end{equation*}
This has the following weight and writhe data,
\begin{center}
\begin{tabular}{|c|c|c|c|c|c|c|}
 \hline
 $\tilde{\varrho}$ & $\alpha_f(\tilde{\varrho})$ & $\alpha_e(\tilde{\varrho})$ & $\alpha_d(\tilde{\varrho})$ & $\alpha_w(\tilde{\varrho})$ & $(-1)^{n'(\tilde{\varrho})} \mq^{\mathfrak{wr}(\tilde{\varrho})} \hat{X}_{\tilde{\varrho}}$ & Total Product \\
 \hline\hline
 $\tilde{\varrho}_{m^2}^{(3)}$ & 1 & $\mq^{-1}-\mq$  &  1 & 1  & $\hat{X}_{\gamma_{m,+}+\gamma_{m,-}}$  & $\left(\mq^{-1}-\mq\right)\hat{X}_{\gamma_{m,+}+\gamma_{m,-}}$   \\
 \hline
\end{tabular}
\end{center}
Summing everything up, we get
\begin{equation}
    \hat \CL^{\FN_-}_{m^2} = \mq^{-1}\hat{X}_{2\gamma_{m,+}} + \mq^{-1}\hat{X}_{2\gamma_{m,-}} +\left(\mq^{-1}-\mq\right)\hat{X}_{\gamma_{m,+}+\gamma_{m,-}}.
    \label{eq:FN-Lm2}
\end{equation}

\subsection{Relations among charts}\label{eq:quantum-chart-relations}
The $\mq$-nonabelianization map undergoes jumps in correspondence of phases of BPS central charges $Z_\eta$. 
Since we are working with the same choice of spectral curve $\Sigma$ as for the previous ${SL}_2$ analysis, the collection of jumps again corresponds to the same BPS spectrum as in \eqref{eq:BPS-spectrum}.

At each critical phase the quantum IR line operators jump by a {motivic Kontsevich-Soibelman transformation} \eqref{eq:qdefKS}.
To define the appropriate transformation we need the protected spin characters for the $SU(2)$ $\CN=2^*$ theory. These have been computed in \cite[Section 4.5]{Longhi:2016wtv} and are as follows
\be\label{eq:BPS-spectrum-motivic}
\begin{split}
	&\Omega(\eta_1 + k(\eta_1+\eta_2);\mq) = \Omega(\eta_2 + k(\eta_1+\eta_2);\mq) = 1
	\qquad(k\geq 0)\,,
	\\
	& \Omega(\eta_p \pm (\eta_1+\eta_2);\mq ) =  1 \,,
	\qquad
	\Omega(\eta_1+\eta_2;\mq) = -\mq-\mq^{-1}\,,
\end{split}
\ee
together with CPT conjugates.
Although all BPS charges belong to the physical sublattice $\fY$ of the ${SL}_2$ local system, they can obviously be regarded as homology classes in the ambient lattice $H_1(\Sigma',\IZ)$.


This entails a straightforward generalization of the relations among Fock-Goncharov and Fenchel-Nielsen charts of the classical ${SL}_2$ local system, namely:
\begin{enumerate}[label=(\arabic*)]
    \item The Fock-Goncharov expressions for $\hat\CL^{\FG}$ are related to those of the Fenchel-Nielsen chart in American resolution $\hat\CL^{\FN_-}$ by 
    \begin{equation}      \label{eq:quantumFGFN}
    \begin{split}
   	    \hat\CL^{\FN_-}
	    & = \prod_{n\geq 0}^{\curvearrowleft} \hat{\mathcal{K}}_{\eta_2+n(\eta_1+\eta_2)} \, \hat\CL^{\FG}
    \end{split}
    \end{equation}
    where $\curvearrowleft$ means that $n$ grows from right to left. Note that unlike \eqref{eq:classFGFN}, this applies to cycles in {the whole lattice} $\gamma\in H_1(\Sigma',\IZ)$. 
We have checked perturbatively that this relation indeed holds, perturbatively.
More precisely we consider an $\epsilon$-expansion based on the assignment $\hat X_{\gamma_{l,\pm}} \to \epsilon^{\pm 1}\hat X_{\gamma_{l,\pm}} $, $\hat X_{\gamma_{m,\pm}} \to \epsilon^{\pm 1}\hat X_{\gamma_{m,\pm}}$ and $X_{\gamma_p}\to \epsilon^3 X_{\gamma_p}$ and checked this relation up to order $\epsilon^{10}$.

    \item 
    The Fenchel-Nielsen expressions $\hat\CL^{\FN_\pm}$ are related to each other by the jump of the network at $\theta=\frac{\pi}{2}$ which induces the following motivic Kontsevich-Soibelman transformation
\be\label{eq:FN-jump-quantum}
\begin{split}
	\hat\CL^{\FN_+}_a & = \hat{\mathcal{K}}_{\eta_3} 
	\hat{\mathcal{K}}_{\eta_1+\eta_2}^{-\mq-\mq^{-1}}
	\hat{\mathcal{K}}_{2\eta_1+2\eta_2+\eta_3}
	\hat\CL^{\FN_-}_a 
        \\
        & \mathop{=}^{\eqref{eq:basis-change-sl2}} \hat{\mathcal{K}}_{\eta_p-2\eta_m} 
        \hat{\mathcal{K}}_{2\eta_m }^{-\mq-\mq^{-1}}
        \hat{\mathcal{K}}_{2\eta_m +\eta_p}
        \hat\CL^{\FN_-}_a 
        \end{split}
\ee
At the level of generators for $\Sk(\Sigma\times I, GL_1)$ this amounts to the following substitutions into $\hat\CL^{\FN_-}_a$ 
\be\label{eq:quantum-FN-jump-explicit}
\begin{split}
    & 
    \hat{X}_{\gamma_{m,\pm}} \to 
    \hat{X}_{\gamma_{m,\pm}},  \qquad
    \hat{X}_{\gamma_p} \to \hat{X}_{\gamma_p} , \\
    &
    \hat{X}_{\gamma_{l,+}} \to
    \frac{1-\mq\hat{X} _{\gamma_{m,-}-\gamma_{m,+}-\gamma_p}}{\left(1-\mq\hat{X} _{\gamma_{m,+}-\gamma_{m,-}-\gamma_p}\right)\left(1-\hat{X}_{\gamma_{m,-}-\gamma_{m,+}} \right)\left(1-\mq^2\hat{X}_{\gamma_{m,-}-\gamma_{m,+}} \right)} \hat{X}_{\gamma_{l,+}} ,\\ 
    & 
    \hat{X}_{\gamma_{l,-}}\to
    \frac{\left(1-\mq^{-1}\hat{X}_{\gamma_{m,-}-\gamma_{m,+}+\gamma_p} \right)\left(1-\hat{X} _{\gamma_{m,+}-\gamma_{m,-}}\right)\left(1-\mq^2\hat{X} _{\gamma_{m,+}-\gamma_{m,-}}\right)}{1-\mq\hat{X}_{\gamma_{m,+}-\gamma_{m,-}+\gamma_p} }\hat{X}_{\gamma_{l,-}} .
    \end{split}
\ee
We have checked that these expressions relate the expansions of UV loop operators $\hat\CL_{m}, \hat \CL_{l}, \hat \CL_{m^2}$ in American and British coordinates summarized below in Section \ref{sec:FG-FN-Results} and obtained directly from $\mq$-nonabelianization.

\subsection{Computation of $\hat{\mathcal{L}}_{l^2}$ in Fenchel-Nielsen charts}
The relations discussed in Section \ref{eq:quantum-chart-relations} allow to compute the $\mq$-nonabelianization for the skein element $[\varrho_l] \in \Sk(C\times I, GL_2)$ in the Fenchel-Nielsen charts, starting from the Fock-Goncharov $\hat\CL^{\FG}_{l^2}$ expression obtained in \eqref{eq:FG-Ll2}
\be
	\hat\CL^{\FN_-}_{l^2}
	= \prod_{n\geq 0}^{\curvearrowleft} \hat{\mathcal{K}}_{\eta_2+n(\eta_1+\eta_2)}  
	\cdot \hat\CL^{\FG}_{l^2}
\ee

While this relation is straightforward conceptually, it is rather challenging to compute in practice. 
We take an alternative route. Observe that in the Fock-Goncharov chart we have the quantum determinant relation 
\be\label{eq:quantum-determinant}
	\hat{\CL}^{\FG}_{l^2}
	=  
	\mq\, (\hat\CL^{\FG}_l)^2
	-\left(\mq+\mq^{-1}\right)\hat{X}_{\gamma_{l,+}+\gamma_{l,-}}\,.
\ee
Moreover, observe that $\gamma_{l,+}+\gamma_{l,-}$ belongs to the \textit{invariant} sublattice under the map on $H_1(\Sigma',\IZ)$ induced by the deck transformation. 
This means in particular that their pairing with any element of the anti-invariant sublattice \eqref{eq:basis-change-sl2} must vanish
\be
	\langle \gamma_{l,+}+\gamma_{l,-} ,\eta_i\rangle = 0\qquad i=1,2,3\,.
\ee
It then follows that the motivic Kontsevich-Soibelman transformations associated with BPS states, whose charges belong to the anti-invariant sublattice spanned by $\eta_i$, leave $\hat{X}_{\gamma_{l,+}+\gamma_{l,-}}$ invariant.
Therefore, the combination $\hat{\CL}_{l^2}(\hat X^{\theta_0}_{\gamma}) - \mq\, \hat\CL_l^2(\hat X^{\theta_0}_{\gamma})$ is also invariant under wall-crossing, and we obtain
\be
	\hat{\CL}^{\FN_\pm}_{l^2}- \mq\, (\hat\CL^{\FN_\pm}_l)^2
	=
	\hat{\CL}^{\FG}_{l^2} - \mq\, (\hat\CL^{\FG}_l)^2
	=
	-\left(\mq+\mq^{-1}\right)\hat{X}_{\gamma_{l,+}+\gamma_{l,\pm}}\,.
\ee
This entails the following results for $\hat \CL^{\FN_\pm}_{l^2}$ in the American and British Fenchel-Nielsen charts, respectively
\begin{align}
\begin{split}\label{eq:FN-Ll2}
	\hat{\CL}^{\FN_-}_{l^2}
	& = \mq\, (\hat\CL^{\FN_-}_l)^2
	-\left(\mq+\mq^{-1}\right)\hat{X}_{\gamma_{l,+}+\gamma_{l,-}} 
	\\
	& = \mq^{-1}\left(1-\mq^{-1}\hat{X} _{\gamma_{m,+}-\gamma_{m,-}-\gamma_p}\right)\left(1-\mq^{-3}\hat{X} _{\gamma_{m,+}-\gamma_{m,-}-\gamma_p}\right)\hat{X} _{2\gamma_{l,+}}
	\\
	& \quad +\frac{\mq\left(1-\mq \hat{X} _{\gamma_{m,+}-\gamma_{m,-}+\gamma_p}\right)\left(1-\mq^3 \hat{X} _{\gamma_{m,+}-\gamma_{m,-}+\gamma_p}\right)}{\left(1- \hat{X} _{\gamma_{m,+}-\gamma_{m,-}}\right)\left(1-\mq^2  \hat{X} _{\gamma_{m,+}-\gamma_{m,-}}\right)^2\left(1-\mq^4  \hat{X} _{\gamma_{m,+}-\gamma_{m,-}}\right)}\hat{X} _{2\gamma_{l,-}}
	\\
	& \quad +\left(-\mq^{-1}+\mq+\frac{\mq(1+\mq^{-2})\left(1+\mq^2-\mq\left(\hat{X} _{\gamma_p}+\hat{X} _{-\gamma_p}\right)\right)\hat{X} _{\gamma_{m,+}-\gamma_{m,-}}}{\left(1-\mq^{-2}\hat{X} _{\gamma_{m,+}-\gamma_{m,-}}\right)\left(1-\mq^2\hat{X} _{\gamma_{m,+}-\gamma_{m,-}}\right)}\right)\hat{X} _{\gamma_{l,+}+\gamma_{l,-}},
\end{split}
\displaybreak[0] \\[2ex]
\begin{split}\label{eq:FN-Ll2-british}
	\hat\CL^{\FN_+}_{l^2}
	& = \mq \,(\hat{\CL}^{\FN_+}_l)^2-\left(\mq+\mq^{-1}\right)\hat{X}_{\gamma_{l,+}+\gamma_{l,-}}
	\\
	& = \frac{\mq\left(1-\mq\,X_{\gamma_{m,-}-\gamma_{m,+}-\gamma_p}\right)
	           \left(1-\mq^{3}X_{\gamma_{m,-}-\gamma_{m,+}-\gamma_p}\right)}
	          {\left(1-X_{\gamma_{m,-}-\gamma_{m,+}}\right)
	           \left(1-\mq^{2}X_{\gamma_{m,-}-\gamma_{m,+}}\right)^{2}
	           \left(1-\mq^{4}X_{\gamma_{m,-}-\gamma_{m,+}}\right)} \hat{X}_{2\gamma_{l,+}} 
	\\
	& \quad +\left( \mq-\left(1+\mq^{-2}\right)X_{\gamma_{m,-}-\gamma_{m,+}+\gamma_p}
	     +\mq^{-3}X_{2\gamma_{m,-}-2\gamma_{m,+}+2\gamma_p}\right)\hat{X}_{2\gamma_{l,-}} 
	\\
	& \quad + \left(\mq-\mq^{-1}
	  \;+\; \frac{\left(\mq^{2}+1\right)\left(X_{\gamma_p}-\mq\right)\left(X_{-\gamma_p}-\mq\right)}
	             {\mq\left(X_{\gamma_{m,-}-\gamma_{m,+}}+X_{\gamma_{m,+}-\gamma_{m,-}}
	                    -\mq^{2}-\mq^{-2}\right)}\right)\hat{X}_{\gamma_{l,+}+\gamma_{l,-}} .
\end{split}
\end{align}
\begin{remark}[Quantum determinant as a fully antisymmetric Wilson line]\label{rmk:q-det}
    Similar to \eqref{eq:quantum-determinant}, it can be checked in either charts that we have the relation
\be\label{eq:quantum-det-equation}
 \frac{\mq^{-1}\, (\hat\CL^{*}_m)^2 -    \hat{\CL}^{*}_{m^2}}{\mq+\mq^{-1}}
	 = \hat{X}_{\gamma_{m,+}+\gamma_{m,-}}\,,
\ee
with $\gamma_{m,+}+\gamma_{m,-}$ also belonging to the invariant sublattice under deck transformations.
In particular, the superscript $*$ here denotes the fact that this expression is invariant under Kontsevich-Soibelman transformations and holds in all charts $\FG, \FN_{\pm}$.
Note that this combination of trace coordinates is the $\mq$-deformation of the determinant functions appearing in \eqref{eq:Fricke-gl2}.
In fact, it also coincides with the expectation value of a Wilson loop in representation $\mathsf{W}_{\ydiagram{1,1}}$ wrapped around the meridian, see \cite[eq. (5.3)]{Ekholm:2024ceb}.

The quantum determinant relation for $\hat\CL_l$ differs from the one of $\hat \CL_m$ by $\mq\to \mq^{-1}$. 
This might seems surprising at first, but can be traced to the difference in the choice of basepoint for the two loops, see Figure \ref{eq:rho-def-gl2}. In our conventions with monotonically decreasing paths, this choice of basepoint means that $\rho_{m^2}$ has a negative crossing, while $\rho_{l^2}$ has a positive crossing.
\end{remark}

\end{enumerate}

\subsection{Summary of results}\label{sec:FG-FN-Results}

Collecting expressions \eqref{eq:FG-Lm}, \eqref{eq:FG-Ll}, \eqref{eq:FG-Lm2} and \eqref{eq:FG-Ll2}, we have the loop operators in Fock-Goncharov coordinates as 
\begin{equation}\label{eq:FG-summary}
\renewcommand{\arraystretch}{1.6}
\setlength{\arraycolsep}{6pt}
\begin{array}{|c|l|}
\hline
\hat{\CL}^{\FG}_{m} &
  \hat{X}_{\gamma_{m,+}}
  + \hat{X}_{\gamma_{m,-}}\bigl(1-\mq^{-1}\hat{X}_{\gamma_{l,-}-\gamma_{l,+}+\gamma_p}\bigr)
\\[.4ex]
\hline
\hat{\CL}^{\FG}_{l} &
  \hat{X}_{\gamma_{l,-}}
  + \bigl(1-\mq^{-1}\hat{X}_{\gamma_{m,+}-\gamma_{m,-}-\gamma_p}\bigr)\hat{X}_{\gamma_{l,+}}
\\[.4ex]
\hline
\hat{\CL}^{\FG}_{m^2} &
\begin{aligned}[t]
  &\mq^{-1}\hat{X}_{2\gamma_{m,+}}
   + \hat{X}_{2\gamma_{m,-}}\Bigl(\mq^{-1}
     - \bigl(\mq^{-2}+\mq^{-4}\bigr)\hat{X}_{\gamma_{l,-}-\gamma_{l,+}+\gamma_p}
     + \mq^{-5}\hat{X}_{\gamma_{l,-}-\gamma_{l,+}+\gamma_p}^{\,2}\Bigr) \\
  &\quad + \hat{X}_{\gamma_{m,+}+\gamma_{m,-}}\Bigl(\mq^{-1}-\mq
     - \bigl(1+\mq^{-2}\bigr)\hat{X}_{\gamma_{l,-}-\gamma_{l,+}+\gamma_p}\Bigr)
\end{aligned}
\\[.4ex]
\hline
\hat{\CL}^{\FG}_{l^2} &
\begin{aligned}[t]
  &\mq\,\hat{X}_{2\gamma_{l,-}}
   + \Bigl(\mq
     - \bigl(1+\mq^{-2}\bigr)\hat{X}_{\gamma_{m,+}-\gamma_{m,-}-\gamma_p}
     + \mq^{-3}\hat{X}_{\gamma_{m,+}-\gamma_{m,-}-\gamma_p}^{\,2}\Bigr)\hat{X}_{2\gamma_{l,+}} \\
  &\quad + \Bigl(\mq-\mq^{-1}
     - \bigl(1+\mq^{2}\bigr)\hat{X}_{\gamma_{m,+}-\gamma_{m,-}-\gamma_p}\Bigr)\hat{X}_{\gamma_{l,+}+\gamma_{l,-}}
\end{aligned}
\\[.4ex]
\hline
\end{array}
\end{equation}
with our earlier conventions of Eq \eqref{eq:cyclebasischange}, i.e.  $\eta_2 =\gamma_p+ \gamma_{l,-}-\gamma_{l,+}$ and $\eta_3 = \gamma_p+\gamma_{m,-}-\gamma_{m,+}$. 
Collecting \eqref{eq:FN-Lm}, \eqref{eq:FN-Ll}, \eqref{eq:FN-Lm2} and \eqref{eq:FN-Ll2}, we have the loop operators in American Fenchel-Nielsen coordinates as
\begin{equation}\label{eq:FN-A-summary}
\small
\renewcommand{\arraystretch}{1.8}
\setlength{\arraycolsep}{6pt}
\begin{array}{|c|l|}
\hline
\hat{\CL}^{\FN_-}_{m} &
  \hat{X} _{\gamma_{m,+}} + \hat{X} _{\gamma_{m,-}}
\\
\hline
\hat{\CL}^{\FN_-}_{l} &
\begin{aligned}[t]
  & \left(1-\mq^{-1}\hat{X} _{\gamma_{m,+}-\gamma_{m,-}-\gamma_p}\right) \hat{X} _{\gamma_{l,+}} + \frac{1-\mq \hat{X} _{\gamma_{m,+}-\gamma_{m,-}+\gamma_p}}{(1-\hat{X} _{\gamma_{m,+}-\gamma_{m,-}})(1-\mq^2\hat{X} _{\gamma_{m,+}-\gamma_{m,-}})}\hat{X} _{\gamma_{l,-}}
\end{aligned}
\\
\hline
\hat{\CL}^{\FN_-}_{m^2} &
  \mq^{-1}\hat{X} _{2\gamma_{m,+}}
  + \mq^{-1}\hat{X} _{2\gamma_{m,-}}
  + \bigl(\mq^{-1}-\mq\bigr)\hat{X} _{\gamma_{m,+}+\gamma_{m,-}}= \mq^{-1} \hat{\CL}_m^2-(\mq+\mq^{-1})\hat{X}_{\gamma_{m,+}+\gamma_{m,-}} 
\\
\hline
\hat{\CL}^{\FN_-}_{l^2} & \mq\, \hat\CL_l^2 -\left(\mq+\mq^{-1}\right)\hat{X}_{\gamma_{l,+}+\gamma_{l,-}}  \qquad \text{see Eq \eqref{eq:FN-Ll2} for a fully expanded form}\\
\hline
\end{array}
\end{equation}

We omitted the explicit non-abelianization calculation for the loop operators in British Fenchel-Nielsen coordinates, but the procedure is exactly the same as in the American case, and we have
\begin{equation}\label{eq:FN-B-summary}
\small
\renewcommand{\arraystretch}{1.8}
\setlength{\arraycolsep}{6pt}
\begin{array}{|c|l|}
\hline
\hat{\CL}^{\FN_+}_{m} &
  \hat{X} _{\gamma_{m,+}} + \hat{X} _{\gamma_{m,-}}
\\
\hline
\hat{\CL}^{\FN_+}_{l} &
\begin{aligned}[t]
  & 
  \frac{1-\mq \hat{X} _{\gamma_{m,-}-\gamma_{m,+}-\gamma_p}}{\left(1-\hat{X} _{\gamma_{m,-} -\gamma_{m,+}}\right)\left(1-\mq^2\hat{X} _{\gamma_{m,-} -\gamma_{m,+}}\right)}\hat{X} _{\gamma_{l,+}} + \left(1-\mq^{-1}\hat{X} _{\gamma_{m,-} - \gamma_{m,+}+\gamma_p}\right)\hat{X} _{\gamma_{l,-}}
\end{aligned}
\\
\hline
\hat{\CL}^{\FN_+}_{m^2} &
  \mq^{-1}\hat{X} _{2\gamma_{m,+}}
  + \mq^{-1}\hat{X} _{2\gamma_{m,-}}
  + \bigl(\mq^{-1}-\mq\bigr)\hat{X} _{\gamma_{m,+}+\gamma_{m,-}} = \mq^{-1} \hat{\CL}_m^2-(\mq+\mq^{-1})\hat{X}_{\gamma_{m,+}+\gamma_{m,-}} 
\\
\hline
\hat{\CL}^{\FN_+}_{l^2} & \mq\, \hat\CL_l^2 -\left(\mq+\mq^{-1}\right)\hat{X}_{\gamma_{l,+}+\gamma_{l,-}}  \qquad \text{see Eq \eqref{eq:FN-Ll2-british} for a fully expanded form}\\
\hline
\end{array}
\end{equation}

\begin{remark}[Specialization to classical ${SL}_2$ holonomies]\label{rmk:gl2-to-sl2-specialization}
Specializing $\mq$ as in \eqref{eq:q-classical} and $\hat X_{\gamma_{l,+}}  = \hat X_{-\gamma_{l,-}}=X_{\eta_l}$, $\hat X_{\gamma_{m,+}}  = \hat X_{-\gamma_{m,-}}=X_{\eta_m}$, $X_{\gamma_p} = X_{\eta_p}$ reduces the above quantum $
{GL}_2$ operators to the classical ${SL}_2$ expressions \eqref{eq:FGclassUVIRmap-lmp}, \eqref{eq:N2-star-FN-UV-IR-classical-american-lmp}, \eqref{eq:N2-star-FN-UV-IR-classical-british-lmp}, respectively.
\end{remark}

\begin{remark}[Massless limit]
    The massless limit of the above expressions is given by specialising to $X_{\gamma_p} = \mq^{-1}$. 
    As a check that the puncture is removed in the massless limit, observe that a UV operator corresponding to the loop $\varrho_p$ around the puncture evaluates through the UV-IR map to $X_{\gamma_p}+X_{-\gamma_p} = \mq+\mq^{-1}$. This indeed corresponds to the expectation value of the unknot in the ${GL}_2$ skein of a surface, see \eqref{eq:skein-rel-HOMFLY} specialized to $N=2$: as expected once the puncture is removed the loop becomes trivial. 

    It is also interesting to consider the nontrivial line operators. 
    Indeed, it is natural to expact that in the massless limit the covering becomes unbranched and that the spectral networks should disappear. This would leave behind no detours at all, and the UV-IR map should simplify significantly. 
    In particular, once-around loop operators should have just trivial lifts to both sheets. The Fenchel-Nielsen expression for $\hat\CL_m$ already has this form. We expect a similar phenomenon to happen for $\hat\CL_l$.
    A caveat is that the deformation to the massless limit would nevertheless deform the network and that may induce additional Kontsevich-Soibelman jumps.
    Indeed, we observe that in the massless limit the Fenchel-Nielsen expressions for $\hat\CL_{l}$ becomes
    \be
\begin{split}
    \hat{\CL}_{l}(\hat X_\gamma & = 
 \left(1-\hat{X}_{\gamma_{m,+}-\gamma_{m,-}}\right) \hat{X}_{\gamma_{l,+}} 
 + \frac{1}{1-\mq^2\hat{X}_{\gamma_{m,+}-\gamma_{m,-}}}\hat{X}_{\gamma_{l,-}}
 \\
 & = 
  (\mq^2\hat{X}_{\gamma_{m,+}-\gamma_{m,-}};\mq^2)_\infty^{-1}
  \cdot 
  \left( \hat{X}_{\gamma_{l,+}} 
 + \hat{X}_{\gamma_{l,-}}\right)
 \cdot(\mq^2\hat{X}_{\gamma_{m,+}-\gamma_{m,-}};\mq^2)_\infty
 \\
\end{split}
    \ee
    This resembles the trivial uplift of the path $\varrho_l$, up to conjugation by a Kontsevich-Soibelman transformation corresponding to \textit{half} of the vector multiplet. 
    \end{remark}

\section{Match with $GL_2$ spherical DAHA}\label{sec:line-ops-DAHA}

In this section we provide an explicit description for the isomorphism between the algebra of loop operators for $GL_2$ local systems on a punctured torus and the spherical DAHA of $GL_2$
\be\label{eq:gl2-skein-daha-isomorphism}
	\Sk(C\times I, GL_{2}) \ \simeq\ \SH_2^{q,t}\,.
\ee

\subsection{Macdonald generators from Fenchel-Nielsen charts}\label{sec:SH2-modules-from-UV-IR}

The definition and properties of spherical DAHA $\SH_N^{{q,t}}$ are collected in Appendix \ref{app:spherical-DAHA}.
This algebra admits a faithful representation on symmetric polynomials with generators $E^\pm_r, D^\pm_r$ defined in \eqref{eq:SHn-generators-expressions} represented, respectively by multiplication by elementary symmetric functions $\sfE_r$ and by Macdonald $q$-difference operators. Specializing the general expressions \eqref{eq:SH-X-generators} and \eqref{eq:Macdonald-ops} to the case $N=2$ gives 
\be
\begin{split}
	\sfE^\pm_1 &= \hat x_1^{\pm 1} +\hat x_2^{\pm 1} \\
	\sfE^\pm_2 &= \hat x_1^{\pm 1} \hat x_2^{\pm 1} \\
\end{split}
\qquad
\begin{split}
\sfD^\pm_1 &=t^{\mp\frac{1}{2}}
	\left(
	\frac{t^{\pm 1} x_{1}-x_2}{x_{1}-x_2} \, \Xi_{1}^{\pm 1}
	+\frac{t^{\pm 1} x_{2}-x_1}{x_{2}-x_1}\, \Xi_{2}^{\pm 1}
	\right)\\
	\sfD^\pm_2 &= \Xi_{1}^{\pm 1}\Xi_{2}^{\pm 1}.
\end{split}
\ee

We next show that this presentation of the algebra agrees with the $GL_2$ skein algebra of the punctured torus computed with the Fenchel-Nielsen quantum UV-IR map. 
More specifically, we will show how the line operators $\hat\CL^{\FN_\pm}_{l}$, $\hat\CL^{\FN_\pm}_{m}$, $\hat\CL^{\FN_\pm}_{l^2}$ and $\hat\CL^{\FN_\pm}_{m^2}$ that generate the latter are related to the generators of the positive half of spherical DAHA of $GL_2$.
Generators with negative winding $\hat\CL^{\FN_\pm}_{-m}$, $\hat\CL^{\FN_\pm}_{-l}$, $\hat\CL^{\FN_\pm}_{-m^2}$ and $\hat\CL_{-l^2}^{\FN_\pm}$  can be calculated in the same way as the positive ones. Here $-m, -l$ etc. denote paths obtained by reversing the orientation of $\varrho_m, \varrho_l$ etc, in $C\times I$. 
The resulting expressions are related to those of positive-winding modes by the replacements $\gamma_{\pm,m} \to -\gamma_{\mp,m}$, $\gamma_{\pm,l}\to -\gamma_{\mp,l}$.

Here we focus on the American chart, and later comment on the British one.
We identify parameters of $\SH_2^{q,t}$ with 
\begin{equation}
\begin{split}
   q^{\frac12} = -\mq^{-1},  \qquad t = \frac{\mq^{-1}}{X_{\gamma_p}},
    \end{split}
    \label{eq:AmToMac}
\end{equation}
and the Heisenberg generators with the quantum torus generators of $\Sk(\Sigma\times I,GL_1)$
\begin{equation}
\begin{split}
   \xh_1 = \hat{X}_{\gamma_{m,+}},
   & \qquad 
   \xh_2 = \hat{X}_{\gamma_{m,-}},
   \qquad
   \yh_1 =  \hat{X}_{\gamma_{l,+}},
   \qquad 
   \hat{y}_2 = \frac{q}{t}\hat{X}_{\gamma_{l,-}}.
    \end{split}
    \label{eq:AmToMac}
\end{equation}
As a check, $\hat{x}_i$ and $\hat{y}_j$ obey the Heisenberg algebra relations
$    \yh_i \xh_j = q^{\delta_{ij}} \xh_j\yh_i $.
With these identifications we reach the following identifications between UV quantum loop operators and generators of $\SH_2^{q,t}$
\be\label{eq:Mac-From-U}
\begin{split}
	\CU\, \hat{\CL}^{\FN_-}_{m}\, \CU^{-1} & = \xh_1 + \xh_2 
	= \sfE_1^+,
    	\\ 
	\CU\,\hat{\CL}^{\FN_-}_{l}\,\CU^{-1}  & = 
	\frac{t \xh_1 - \xh_2}{ \xh_1 - \xh_2} \yh_1
	+\frac{t \xh_2 - \xh_1}{ \xh_2 - \xh_1} \yh_2 
	= t^{\frac{1}{2}}\sfD_1^+,
	\\
	\CU\,\left(
	\frac{\hat{\CL}^{\FN_-}_{m^2} + q^{\frac{1}{2}} (\hat\CL^{\FN_-}_m)^2}{q^{-\frac{1}{2}}+q^{\frac{1}{2}}}
	 \right)\,\CU^{-1}
	& = \hat{x}_1\hat{x}_2
	= \sfE_2^+
	\\
	\CU\,\left(
	\frac{\hat{\CL}^{\FN_-}_{l^2} + q^{-\frac{1}{2}} (\hat\CL^{\FN_-}_l)^2}{q^{-\frac{1}{2}}+q^{\frac{1}{2}}}
	 \right)\,\CU^{-1}
	& = t\,\hat{y}_1\hat{y}_2
	= t\, \sfD_2^+
\end{split}
\ee
up to the similarity transformation 
\be\label{eq:conjugation}
	\CU = 
	\frac{1}{\xh_1- \xh_2}\cdot\left(q\frac{\xh_1}{\xh_2};q\right)_\infty^{-1}\,.
\ee
Note that the operators mapping to $\sfE_2^+, \sfD_2^+$ are the quantum deformations of determinants $\CD_l, \CD_m$ appearing in the $GL_2$ trace-determinant relations \eqref{eq:Fricke-gl2}. In fact they resemble the Schur function $s_{\ydiagram{1,1}}$ with variables $\hat x_i, \hat y_i$ respectively, for reasons explained in Remark \ref{rmk:q-det}.

Relations \eqref{eq:Mac-From-U} show that the UV quantum loop operators, i.e. the elements $m,l,m^2,l^2\in \Sk(C\times I, GL_2)$, correspond through $\mq$-nonabelianization to generators of the positive half of $\SH_{2}^{q,t}$. 
We have checked that the operators $\hat\CL^{\FN_{-}}_{-l}$, $\hat\CL^{\FN_{-}}_{-m}$, $\hat\CL^{\FN_{-}}_{-l^2}$ and $\hat\CL^{\FN_{-}}_{-m^2}$ computed via $\mq$-nonabelianization give the negative generators $\sfE_1^-$, $\sfE_2^-$, $t^{-\frac12}\sfD_1^-$ and $t^{-1}\sfD_2^-$. 
Together these give a fully explicit realization of the isomorphism \eqref{eq:gl2-skein-daha-isomorphism} between the $GL_2$ skein algebra of the punctured torus and the spherical DAHA of type ${GL}_2$.

Before moving on to other instances of the isomorphism \eqref{eq:gl2-skein-daha-isomorphism} we pause to comment on the nature of the conjugation operator $\CU$.
Each of its two contributions has physical / geometric interpretations
\begin{enumerate}
\item
The Vandermonde factor $(\hat x_1-\hat x_2)^{-1}$ corresponds to the open topological string partition function of a stack of $N=2$ branes on a solid torus $S^1\times D^2$. This arises as follows. The symmetric polynomial module $\CP_2^{S_2}$ of $\SH_2^{q,t}$ is isomorphic to the $GL_2$ skein module of $S^1\times D^2$. 
In the literature on skein theory, and correspondingly in the literature on polynomial representations of spherical DAHAs, the vacuum partition function is conventionally normalized to be the Macdonald polynomial of the empty partition $P_{\emptyset}^{(q,t)} = 1$. However in string theory the vacuum partition function is naturally normalized to be the partition function of all holomorphic curves on a stack of $N$ branes on $S^1\times D^2$ viewed as the zero-section of $T^*(S^1\times D^2)$.
In this setting, the only curves that arise are $N(N-1)/2$ annuli stretching between the $N$ branes. Their partition function is precisely the Vandermonde determinant, where $x_i$ encode the open string moduli of the branes.

\item
The factor $(q\frac{\xh_1}{\xh_2};q)_\infty$ also has a geometric interpretation: it corresponds to the partition function of a \emph{disk}. 
In our setting, this factor corresponds to \emph{half} the contribution of the vector multiplet \eqref{eq:BPS-spectrum-motivic} that relates the two Fenchel-Nielsen charts. 
The appearance of this factor may appear somewhat surprising for two reasons. First, vector multiplet contributions are known to arise from \textit{annuli} instead of disks \cite{Gaiotto:2009hg} (the partition function of an annulus would be a Vandermonde). Second, in the context of 4d $\CN=2$ wall-crossing the two contributions \eqref{eq:BPS-spectrum-motivic} by the vector multiplet come with the same central charge, and therefore should appear always together. 
Both of these apparent issues have a natural resolution. 
\begin{enumerate}
\item
The apparent tension between disk and annulus is resolved by recalling that a vectormultiplet corresponds to a \textit{1-parameter family} of annuli, and that BPS {states} arise from quantization of the moduli space of sheaves.
It is only the boundaries of moduli space that contribute to the space of BPS states: in the case of vector multiplets the moduli space is $I=[0,1]$ and there are two contributions.
In fact each boundary corresponds to a degeneration of the annulus into a disk, see \cite{Banerjee:2022oed} for this specific example and for an extended discussion.

\item
To understand how two contributions from the vector multiplet can be separated from each other, recall that $q$-nonabelianization is really a map between \textit{skein algebras of surfaces}. In particular this means that every contribution should be regarded as a loop in three dimensions, and in this context the two disjoint loops can be considered separately. 
In fact, in the recent work \cite[Section 11]{Ekholm:2025anq} it was shown that the vector multiplet contribution to the motivic Kontsevich-Soibelman wall-crossing formula splits into two distinct disk contributions, and that it is natural to consider each of them separately in the setting of skein algebras.

\end{enumerate}
\end{enumerate}

\begin{remark}[Macdonald polynomial module from Fenchel-Nielsen British chart.]
\label{eq:macdonald-british}
For completeness, we remark that another isomorphism between $\Sk(C\times I, GL_2)$ and $\SH_2^{q,t}$  can be established in the British Fenchel-Nielsen chart. In this case we identify
\be\label{eq:BritToMac}
    q^\frac12 = -\mq^{-1}, 
    \qquad 
    t = \mq^{-1}\, X_{\gamma_p}, \
\ee
Comparing with \eqref{eq:AmToMac} with American and British parameters denoted $q_{A,B}$ and $t_{A,B}$ respectively, the two isomorphisms are related by the involution
\be
	q_A=q_B\,,
	\qquad
	t_B = -\frac{q_A}{t_A}\,.
\ee
This is a known symmetry of $GL_2$ spherical DAHA, related to bispectral duality of the associated quantum integrable system \cite{di2024ruijsenaars}.\footnote{In the reference this is realized by $\tau\to -\tau$, where $t = e^{\pi b (2\tau +c)}, q=e^{\pi i b^2}$ with $c = \frac{i}{2}(b+b^{-1})$. We thank Misha Bershtein for bringing this to our attention.}

We then denote
\be    \label{eq:BritToMac}
\begin{split}
	\xh_1 = \hat{X}_{\gamma_{m,+}},  \qquad
	\xh_2 = \hat{X}_{\gamma_{m,-}} ,\qquad
	\yh_1 = \frac{q}{t} \hat{X}_{\gamma_{l,+}}, \qquad
	\hat{y}_2 = \hat{X}_{\gamma_{l,-}},
\end{split}
\ee
which again obeys the Heisenberg algebra
$    \yh_i \xh_j = q^{\delta_{ij}} \xh_j\yh_i$.
Using these, we have the following identifications between UV quantum loop operators and generators of spherical DAHA
\begin{equation}
    \begin{split}
    \CU \, \hat\CL^{\FN_+}_{m} \, \CU^{-1} 
    & =  \xh_1 + \xh_2 = \sfE_1^+ \,,
    \\
    \CU \,\hat\CL_{l}^{\FN_+}  \, \CU^{-1}
    & = \frac{t \xh_1 - \xh_2}{\xh_1 - \xh_2} \yh_1
    +\frac{t\xh_2 - \xh_1}{ \xh_2 - \xh_1} \yh_2 
     = t^{\frac{1}{2}}\,\sfD_1^+
    \,, \\
    \CU \, \left(\frac{\hat\CL^{\FN_+}_{m^2} + q^{\frac{1}{2}} (\hat\CL^{\FN_+}_m)^2}{q^{-\frac{1}{2}} + q^{\frac{1}{2}}} \right) \, \CU^{-1}
    & = \hat{x}_1\hat{x}_2 
    = \sfE_2^+ \,,\\
    \CU\,\left(
	\frac{\hat{\CL}^{\FN_+}_{l^2} + q^{-\frac{1}{2}} (\hat\CL^{\FN_+}_l)^2}{q^{-\frac{1}{2}}+q^{\frac{1}{2}}}
	 \right)\,\CU^{-1}
	& = t\,\hat{y}_1\hat{y}_2
	= t\,\sfD_2^+ \,,
    \end{split}
    \label{eq:changetobritish}
\end{equation}
where
\be
	\CU =  \frac{1}{\hat x_1 - \hat x_2}\cdot \left(q\frac{\hat{x}_2}{\hat{x}_1};q\right)_\infty^{-1}\,.
\ee
\end{remark}

\subsection{Cluster modules from Fock-Goncharov charts}\label{sec:SH2-modules-from-UV-IR-FG}

The relations \eqref{eq:Mac-From-U} between generators of $\Sk(C\times I, GL_2)$ and those of $\SH_2^{q,t}$ hinge on the computation of the $\mq$-nonabealianization in a Fenchel-Nielsen chart. 
Indeed the output are expressions \eqref{eq:FN-A-summary} for the UV quantum loop operators expressed directly in terms of quantum torus variables $\hat X_\gamma$, which are directly related to the quantum torus variables that arise in the Macdonald polynomial module, see \eqref{eq:AmToMac}.

Here we consider what happens when the $\mq$-nonabelianization map is computed in a Fock-Goncharov chart. 
Since Fock-Goncharov charts provide cluster coordinates, we refer to the corresponding modules as cluster modules.
For example, let us adopt identifications of $q,t$ as in \eqref{eq:AmToMac} and let us consider the quantum UV-IR map in the Fock-Goncharov chart studied above. We define $\hat x_i, \hat y_i$ operators according to
\be
	\hat x_{1} = \hat X_{\gamma_{m,+}},\qquad
	\hat x_{2} = \hat X_{\gamma_{m,-}},\qquad
	\hat y_{1} = \hat X_{\gamma_{l,+}},\qquad
	\hat y_{2} = \hat X_{\gamma_{l,-}},
\ee
where $\hat X_{\gamma} \equiv \hat X_{\gamma}^{\theta_0}$ here is understood.
We then obtain the following expressions for generators of $\SH_2^{q,t}$ in terms of multiplicative operators $\hat x_i = x_i$ and $q$-shift operators $\hat y_i = \Xi_i$
\be
\begin{split}
	\tilde \sfE_1^+
	& =
	\hat{\CL}^{\FG}_{m} 
	= \xh_1
    + \xh_2\left(1  - \frac{q}{t} \,\frac{\yh_2}{\yh_1} \right),
    	\\ 
	\tilde \sfD_1^+
	& = 
	\hat{\CL}^{\FG}_{l}
	= 
	\yh_2 + \left(1
    - t \,\frac{\xh_1}{\xh_2} \right) \yh_1,
	\\
	\tilde \sfE_2^+
	& =
	\frac{\hat{\CL}^{\FG}_{m^2} + q^{\frac{1}{2}} (\hat\CL^{\FG}_m)^2}{q^{-\frac{1}{2}}+q^{\frac{1}{2}}}
	= \hat{x}_1 \hat{x}_2,
	\\
	\tilde \sfD_2^+
	& = 
	\frac{\hat{\CL}^{\FG}_{l^2} + q^{-\frac{1}{2}} (\hat\CL^{\FG}_l)^2}{q^{-\frac{1}{2}}+q^{\frac{1}{2}}}
	= \hat{y}_1 \hat{y}_2.
\end{split}
\ee
The generators $\tilde \sfE_{i}^{\pm},\tilde \sfD_{i}^{\pm}$ are related to $\sfE_{i}^{\pm}, \sfD_{i}^{\pm}$ by a similarity transformation which in this case is a composition of the quantum FG-FN change of variables \eqref{eq:quantumFGFN} and the operator $\CU$ in \eqref{eq:conjugation}. 
In particular, this means that the action of $\tilde \sfD_i$ can be simultaneously diagonalized on a module generated by  acting with the corresponding sequence of motivic Kontsevich-Soibelman transformations on the module generated by Macdonald polynomials.

\begin{remark}[Framed quivers and unphysical cycles]
After this work was completed we became aware of \cite{di2024ruijsenaars} whose results appear directly related to the realization of $\SH_2^{q,t}$ in Fock-Goncharov charts studied here.
The reference studies a cluster realization of $GL_2$ spherical DAHA through the combinatorics of a framed quiver. 
The relation to our work should be understood through a generalization of the correspondence between quivers and networks \cite{Alim:2011ae, 2013arXiv1302.7030B, Gabella:2017hpz}.
The standard dictionary states that vanilla (i.e. unframed) BPS quivers feature a number of vertices corresponding to generators the physical sublattice $\fY$, which in our case has rank 3. 
To capture the full $GL_2$ skein algebra it is essential to extend this dictionary to framed quivers, where additional nodes correspond to missing generators for the entire homology lattice of $\Sigma$, which in our case has rank 5.
Indeed, in the reference the resulting framed quiver has 5 nodes: 3 for the usual Markov quiver, plus two framed nodes. We can match the charges of each node by comparing operators $L_{(1,0)}$ and $\Delta_{(1,0)}$ in 
\cite[(4.1)]{di2024ruijsenaars} with our expressions for $\hat\CL^{\operatorname{FG}}_{m}$ in  \eqref{eq:FG-summary} and for the quantum determinant in \eqref{eq:quantum-det-equation}.
The two match exactly, upon identification of $Y_{e_4}$ with $\hat{X}_{\gamma_{m,-}}$, $Y_{e_2+e_4}$ with $\hat{X}_{\gamma_{m,-}+\gamma_{l,-}-\gamma_{l,+}+\gamma_p}$ and $Y_{e_1+e_2+e_4}$ with $\hat{X}_{\gamma_{m,+}}$.
We therefore conclude that the framing nodes correspond to 'unphysical' generators of $H_1(\Sigma,\IZ)$ in this case.

Moreover, the authors of \cite{di2024ruijsenaars} exploit $SL_2(\IZ)$ equivariance of the cluster algebra construction to deduce the form of other generators such as our $\hat \CL_l^{\FG}, \hat \CL_{l^2}^{\FG}$.
In the setting of spectral networks, the corresponding realization of mutations that generate the mapping class group of the punctured torus has been described in terms of Kontsevich-Soibelman transitions of the spectral network in \cite{Gang:2017ojg}.
\end{remark}

\appendix

\section{Spherical double affine Hecke algebras}\label{app:spherical-DAHA}

We collect the definition of spherical double affine Hecke algebras and their polynomial module.

The Double Affine Hecke Algebra (DAHA) $\ddot \H_N^{q,t}$ of type $GL_N$ is generated by
\be
	T_1,\dots, T_{N-1}\,,
	\qquad
	X_1^{\pm 1},\dots, X_N^{\pm 1}\,,
	\qquad 
	Y_1^{\pm 1},\dots, Y_N^{\pm 1}
\ee
where subject to the following relations
\be\label{eq:DAHA-rel-1}
\begin{split}
	&\prod_{\epsilon = \pm}(T_i- \epsilon t^{\epsilon \frac{1}{2}}) = 0 ,\\
	&T_i T_{i+1} T_i = T_{i+1} T_i T_{i+1}  ,\\
	&T_i T_{j} = T_{j} T_i  \qquad |i-j| > 1,  \\
\end{split}
\qquad
\begin{split}
&X_i X_j = X_j X_i, 
\\& Y_i Y_j = Y_j Y_i, 
\\
&T_i X_i T_i = X_{i+1}, 
\end{split}
\qquad
\begin{split}
&T_i Y_{i+1} T_i = Y_{i}, 
\\&X_1^{-1}Y_2 = Y_2 X_1^{-1}T_1^{-2},
\\&Y_1 X_1 X_2 \dots X_N = q X_1 X_2 \dots X_N Y_1 .
\end{split}
\ee

The \textit{spherical DAHA} is defined by conjugation 
\be\label{eq:SHn-def}
	\SH_N^{q,t}= e_N \, \ddot \H_N^{q,t}\, e_N\,.
\ee
by the spherical projector
\be\label{eq:symmetric-idempotent}
	e_N := \frac{ 1}{ [n]_{t}!}\sum_{w \in S_N} t^{\frac{|w|}{2}} \, T_w,
\ee
where $[N]_t  = \frac{t^N-1}{t-1}$, $T_w = T_{i_1}\cdots T_{i_m}$ and $|w| = m$ for $w = \sigma_{i_1}\cdots \sigma_{i_m}$. Note in particular that this is an idempotent $e_N^2 = e_N$, and that $e_N T_i e_N = t^{\frac{1}{2}}$. 
Generators of the spherical DAHA in fact correspond to symmetric functions of the generators $X_i, Y_i$. A set of generators can be chosen as follows \cite{schiffmann2013elliptic, cherednik2005double}
\be\label{eq:SHn-generators}
	\SH_N^{q,t} \simeq \langle E_1^{\pm}\dots, E^{\pm}_N , D^{\pm}_1,\dots, D^{\pm}_N \hspace{0.2cm} | \hspace{0.2cm} \text{relations induced by \eqref{eq:DAHA-rel-1}}\rangle,
\ee
with
\be\label{eq:SHn-generators-expressions}
\begin{split}
	E_r^\pm & = e_N\, e_r(X_1^{\pm 1},\dots, X_N^{\pm 1})\, e_N, \\
	D_r^\pm & = e_N\, e_r(Y_1^{\pm 1},\dots, Y_N^{\pm 1})\, e_N, \\
\end{split}	
\qquad
\begin{split}
	r=1,\dots, N
\end{split}	
\ee
where $e_r(z_1, \dots, z_N):=  \sum_{{i_1}<i_2\dots< i_r} z_{i_1}z_{i_2}\dots z_{i_r} $ denote the $r$-th elementary symmetric polynomials.

Spherical DAHA  $\SH_N^{q,t}$ admits a faithful representation on the space of symmetric Laurent polynomials
\be\label{eq:DAHA-P-mod}
	\CP_N^{S_N} = \IC[x_1^{\pm 1}, \dots, x_{N}^{\pm1}]^{S_N},
\ee
Denoting the action of $E_{r}^{\pm}$ and $D_r^{\pm}$  by $\sfE_i$ and $\sfD_i$ respectively, they correspond to
\begin{equation}
\label{eq:SH-X-generators}
    \sfE_r^{\pm} = e_r(x_1,\dots,x_N),
\end{equation}
\begin{equation}
    \label{eq:Macdonald-ops}
    \sfD_r^\pm  = t^{\pm\frac{r(r-N)}{2}}\sum_{1\leq i_1 < \cdots<i_r\leq N}\prod_{k=1}^r\left(\prod_{j \notin \{i_k\}_{k=1}^r}\frac{t^{\pm 1} x_{i_k}-x_j}{x_{i_k}-x_j}\right)\Xi_{i_k}^{\pm 1},
\end{equation}
where $\Xi_i$ are $q$-shift operators
\be
	\Xi_i f(x_1,\dots, x_i,\dots,x_N) = f(x_1,\dots, q\,x_i,\dots,x_N).
\ee
The operators $\sfD_r^\pm$ are also known as Macdonald $q$-difference operators \cite{macdonald1998symmetric}. Their most important property is that they are in mutual involution, and they are simultaneously diagonalized by Macdonald polynomials as a basis for $\CP_N^{S_N,+} = \IC[x_1, \dots, x_{N}]^{S_N}$.

\section{Weights of lifted paths for $\mq$-nonabelianization}\label{app:q-nonabelianization-weights}
We review the weights associated to lifted paths in the $\mq$-nonabelianization map, following \cite{Neitzke:2020jik}.
\paragraph{Weights for lifted paths.} There are four contributions to the weight $\alpha(\tilde\varrho)$ of each lift $\tilde\varrho$ 
\be
	\alpha(\tilde\varrho) = \alpha_f(\tilde\varrho) \cdot \alpha_d(\tilde\varrho)\cdot \alpha_e(\tilde\varrho) \cdot \alpha_w(\tilde\varrho)
\ee
defined as follows.
\begin{itemize}
\item {\bf Framing factors $\alpha_f$}. These depend only on the underlying path $\varrho$ and is the same for all lifts. It is the product of $\mq^{\pm \frac{1}{2}}$ 
for each point where $\varrho$ (projected to $C$) is tangent to the foliation on $C$. Since our paths are monotonically decreasing there are two relevant cases to consider, they are summarized below\footnote{These correspond to the last two frames of \cite[Figure 17]{Neitzke:2020jik} since our paths are monotonically decreasing.},
\be
	\pic{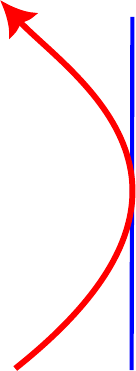}{.4} = \mq^{-\frac12},\qquad \pic{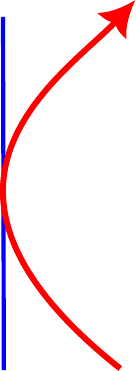}{.4} = \mq^{\frac12},
\ee
where the above diagrams are drawn on the UV curve $C$.

\item {\bf Detour factors $\alpha_d$}. Every detour contributing to $\tilde\varrho$ introduces a factor $\mq^{\pm \frac{1}{2}}$. The sign depends on the relative orientation of the path $\varrho$ and of the trajectory generating the detour (oriented away from the branch point), according to the following rules,
\be
	\pic{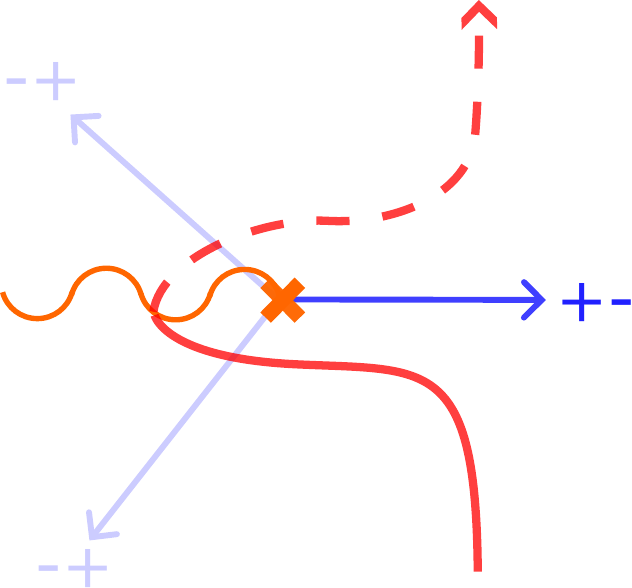}{.25} = \pic{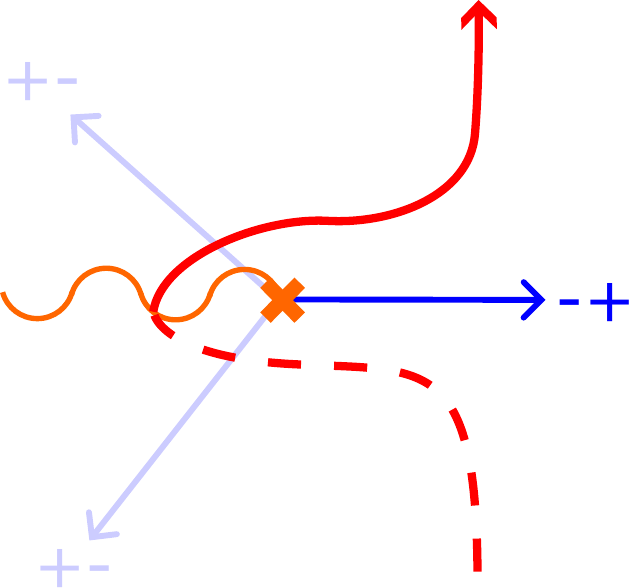}{.25} = \mq^{\frac12}, \qquad \pic{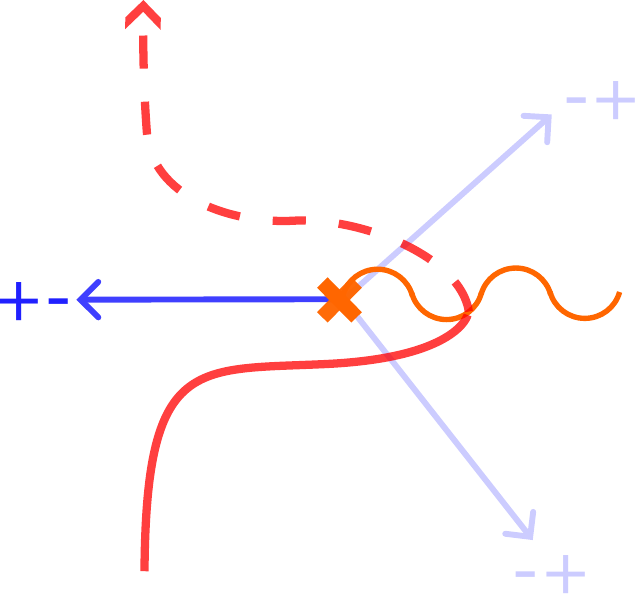}{.25} = \pic{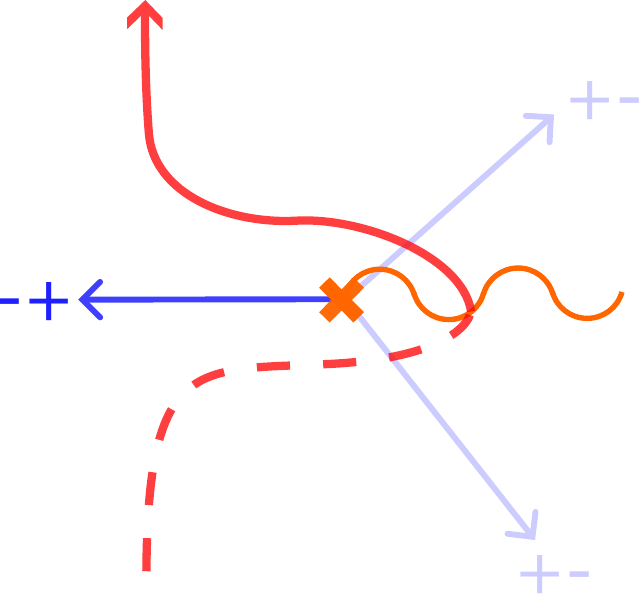}{.25} = \mq^{-\frac12},
\ee
where the above diagrams are drawn on the IR curve $\Sigma$.

\item {\bf Exchange factors $\alpha_e$}. An exchange factor contribution depends on two properties. First, it depends on whether the two segments of $\varrho$ connected by the flow line cross that line in the same direction or opposite directions. Second, consider the projection of the two strands to the plane perpendicular to the flow line,\footnote{That is, a local patch of the leaf space, in the terminology of \cite{Neitzke:2020jik}.} in this plane the two oriented segments form either a positive crossing or a negative one, and the contribution depends on this. The four cases correspond to the following contributions.
\begin{equation}
  \setlength{\arraycolsep}{8pt}
  \begin{array}{c|cccc}
    \pic{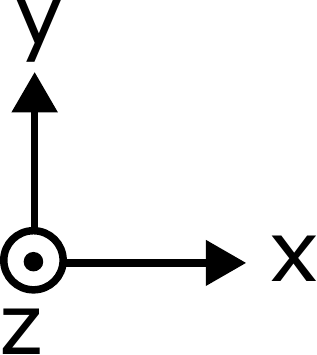}{.2} &
      \pic{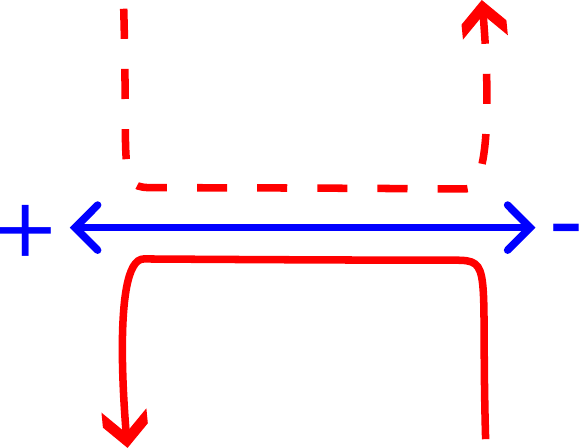}{.22} &
      \pic{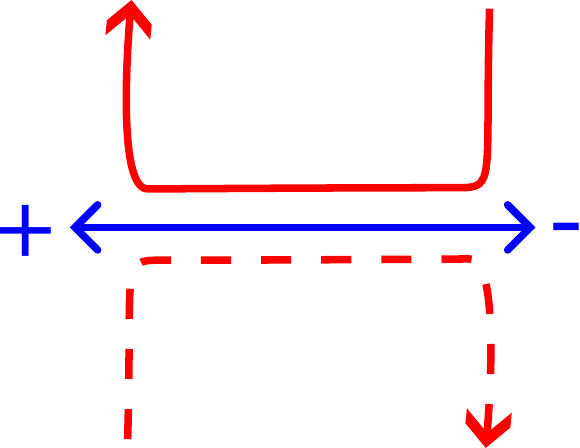}{.22} &
      \pic{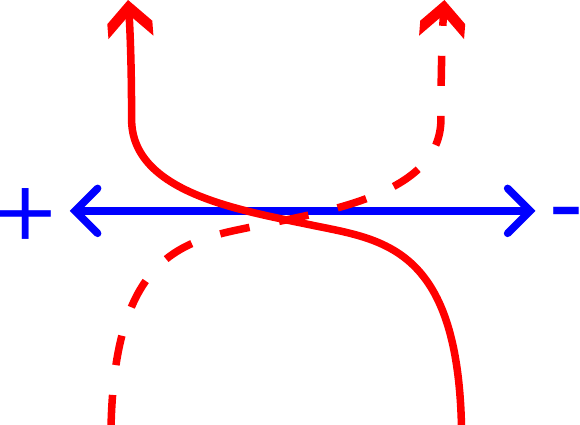}{.22} &
      \pic{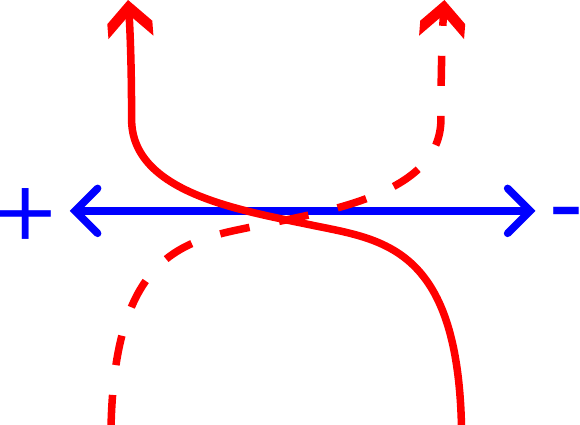}{.22} \\[3em]
    \pic{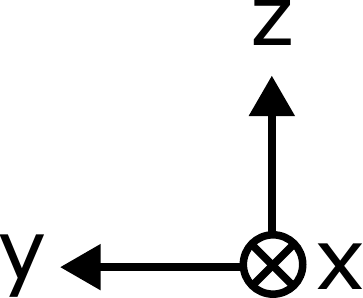}{.2} &
      \pic{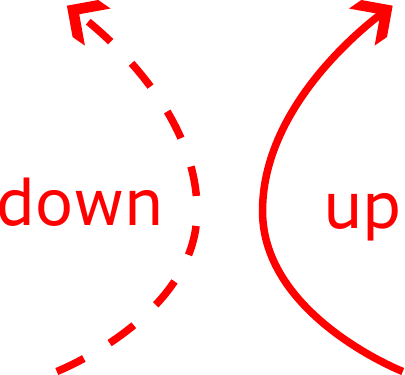}{.22} &
      \pic{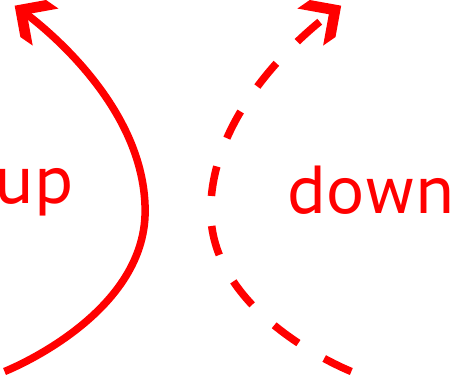}{.22} &
      \pic{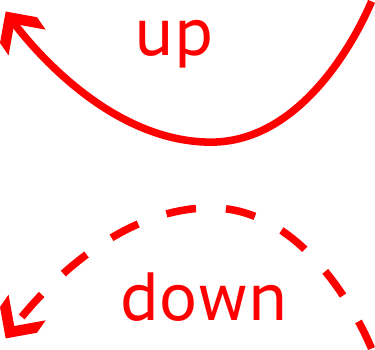}{.22} &
      \pic{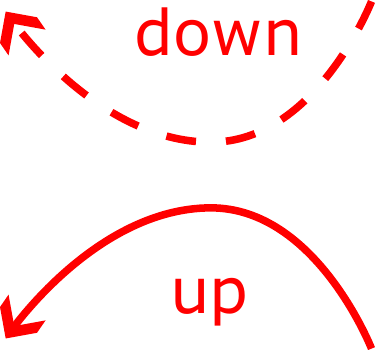}{.22} \\[2.5em]
    \hline
    \rule{0pt}{2.5ex}
    \alpha_e & 1-\mq^{-2} & 1-\mq^{2} & \mq-\mq^{-1} & \mq^{-1}-\mq
  \end{array}
  \label{eq:exchange-rules}
\end{equation}
In the above diagrams, ``up'' and ``down'' refer to the direction out of the page and into the page respectively, and branches of $\Sigma$ are aligned along the x-y plane, and hence the interval $I$ is along the z-axis.

\item {\bf Winding factors $\alpha_w$}. These are powers of $\mq^{\pm \frac{1}{2}}$ contributed by points wherever the \emph{projection} of $\tilde\varrho$ is tangent to the foliation. The sign depends on the sheet to which $\tilde\varrho$ is lifted and there are four possible cases summarized below\footnote{These correspond to the first two frames of \cite[Figure 20]{Neitzke:2020jik} only, since our paths are monotonically decreasing.}
\begin{equation}
    \pic{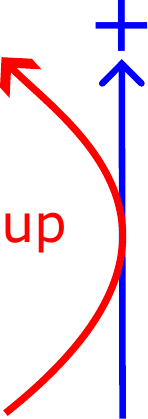}{.3}\; =\; \pic{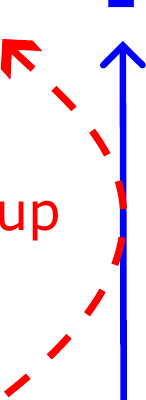}{.3} \;= \;\pic{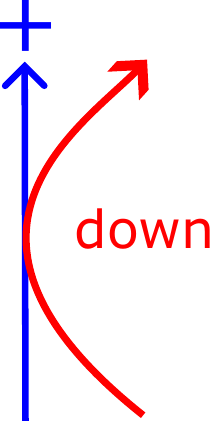}{.3} \;= \;\pic{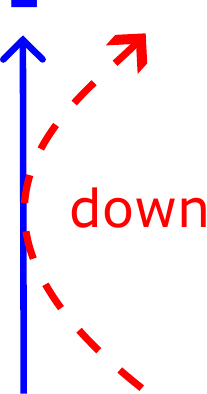}{.3}\; =\; \mq^{\frac12}, \qquad \pic{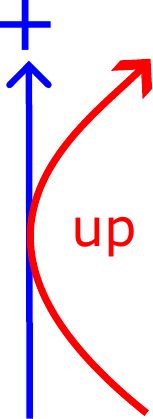}{.3} \;= \;\pic{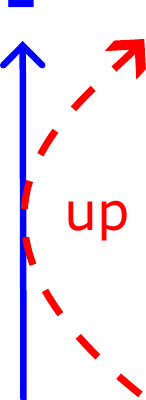}{.3} \;=\; \pic{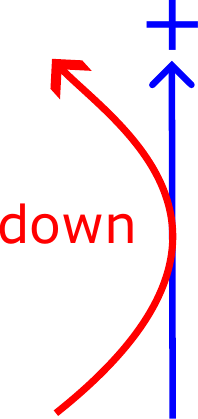}{.3} \;= \;\pic{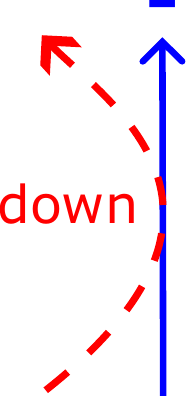}{.3} \;= \;\mq^{-\frac12}
\end{equation}
The above diagrams are drawn on the branches of $\Sigma$. We note that unlike framing, winding factors are contributed also by tangencies arising from detours and exchange paths.

\end{itemize}

\newpage
\bibliographystyle{unsrt}
\bibliography{bibliography.bib}

@article{Witten:1988hf,
    author = "Witten, Edward",
    editor = "Mitra, Asoke N.",
    title = "{Quantum Field Theory and the Jones Polynomial}",
    reportNumber = "IASSNS-HEP-88-33",
    doi = "10.1007/BF01217730",
    journal = "Commun. Math. Phys.",
    volume = "121",
    pages = "351--399",
    year = "1989"
}

@article{Gaiotto:2009hg,
    author = "Gaiotto, Davide and Moore, Gregory W. and Neitzke, Andrew",
    title = "{Wall-crossing, Hitchin systems, and the WKB approximation}",
    eprint = "0907.3987",
    archivePrefix = "arXiv",
    primaryClass = "hep-th",
    doi = "10.1016/j.aim.2012.09.027",
    journal = "Adv. Math.",
    volume = "234",
    pages = "239--403",
    year = "2013"
}

@article{Gaiotto:2012rg,
    author = "Gaiotto, Davide and Moore, Gregory W. and Neitzke, Andrew",
    title = "{Spectral networks}",
    eprint = "1204.4824",
    archivePrefix = "arXiv",
    primaryClass = "hep-th",
    doi = "10.1007/s00023-013-0239-7",
    journal = "Annales Henri Poincare",
    volume = "14",
    pages = "1643--1731",
    year = "2013"
}

@article{Galakhov:2014xba,
    author = "Galakhov, Dmitry and Longhi, Pietro and Moore, Gregory W.",
    title = "{Spectral Networks with Spin}",
    eprint = "1408.0207",
    archivePrefix = "arXiv",
    primaryClass = "hep-th",
    reportNumber = "RUNHETC-2014-14, ITEP-TH-10-14",
    doi = "10.1007/s00220-015-2455-0",
    journal = "Commun. Math. Phys.",
    volume = "340",
    number = "1",
    pages = "171--232",
    year = "2015"
}

@inproceedings{Seiberg:1996nz,
    author = "Seiberg, Nathan and Witten, Edward",
    title = "{Gauge dynamics and compactification to three-dimensions}",
    booktitle = "{Conference on the Mathematical Beauty of Physics (In Memory of C. Itzykson)}",
    eprint = "hep-th/9607163",
    archivePrefix = "arXiv",
    reportNumber = "IASSNS-HEP-96-78, RU-96-55",
    pages = "333--366",
    month = "6",
    year = "1996"
}

@article{Gaiotto:2010be,
    author = "Gaiotto, Davide and Moore, Gregory W. and Neitzke, Andrew",
    title = "{Framed BPS States}",
    eprint = "1006.0146",
    archivePrefix = "arXiv",
    primaryClass = "hep-th",
    doi = "10.4310/ATMP.2013.v17.n2.a1",
    journal = "Adv. Theor. Math. Phys.",
    volume = "17",
    number = "2",
    pages = "241--397",
    year = "2013"
}

@article{Kapustin:2007wm,
    author = "Kapustin, Anton and Saulina, Natalia",
    title = "{The Algebra of Wilson-'t Hooft operators}",
    eprint = "0710.2097",
    archivePrefix = "arXiv",
    primaryClass = "hep-th",
    doi = "10.1016/j.nuclphysb.2009.02.004",
    journal = "Nucl. Phys. B",
    volume = "814",
    pages = "327--365",
    year = "2009"
}

@article{Kapustin:2006hi,
    author = "Kapustin, Anton",
    title = "{Holomorphic reduction of N=2 gauge theories, Wilson-'t Hooft operators, and S-duality}",
    eprint = "hep-th/0612119",
    archivePrefix = "arXiv",
    reportNumber = "CALT-68-2623",
    month = "12",
    year = "2006"
}

@article{Kapustin:2006pk,
    author = "Kapustin, Anton and Witten, Edward",
    title = "{Electric-Magnetic Duality And The Geometric Langlands Program}",
    eprint = "hep-th/0604151",
    archivePrefix = "arXiv",
    doi = "10.4310/CNTP.2007.v1.n1.a1",
    journal = "Commun. Num. Theor. Phys.",
    volume = "1",
    pages = "1--236",
    year = "2007"
}

@article{Kapustin:2005py,
    author = "Kapustin, Anton",
    title = "{Wilson-'t Hooft operators in four-dimensional gauge theories and S-duality}",
    eprint = "hep-th/0501015",
    archivePrefix = "arXiv",
    reportNumber = "CALT-68-2536",
    doi = "10.1103/PhysRevD.74.025005",
    journal = "Phys. Rev. D",
    volume = "74",
    pages = "025005",
    year = "2006"
}

@article{Drukker:2009tz,
    author = "Drukker, Nadav and Morrison, David R. and Okuda, Takuya",
    title = "{Loop operators and S-duality from curves on Riemann surfaces}",
    eprint = "0907.2593",
    archivePrefix = "arXiv",
    primaryClass = "hep-th",
    reportNumber = "HU-EP-09-29, NSF-KITP-09-111",
    doi = "10.1088/1126-6708/2009/09/031",
    journal = "JHEP",
    volume = "09",
    pages = "031",
    year = "2009"
}

@article{Xie:2013vfa,
    author = "Xie, Dan",
    title = "{Aspects of line operators of class S theories}",
    eprint = "1312.3371",
    archivePrefix = "arXiv",
    primaryClass = "hep-th",
    month = "12",
    year = "2013"
}

@article{Tachikawa:2013hya,
    author = "Tachikawa, Yuji",
    title = "{On the 6d origin of discrete additional data of 4d gauge theories}",
    eprint = "1309.0697",
    archivePrefix = "arXiv",
    primaryClass = "hep-th",
    reportNumber = "IPMU-13-0163, UT-13-31",
    doi = "10.1007/JHEP05(2014)020",
    journal = "JHEP",
    volume = "05",
    pages = "020",
    year = "2014"
}

@article{Aharony:2013hda,
    author = "Aharony, Ofer and Seiberg, Nathan and Tachikawa, Yuji",
    title = "{Reading between the lines of four-dimensional gauge theories}",
    eprint = "1305.0318",
    archivePrefix = "arXiv",
    primaryClass = "hep-th",
    reportNumber = "WIS-03-13-APR-DPPA, WIS/03/13-APR-DPPA, UT-13-15, IPMU13-0081",
    doi = "10.1007/JHEP08(2013)115",
    journal = "JHEP",
    volume = "08",
    pages = "115",
    year = "2013"
}

@article{Coman:2015lna,
    author = "Coman, Ioana and Gabella, Maxime and Teschner, Joerg",
    title = "{Line operators in theories of class $\mathcal{S}$, quantized moduli space of flat connections, and Toda field theory}",
    eprint = "1505.05898",
    archivePrefix = "arXiv",
    primaryClass = "hep-th",
    reportNumber = "DESY-15-083",
    doi = "10.1007/JHEP10(2015)143",
    journal = "JHEP",
    volume = "10",
    pages = "143",
    year = "2015"
}

@article{Bhardwaj:2021pfz,
    author = "Bhardwaj, Lakshya and Hubner, Max and Schafer-Nameki, Sakura",
    title = "{1-form Symmetries of 4d N=2 Class S Theories}",
    eprint = "2102.01693",
    archivePrefix = "arXiv",
    primaryClass = "hep-th",
    doi = "10.21468/SciPostPhys.11.5.096",
    journal = "SciPost Phys.",
    volume = "11",
    pages = "096",
    year = "2021"
}

@article{Gaiotto:2009we,
    author = "Gaiotto, Davide",
    title = "{N=2 dualities}",
    eprint = "0904.2715",
    archivePrefix = "arXiv",
    primaryClass = "hep-th",
    doi = "10.1007/JHEP08(2012)034",
    journal = "JHEP",
    volume = "08",
    pages = "034",
    year = "2012"
}

@article{Braverman:2016pwk,
    author = "Braverman, Alexander and Finkelberg, Michael and Nakajima, Hiraku",
    title = "{Coulomb branches of $3d$ $\mathcal{N}=4$ quiver gauge theories and slices in the affine Grassmannian}",
    eprint = "1604.03625",
    archivePrefix = "arXiv",
    primaryClass = "math.RT",
    doi = "10.4310/ATMP.2019.v23.n1.a3",
    journal = "Adv. Theor. Math. Phys.",
    volume = "23",
    pages = "75--166",
    year = "2019"
}

@article{Braverman:2016wma,
    author = "Braverman, Alexander and Finkelberg, Michael and Nakajima, Hiraku",
    title = "{Towards a mathematical definition of Coulomb branches of $3$-dimensional $\mathcal{N} = 4$ gauge theories, II}",
    eprint = "1601.03586",
    archivePrefix = "arXiv",
    primaryClass = "math.RT",
    doi = "10.4310/ATMP.2018.v22.n5.a1",
    journal = "Adv. Theor. Math. Phys.",
    volume = "22",
    pages = "1071--1147",
    year = "2018"
}

@article{Nakajima:2015txa,
    author = "Nakajima, Hiraku",
    title = "{Towards a mathematical definition of Coulomb branches of $3$-dimensional $\mathcal{N}=4$ gauge theories, I}",
    eprint = "1503.03676",
    archivePrefix = "arXiv",
    primaryClass = "math-ph",
    doi = "10.4310/ATMP.2016.v20.n3.a4",
    journal = "Adv. Theor. Math. Phys.",
    volume = "20",
    pages = "595--669",
    year = "2016"
}

@article{Cremonesi:2013lqa,
    author = "Cremonesi, Stefano and Hanany, Amihay and Zaffaroni, Alberto",
    title = "{Monopole operators and Hilbert series of Coulomb branches of $3d$  $\mathcal{N} = 4$ gauge theories}",
    eprint = "1309.2657",
    archivePrefix = "arXiv",
    primaryClass = "hep-th",
    reportNumber = "IMPERIAL-TP-13-AH-03",
    doi = "10.1007/JHEP01(2014)005",
    journal = "JHEP",
    volume = "01",
    pages = "005",
    year = "2014"
}

@article{Ekholm:2024ceb,
    author = "Ekholm, Tobias and Longhi, Pietro and Nakamura, Lukas",
    title = "{The worldsheet skein D-module and basic curves on Lagrangian fillings of the Hopf link conormal}",
    eprint = "2407.09836",
    archivePrefix = "arXiv",
    primaryClass = "math.SG",
    month = "7",
    year = "2024"
}

@article{Gabella:2016zxu,
    author = "Gabella, Maxime",
    title = "{Quantum Holonomies from Spectral Networks and Framed BPS States}",
    eprint = "1603.05258",
    archivePrefix = "arXiv",
    primaryClass = "hep-th",
    doi = "10.1007/s00220-016-2729-1",
    journal = "Commun. Math. Phys.",
    volume = "351",
    number = "2",
    pages = "563--598",
    year = "2017"
}

@article{Neitzke:2021gxr,
    author = "Neitzke, Andrew and Yan, Fei",
    title = "{The quantum UV-IR map for line defects in $ \mathfrak{gl} $(3)-type class S theories}",
    eprint = "2112.03775",
    archivePrefix = "arXiv",
    primaryClass = "hep-th",
    doi = "10.1007/JHEP09(2022)081",
    journal = "JHEP",
    volume = "09",
    pages = "081",
    year = "2022"
}

@article{Neitzke:2020jik,
    author = "Neitzke, Andrew and Yan, Fei",
    title = "{$q$-nonabelianization for line defects}",
    eprint = "2002.08382",
    archivePrefix = "arXiv",
    primaryClass = "hep-th",
    doi = "10.1007/JHEP09(2020)153",
    journal = "JHEP",
    volume = "09",
    pages = "153",
    year = "2020"
}

@article{Ekholm:2025anq,
    author = "Ekholm, Tobias and Longhi, Pietro and Park, Sunghyuk and Shende, Vivek",
    title = "{Skein traces from curve counting}",
    eprint = "2510.19041",
    archivePrefix = "arXiv",
    primaryClass = "math.SG",
    month = "10",
    year = "2025"
}

@article{morton2021dahas,
  title={DAHAs and skein theory},
  author={Morton, Hugh R and Samuelson, Peter},
  journal={Communications in Mathematical Physics},
  volume={385},
  number={3},
  pages={1655--1693},
  year={2021},
  publisher={Springer}
}

@article{morton2017homflypt,
  title={THE HOMFLYPT SKEIN ALGEBRA OF THE TORUS AND THE ELLIPTIC HALL ALGEBRA},
  author={Morton, Hugh and Samuelson, Peter},
  journal={DUKE MATHEMATICAL JOURNAL},
  volume={166},
  number={5},
  pages={801--854},
  year={2017},
  publisher={Duke University Press}
}

@article{schiffmann2013elliptic,
  title={The elliptic Hall algebra and the K-theory of the Hilbert scheme of A 2},
  author={Schiffmann, Olivier and Vasserot, Eric},
  year={2013}
}

@book{cherednik2005double,
  title={Double affine Hecke algebras},
  author={Cherednik, Ivan},
  volume={319},
  year={2005},
  publisher={Cambridge University Press}
}

@ARTICLE{2017arXiv170806024J,
       author = {{Jordan}, David and {Vazirani}, Monica},
        title = "{The rectangular representation of the double affine Hecke algebra via elliptic Schur-Weyl duality}",
      journal = {arXiv e-prints},
         year = 2017,
        month = aug,
          eid = {arXiv:1708.06024},
        pages = {arXiv:1708.06024},
          doi = {10.48550/arXiv.1708.06024},
archivePrefix = {arXiv},
       eprint = {1708.06024},
 primaryClass = {math.RT},
       adsurl = {https://ui.adsabs.harvard.edu/abs/2017arXiv170806024J}
}

@book{macdonald1998symmetric,
  title={Symmetric functions and Hall polynomials},
  author={Macdonald, Ian Grant},
  year={1998},
  publisher={Oxford university press}
}

@article{Donagi:1995cf,
    author = "Donagi, Ron and Witten, Edward",
    title = "{Supersymmetric Yang-Mills theory and integrable systems}",
    eprint = "hep-th/9510101",
    archivePrefix = "arXiv",
    reportNumber = "IASSNS-HEP-95-78",
    doi = "10.1016/0550-3213(95)00609-5",
    journal = "Nucl. Phys. B",
    volume = "460",
    pages = "299--334",
    year = "1996"
}

@article{Gorsky:1995zq,
    author = "Gorsky, A. and Krichever, I. and Marshakov, A. and Mironov, A. and Morozov, A.",
    title = "{Integrability and Seiberg-Witten exact solution}",
    eprint = "hep-th/9505035",
    archivePrefix = "arXiv",
    reportNumber = "UUITP-6-95, ITEP-M3-95, FIAN-TD-9-95",
    doi = "10.1016/0370-2693(95)00723-X",
    journal = "Phys. Lett. B",
    volume = "355",
    pages = "466--474",
    year = "1995"
}

@article{Martinec:1995by,
    author = "Martinec, Emil J. and Warner, Nicholas P.",
    title = "{Integrable systems and supersymmetric gauge theory}",
    eprint = "hep-th/9509161",
    archivePrefix = "arXiv",
    reportNumber = "EFI-95-61, USC-95-025",
    doi = "10.1016/0550-3213(95)00588-9",
    journal = "Nucl. Phys. B",
    volume = "459",
    pages = "97--112",
    year = "1996"
}

@article{Gaiotto:2009gz,
    author = "Gaiotto, Davide and Maldacena, Juan",
    title = "{The Gravity duals of N=2 superconformal field theories}",
    eprint = "0904.4466",
    archivePrefix = "arXiv",
    primaryClass = "hep-th",
    doi = "10.1007/JHEP10(2012)189",
    journal = "JHEP",
    volume = "10",
    pages = "189",
    year = "2012"
}

@article{DHoker:1997hut,
    author = "D'Hoker, Eric and Phong, D. H.",
    title = "{Calogero-Moser systems in SU(N) Seiberg-Witten theory}",
    eprint = "hep-th/9709053",
    archivePrefix = "arXiv",
    reportNumber = "UCLA-97-TEP-18",
    doi = "10.1016/S0550-3213(97)00763-3",
    journal = "Nucl. Phys. B",
    volume = "513",
    pages = "405--444",
    year = "1998"
}

@article{Gang:2017ojg,
    author = "Gang, Dongmin and Longhi, Pietro and Yamazaki, Masahito",
    title = "{$\mathrm{S}$ duality and framed BPS states via BPS graphs}",
    eprint = "1711.04038",
    archivePrefix = "arXiv",
    primaryClass = "hep-th",
    reportNumber = "UUITP-42/17, IPMU-17-0152",
    doi = "10.4310/ATMP.2019.v23.n5.a4",
    journal = "Adv. Theor. Math. Phys.",
    volume = "23",
    number = "5",
    pages = "1361--1410",
    year = "2019"
}

@article{Hollands:2013qza,
    author = "Hollands, Lotte and Neitzke, Andrew",
    title = "{Spectral Networks and Fenchel{\textendash}Nielsen Coordinates}",
    eprint = "1312.2979",
    archivePrefix = "arXiv",
    primaryClass = "math.GT",
    doi = "10.1007/s11005-016-0842-x",
    journal = "Lett. Math. Phys.",
    volume = "106",
    number = "6",
    pages = "811--877",
    year = "2016"
}

@phdthesis{Longhi:2015ivt,
    author = "Longhi, Pietro",
    title = "{The structure of BPS spectra}",
    doi = "10.7282/T3FQ9ZMF",
    school = "Rutgers U., Piscataway",
    year = "2015"
}

@article{Gaiotto:2010okc,
    author = "Gaiotto, Davide and Moore, Gregory W. and Neitzke, Andrew",
    title = "{Four-dimensional wall-crossing via three-dimensional field theory}",
    eprint = "0807.4723",
    archivePrefix = "arXiv",
    primaryClass = "hep-th",
    doi = "10.1007/s00220-010-1071-2",
    journal = "Commun. Math. Phys.",
    volume = "299",
    pages = "163--224",
    year = "2010"
}

@article{Dimofte:2009bv,
    author = "Dimofte, Tudor and Gukov, Sergei",
    title = "{Refined, Motivic, and Quantum}",
    eprint = "0904.1420",
    archivePrefix = "arXiv",
    primaryClass = "hep-th",
    reportNumber = "CALT-68-2725",
    doi = "10.1007/s11005-009-0357-9",
    journal = "Lett. Math. Phys.",
    volume = "91",
    pages = "1",
    year = "2010"
}

@article{Dimofte:2009tm,
    author = "Dimofte, Tudor and Gukov, Sergei and Soibelman, Yan",
    title = "{Quantum Wall Crossing in N=2 Gauge Theories}",
    eprint = "0912.1346",
    archivePrefix = "arXiv",
    primaryClass = "hep-th",
    reportNumber = "CALT-68-2766",
    doi = "10.1007/s11005-010-0437-x",
    journal = "Lett. Math. Phys.",
    volume = "95",
    pages = "1--25",
    year = "2011"
}

@article{ruter2021novel,
  title={Novel wall-crossing behaviour in rank one $\mathcal{N}= 2^*$ gauge theory},
  author={R{\"u}ter, Philipp and Szabo, Richard J},
  journal={Journal of High Energy Physics},
  volume={2021},
  number={9},
  pages={1--48},
  year={2021},
  publisher={Springer}
}

@article{Longhi:2016wtv,
    author = "Longhi, Pietro",
    title = "{Wall-Crossing Invariants from Spectral Networks}",
    eprint = "1611.00150",
    archivePrefix = "arXiv",
    primaryClass = "hep-th",
    doi = "10.1007/s00023-017-0635-5",
    journal = "Annales Henri Poincare",
    volume = "19",
    number = "3",
    pages = "775--842",
    year = "2018"
}

@incollection{goldman2009trace,
  title={Trace coordinates on Fricke spaces of some simple hyperbolic surfaces},
  author={Goldman, William M},
  booktitle={Handbook of Teichm{\"u}ller Theory, Volume II},
  pages={611--684},
  year={2009},
  publisher={European Mathematical Society-EMS-Publishing House GmbH}
}

@article{Kontsevich:2008fj,
    author = "Kontsevich, Maxim and Soibelman, Yan",
    title = "{Stability structures, motivic Donaldson-Thomas invariants and cluster transformations}",
    eprint = "0811.2435",
    archivePrefix = "arXiv",
    primaryClass = "math.AG",
    month = "11",
    year = "2008"
}

@article{Banerjee:2022oed,
    author = "Banerjee, Sibasish and Longhi, Pietro and Romo, Mauricio",
    title = "{A-branes, Foliations and Localization}",
    eprint = "2201.12223",
    archivePrefix = "arXiv",
    primaryClass = "hep-th",
    doi = "10.1007/s00023-022-01231-8",
    journal = "Annales Henri Poincare",
    volume = "24",
    number = "4",
    pages = "1077--1136",
    year = "2023"
}

@article{Moore:2015szp,
    author = "Moore, Gregory W. and Royston, Andrew B. and Van den Bleeken, Dieter",
    title = "{Semiclassical framed BPS states}",
    eprint = "1512.08924",
    archivePrefix = "arXiv",
    primaryClass = "hep-th",
    reportNumber = "MI-TH-1603",
    doi = "10.1007/JHEP07(2016)071",
    journal = "JHEP",
    volume = "07",
    pages = "071",
    year = "2016"
}

@article{Moore:2015qyu,
    author = "Moore, Gregory W. and Royston, Andrew B. and Van den Bleeken, Dieter",
    editor = "Li, Si and Lian, Bong H. and Song, Wei and Yau, Shing-Tung",
    title = "{$L^2$-Kernels Of Dirac-Type Operators On Monopole Moduli Spaces}",
    eprint = "1512.08923",
    archivePrefix = "arXiv",
    primaryClass = "hep-th",
    reportNumber = "MI-TH-1602",
    journal = "Proc. Symp. Pure Math.",
    pages = "169--182",
    year = "2015"
}

@article{Brennan:2018ura,
    author = "Brennan, T. Daniel and Moore, Gregory W. and Royston, Andrew B.",
    title = "{Wall Crossing from Dirac Zeromodes}",
    eprint = "1805.08783",
    archivePrefix = "arXiv",
    primaryClass = "hep-th",
    reportNumber = "MI-TH-1884",
    doi = "10.1007/JHEP09(2018)038",
    journal = "JHEP",
    volume = "09",
    pages = "038",
    year = "2018"
}

@article{Alday:2009fs,
    author = "Alday, Luis F. and Gaiotto, Davide and Gukov, Sergei and Tachikawa, Yuji and Verlinde, Herman",
    title = "{Loop and surface operators in N=2 gauge theory and Liouville modular geometry}",
    eprint = "0909.0945",
    archivePrefix = "arXiv",
    primaryClass = "hep-th",
    reportNumber = "CALT-68-2741, PUPT-2311",
    doi = "10.1007/JHEP01(2010)113",
    journal = "JHEP",
    volume = "01",
    pages = "113",
    year = "2010"
}

@article{Drukker:2009id,
    author = "Drukker, Nadav and Gomis, Jaume and Okuda, Takuya and Teschner, Joerg",
    title = "{Gauge Theory Loop Operators and Liouville Theory}",
    eprint = "0909.1105",
    archivePrefix = "arXiv",
    primaryClass = "hep-th",
    reportNumber = "DESY-09-169, HU-EP-09-40, PI-STRINGS-144",
    doi = "10.1007/JHEP02(2010)057",
    journal = "JHEP",
    volume = "02",
    pages = "057",
    year = "2010"
}

@article{Hollands:2026rhz,
    author = "Hollands, Lotte and Murugesan, Subrabalan",
    title = "{Liouville Blocks from Spectral Networks}",
    eprint = "2604.25463",
    archivePrefix = "arXiv",
    primaryClass = "hep-th",
    month = "4",
    year = "2026"
}

@article{gunningham2024skeins,
  title={Skeins on tori},
  author={Gunningham, Sam and Jordan, David and Vazirani, Monica},
  journal={arXiv preprint arXiv:2409.05613},
  year={2024}
}

@article{gunningham2023finiteness,
  title={The finiteness conjecture for skein modules: S. Gunningham et al.},
  author={Gunningham, Sam and Jordan, David and Safronov, Pavel},
  journal={Inventiones mathematicae},
  volume={232},
  number={1},
  pages={301--363},
  year={2023},
  publisher={Springer}
}

@article{di2024ruijsenaars,
  title={Ruijsenaars wavefunctions as modular group matrix coefficients},
  author={Di Francesco, Philippe and Kedem, Rinat and Khoroshkin, Sergey and Schrader, Gus and Shapiro, Alexander},
  journal={Letters in mathematical physics},
  volume={114},
  number={6},
  pages={136},
  year={2024},
  publisher={Springer}
}

@article{Alim:2011ae,
    author = "Alim, Murad and Cecotti, Sergio and Cordova, Clay and Espahbodi, Sam and Rastogi, Ashwin and Vafa, Cumrun",
    title = "{BPS Quivers and Spectra of Complete N=2 Quantum Field Theories}",
    eprint = "1109.4941",
    archivePrefix = "arXiv",
    primaryClass = "hep-th",
    doi = "10.1007/s00220-013-1789-8",
    journal = "Commun. Math. Phys.",
    volume = "323",
    pages = "1185--1227",
    year = "2013"
}

@ARTICLE{2013arXiv1302.7030B,
       author = {{Bridgeland}, Tom and {Smith}, Ivan},
        title = "{Quadratic differentials as stability conditions}",
      journal = {arXiv e-prints},
         year = 2013,
        month = feb,
          eid = {arXiv:1302.7030},
        pages = {arXiv:1302.7030},
          doi = {10.48550/arXiv.1302.7030},
archivePrefix = {arXiv},
       eprint = {1302.7030},
 primaryClass = {math.AG},
       adsurl = {https://ui.adsabs.harvard.edu/abs/2013arXiv1302.7030B}
}

@article{Gabella:2017hpz,
    author = "Gabella, Maxime and Longhi, Pietro and Park, Chan Y. and Yamazaki, Masahito",
    title = "{BPS Graphs: From Spectral Networks to BPS Quivers}",
    eprint = "1704.04204",
    archivePrefix = "arXiv",
    primaryClass = "hep-th",
    reportNumber = "UUITP-11-17, IPMU17-0055",
    doi = "10.1007/JHEP07(2017)032",
    journal = "JHEP",
    volume = "07",
    pages = "032",
    year = "2017"
}

@article{frohman2000skein,
  title={Skein modules and the noncommutative torus},
  author={Frohman, Charles and Gelca, R{\u{a}}zvan},
  journal={Transactions of the American Mathematical Society},
  volume={352},
  number={10},
  pages={4877--4888},
  year={2000}
}

@article{terwilliger2011universal,
  title={The universal Askey-Wilson algebra},
  author={Terwilliger, Paul and others},
  journal={SIGMA. Symmetry, Integrability and Geometry: Methods and Applications},
  volume={7},
  pages={069},
  year={2011},
  publisher={SIGMA. Symmetry, Integrability and Geometry: Methods and Applications}
}

@article{koornwinder2008zhedanov,
  title={Zhedanov's algebra AW (3) and the double affine Hecke algebra in the rank one case. II. The spherical subalgebra},
  author={Koornwinder, Tom H and others},
  journal={SIGMA. Symmetry, Integrability and Geometry: Methods and Applications},
  volume={4},
  pages={052},
  year={2008},
  publisher={SIGMA. Symmetry, Integrability and Geometry: Methods and Applications}
}

@article{bullock2000multiplicative,
  title={Multiplicative structure of Kauffman bracket skein module quantizations},
  author={Bullock, Doug and Przytycki, J{\'o}zef H},
  journal={PROCEEDINGS-AMERICAN MATHEMATICAL SOCIETY},
  volume={128},
  number={3},
  pages={923--932},
  year={2000},
  publisher={American Mathematical Society}
}

@book{Gukov:2022gei,
    author = "Gukov, Sergei and Koroteev, Peter and Nawata, Satoshi and Pei, Du and Saberi, Ingmar",
    title = "{Branes and DAHA Representations}",
    eprint = "2206.03565",
    archivePrefix = "arXiv",
    primaryClass = "hep-th",
    reportNumber = "CALT-TH 2022-021",
    doi = "10.1007/978-3-031-28154-9",
    isbn = "978-3-031-28153-2, 978-3-031-28154-9",
    publisher = "Springer",
    series = "SpringerBriefs in Mathematical Physics",
    volume = "48",
    year = "2023"
}

@article{Bousseau:2020qgo,
    author = "Bousseau, Pierrick",
    title = "{Strong Positivity for the Skein Algebras of the 4-Punctured Sphere and of the 1-Punctured Torus}",
    eprint = "2009.02266",
    archivePrefix = "arXiv",
    primaryClass = "math.GT",
    doi = "10.1007/s00220-022-04512-9",
    journal = "Commun. Math. Phys.",
    volume = "398",
    number = "1",
    pages = "1--58",
    year = "2023"
}

@article{Allegretti:2024svn,
    author = "Allegretti, Dylan G. L. and Shan, Peng",
    title = "{Skein algebras and quantized Coulomb branches}",
    eprint = "2401.06737",
    archivePrefix = "arXiv",
    primaryClass = "math.RT",
    month = "1",
    year = "2024"
}

@article{Allegretti:2024idu,
    author = "Allegretti, Dylan G. L. and Shan, Peng",
    title = "{Dualities of $K$-theoretic Coulomb branches from a once-punctured torus}",
    eprint = "2411.17378",
    archivePrefix = "arXiv",
    primaryClass = "math.RT",
    month = "11",
    year = "2024"
}

@article{Arthamonov:2017oxw,
    author = "Arthamonov, S. and Shakirov, Sh.",
    title = "{Genus two generalization of \(A_1\) spherical DAHA}",
    eprint = "1704.02947",
    archivePrefix = "arXiv",
    primaryClass = "math.QA",
    doi = "10.1007/s00029-019-0447-1",
    journal = "Selecta",
    volume = "Math.",
    number = "25",
    pages = "1--29",
    year = "2019"
}

@article{Hikami:2019jaw,
    author = "Hikami, Kazuhiro",
    title = "{DAHA and skein algebra of surfaces: double-torus knots}",
    eprint = "1901.02743",
    archivePrefix = "arXiv",
    primaryClass = "math-ph",
    doi = "10.1007/s11005-019-01189-5",
    journal = "Lett. Math. Phys.",
    volume = "109",
    number = "10",
    pages = "2305--2358",
    year = "2019"
}

\end{document}